\documentclass[reqno,12pt]{amsart}
\usepackage{amssymb}
 \numberwithin{equation}{section}
\newcommand{\nc}{\newcommand}
\newcommand{\Z}{\mathbb Z}
\nc{\beq}{\begin{equation}}
\nc{\eeq}{\end{equation}} \nc{\beqa}{\begin{eqnarray}}
\nc{\eeqa}{\end{eqnarray}} \renewcommand{\a}{\alpha}
\renewcommand{\b}{\beta} \nc{\g}{\gamma} \renewcommand{\d}{\delta}
\nc{\al}{\dot{\alpha}} \nc{\be}{\dot{\beta}}
\nc{\ga}{\dot{\gamma}}
\nc{\de}{\dot{\delta}} \nc{\M}{\mathcal{M}} \nc{\A}{\mathcal{A}}
\nc{\B}{\mathcal{B}}
\nc{\D}{\mathcal{D}}
\nc{\Q}{\mathcal{Q}} \nc{\Qa}{\mathcal{\overline{Q}}}
\nc{\R}{\mathcal{R}}\renewcommand{\H}{\mathcal{H}}
\renewcommand{\P}{\mathcal{P}} \nc{\Pa}{\overline{\mathcal{P}}}
\renewcommand{\S}{\overline{S}}

\def \z{\underline{z}}
\def \y{\underline{y}}

\def \t{\underline{t}}
\def \s{\underline{s}}
\def \T{\theta}
\def \I{a}
\def \J{b}
\def \K{c}
\def \L{d}
\def \a{\alpha}
\def \b{\beta}

\def \d{\delta}

\newtheorem{theorem}{Theorem}
\newtheorem*{conjecture}{Conjecture}
\newtheorem{defn}{Definition}

\newcommand{\C}{\mathbb C}
\newcommand{\Te}{\mathbb T}
\newcommand{\m}{\omega}
\newcommand{\eg}{\Gamma}

\newcommand{\ee}{\end{equation}}
\newcommand{\ba}{\begin{eqnarray}}\newcommand{\ea}{\end{eqnarray}}
\newcommand{\lab}[1]{\label{#1}}

\begin{document}

 \begin{flushright} AEI-2009-106 \end{flushright}

\title[Elliptic hypergeometry of supersymmetric dualities]
{\bf Elliptic hypergeometry of \\ supersymmetric dualities}

\author{V. P. Spiridonov}
\address{Bogoliubov Laboratory of Theoretical Physics,
JINR, Dubna, Moscow Region 141980, Russia and Theory Division, INR RAS,
Moscow, Russia;
e-mail address: spiridon@theor.jinr.ru}

\author{G. S. Vartanov}
\address{Max-Planck-Institut f\"ur Gravitationsphysik, Albert-Einstein-Institut
14476 Golm, Germany; e-mail address: vartanov@aei.mpg.de}


\begin{abstract}
We give a full list of known $\mathcal{N}=1$ supersymmetric quantum field
theories related by the Seiberg electric-magnetic
duality conjectures for $SU(N), SP(2N)$ and $G_2$ gauge groups.
Many of the presented
dualities are new, not considered earlier in the literature. For all
these theories we construct superconformal indices and express them
in terms of elliptic hypergeometric integrals. This gives a
systematic extension of the related R\"omelsberger and Dolan-Osborn
results. Equality of indices in dual theories leads to various
identities for elliptic hypergeometric integrals. About half of them
were proven earlier, and another half represents new challenging
conjectures. In particular, we conjecture a dozen new elliptic
beta integrals on root systems extending the univariate elliptic
beta integral discovered by the first author.
\end{abstract}

\maketitle

\tableofcontents

\section{Introduction}

The main goal of this work consists in merging two fields of
recent active research in mathematical physics---the Seiberg
duality in supersymmetric field theories \cite{Seiberg0,Seiberg} and
the theory of elliptic hypergeometric functions \cite{spi:thesis}.
Seiberg duality is an electric-magnetic duality of certain four
dimensional quantum field theories with the symmetry group
$G_{st}\times G\times F$, where the superconformal group
$G_{st}=SU(2,2|1)$ describes properties of the space-time, $G$ is
a local gauge invariance group, and $F$ is a global symmetry
flavor group. Conjecturally, such theories are equivalent to each
other at their infrared fixed points, existence of which follows
from a deeply nontrivial nonperturbative dynamics
\cite{IN,Shifman:1995ua}.

The simplest topological characteristics of supersymmetric theories
is the Witten index \cite{Witten}. Its highly nontrivial superconformal
generalization was proposed recently by R\"omelsberger
\cite{Romelsberger1,Romelsberger2}
(for $\mathcal{N}=1$ theories) and Kinney et al \cite{Kinney} (for extended
supersymmetric theories). These superconformal indices describe the
structure of BPS states protected by one supercharge and its conjugate.
They can be considered as a kind of partition functions in the corresponding
Hilbert space. Starting from early work \cite{ska,Sun}, it is known that such partition
functions are described by matrix integrals over the classical groups.
The central conjecture of R\"omelsberger \cite{Romelsberger2} claims the
equality of superconformal indices in the Seiberg dual theories.
In an interesting work \cite{Dolan}, Dolan and Osborn have found an explicit
form of these indices for a number of theories and discovered that they
coincide with particular examples of the elliptic hypergeometric integrals
\cite{Spiridonov1}. This identification allowed them to prove R\"omelsberger's
conjecture for several dualities either on the basis of known exact computability
of these integrals or from the existence of non-trivial symmetry transformations
for them.

The general notion of elliptic hypergeometric integrals was introduced
by the first author in \cite{Spiridonov2,Spiridonov3}. First
example of such integrals discovered in \cite{Spiridonov2} has formed
a new class of exactly computable integrals of hypergeometric type
called elliptic beta integrals. Such a name was chosen because these integrals
can be considered as a top level generalization of the well-known
Euler beta integral \cite{aar}:
\begin{equation}
\int_0^1x^{\alpha-1}(1-x)^{\beta-1}dx=\frac{\Gamma(\alpha)\Gamma(\beta)}
{\Gamma(\alpha+\beta)}, \qquad \text{Re}\, \alpha,\; \text{Re}\, \beta>0,
\label{beta-int}\end{equation}
where $\Gamma(x)$ is the Euler gamma function.
Elliptic hypergeometric functions generalize known plain hypergeometric
functions and their $q$-analogues \cite{aar}. Moreover, their properties
have clarified the origins of many old notions of the hypergeometric
world \cite{spi:theta1}.
Limits of the elliptic hypergeometric
integrals (or of the elliptic hypergeometric series hidden
behind them) matched with the elliptic curve degenerations
brought to light new types of $q$-hypergeometric functions as well
\cite{rai:abelian,rai:limits} (see also \cite{BR}).

In the present work (which was initiated in August 2008 after the
first author has known \cite{Dolan}), we extend
systematically the R\"omelsberger and Dolan-Osborn results. More precisely,
we present a full list of known $\mathcal{N}=1$ superconformal
field theories related
by the duality conjecture for simple gauge groups $G=SU(N), SP(2N), G_2$.
For all of them we express superconformal
indices in terms of the elliptic hypergeometric integrals.
Using Seiberg dualities established earlier in the literature (see references below)
we come to a large number of identities for elliptic hypergeometric integrals.
About half of them were proven earlier, which yields a justification of the
corresponding dualities. A part of the emerging relations for indices
was described in \cite{Dolan}, and we prove equalities of superconformal
indices for many other dualities. Another half of the constructed identities
represents new challenging conjectures requiring rigorous mathematical proof.
We give indications how some of them can be proved with the help of
hypergeometric techniques.

Remarkably, from known relations for elliptic
hypergeometric integrals we find many new dualities not considered earlier in
the literature. Thus we describe both new elliptic hypergeometric identities
and new $\mathcal{N}=1$ supersymmetric theories obeying an electric-magnetic duality.
In particular, we conjecture more than ten new elliptic beta integrals on root
systems, which extend the univariate elliptic beta integral of \cite{Spiridonov2}.

Analyzing the general structure of all relations for integrals in this paper, we
formulate two universal conjectures. Namely, we argue that for the existence of a
non-trivial identity for an elliptic hypergeometric integral it is necessary
to have a so-called totally elliptic hypergeometric term
\cite{spi:theta1,spi:short,spi:cirm}. The second conjecture claims
that the same total ellipticity (and related modular invariance) is responsible
for the validity of ${}$ 't Hooft anomaly matching conditions \cite{Hooft},
which are fulfilled for all our dualities (the old and new ones).

A detailed consideration of the multiple duality phenomenon for
$G=SP(2N)$ gauge group case and a brief announcement
of other results of this work were given in paper \cite{SV}.
Our results were reported also at IV-th Sakharov Conference on Physics
(Lebedev Institute, Moscow, May 2009), Conformal Field Theory Workshop (Landau
Institute, Chernogolovka, June 2009),  XVI-th International Congress on
Mathematical Physics (Prague, August 2009), and about ten seminars at
different institutes. We thank the organizers of these
meetings and seminars for invitations and kind hospitality.

\section{General structure of the elliptic hypergeometric integrals}

We start our consideration from the description of the general structure of
elliptic hypergeometric integrals. For any $x\in\C$ and a base
$p\in\C, |p|<1$, we define the infinite product
$$
(x;p)_\infty=\prod_{j=0}^\infty (1-xp^j).
$$
Then the theta function is defined as
$$
\theta(x;p)=(x;p)_\infty (px^{-1};p)_\infty,
$$
where $x\in\C^*$. This function obeys the symmetry properties
$$
\theta(x^{-1};p)=\theta(px;p)=-x^{-1}\theta(x;p)
$$
and the addition law
$$
\theta(xw^{\pm 1},yz^{\pm 1};p) -\theta(xz^{\pm 1},yw^{\pm 1};p)
=yw^{-1}\theta(xy^{\pm 1},wz^{\pm 1};p),
$$
where $x,y,w,z\in\C^*$ and we use the convention
$$
\theta(x_1,\ldots,x_k;p):=\theta(x_1;p)\dots \theta(x_k;p),\qquad
\theta(tx^{\pm1};p):=\theta(tx,tx^{-1};p).
$$
The Jacobi triple product identity for the standard theta series yields
$$
\theta(x;p)=\frac{1}{(p;p)_\infty}\sum_{n\in \Z} p^{n(n-1)/2}(-x)^n.
$$

For arbitrary $q\in \C$ and $n\in\Z$, we introduce the
elliptic shifted factorials
$$
\theta(x;p;q)_n:=\begin{cases}
 \prod_{j=0}^{n-1}\theta(xq^j;p), & \text{for $n>0$}, \\
\prod_{j=1}^{-n}\theta(xq^{-j};p)^{-1}, & \text{for $n<0$},
\end{cases}
$$
with the normalization $\theta(x;p;q)_0=1$.
For $p=0$ we have $\theta(x;0)=1-x$ and
$$
\theta(x;0;q)_n = (x;q)_n =(1-x)(1-qx)\cdots (1-q^{n-1}x),
$$
the standard $q$-Pochhammer symbol \cite{aar}.

For arbitrary $m\in\Z$, we have the quasiperiodicity relations
\begin{eqnarray*}
&& \theta(p^mx;p)=(-x)^{-m}p^{-\frac{m(m-1)}{2}}\theta(x;p), \quad
\\ && \theta(p^mx;p;q)_k=(-x)^{-mk}q^{-\frac{mk(k-1)}{2}}
p^{-\frac{km(m-1)}{2}}\theta(x;p;q)_k, \quad
\\ &&
\theta(x;p;p^mq)_k=(-x)^{-\frac{mk(k-1)}{2}}q^{-\frac{mk(k-1)(2k-1)}{6}}
p^{-\frac{mk(k-1)}{4}(\frac{m(2k-1)}{3}-1)}\theta(x;p;q)_k.
\end{eqnarray*}

We relate bases $p,q$ and $r$ with three complex numbers $\omega_{1,2,3}\in\C$
in the following way
$$
q= e^{2\pi \textup{i}\frac{\omega_1}{\omega_2}}, \quad
p=e^{2\pi \textup{i}\frac{\omega_3}{\omega_2}}, \quad
r=e^{2\pi \textup{i}\frac{\omega_3}{\omega_1}}.
$$
Their ``$\tau\to -1/\tau$" modular transformed partners are
$$
\tilde q= e^{-2\pi \textup{i}\frac{\omega_2}{\omega_1}}, \quad
\tilde p=e^{-2\pi \textup{i}\frac{\omega_2}{\omega_3}},   \quad
\tilde r=e^{-2\pi \textup{i}\frac{\omega_1}{\omega_3}}.
$$
Modular parameters $\tau_1=\omega_1/\omega_2,$ $\tau_2=\omega_3/\omega_2$,
$\tau_3=\omega_3/\omega_1$ define three elliptic curves
constrained by the condition $\tau_3=\tau_2/\tau_1$.

Elliptic gamma functions are defined as appropriate meromorphic solutions
of the following finite difference equation
\beq
f(u+\omega_1)=\theta(e^{2\pi \textup{i} u/\omega_2};p)f(u), \quad u\in\C.
\lab{e-gamma-eq}\end{equation}
Its particular solution, called the standard elliptic gamma function,
has the form
\beq
f(u)=\Gamma(e^{2\pi \textup{i} u/\omega_2};p,q),\qquad
 \Gamma(z;p,q)
=\prod_{j,k=0}^\infty\frac{1-z^{-1}p^{j+1}q^{k+1}}{1-zp^{j}q^{k}},
\label{ell-gamma}\end{equation}
where $ |q|, |p|<1,  z\in\C^*$ (note that the equation itself does not
demand $|q|<1$).
For incommensurate $\omega_{1,2,3}$, it can be defined uniquely
as the meromorphic solution of (\ref{e-gamma-eq}) satisfying simultaneously
two more equations:
$$
f(u+\omega_2)=f(u), \qquad
f(u+\omega_3)=\theta(e^{2\pi \textup{i} u/\omega_2};q)f(u)
$$
and the normalization condition $f(\sum_{k=1}^3\omega_k/2)=1$.

The modified elliptic gamma function has the form
\beq
G(u;\mathbf{\omega})=
\Gamma(e^{2\pi \textup{i} \frac{u}{\omega_2}};p,q)
\Gamma(re^{-2\pi \textup{i} \frac{u}{\omega_1}};\tilde q,r).
\lab{unit-e-gamma}\end{equation}
It defines the unique simultaneous solution of equation (\ref{e-gamma-eq})
and two other equations:
$$
f(u+\omega_2) =\theta(e^{2\pi \textup{i} u/\omega_1};r) f(u),
\qquad   f(u+\omega_3) = \frac{\theta(e^{2\pi \textup{i} \frac{u}{\omega_2}};q)}
{\theta(e^{-2\pi \textup{i} \frac{u}{\omega_1}};\tilde q)}f(u)
$$
with the same normalization condition $f(\sum_{k=1}^3\omega_k/2)=1$.
Here the third equation can be simplified using the modular transformation
for theta functions
\begin{equation}
\theta(e^{-2\pi \textup{i} \frac{u}{\omega_1}};\tilde q)
=e^{\pi \textup{i}B_{2,2}(u|\m_1,\m_2)}\theta(e^{2\pi \textup{i} \frac{u}{\omega_2}};q),
\label{mod-tr}\end{equation}
where
$$
B_{2,2}(u|\m_1,\m_2)=\frac{1}{\m_1\m_2}\left(u^2-(\m_1+\m_2)u+\frac{\m_1^2+\m_2^2}{6}+
\frac{\m_1\m_2}{2}\right)
$$
is the second Bernoulli polynomial.
These statements are based on the Jacobi theorem stating
that if a meromorphic $\varphi(u)$ satisfies the system of equations
$$
\varphi(u+\omega_1)=\varphi(u+\omega_2)=\varphi(u+\omega_3)=\varphi(u)
$$
for $\omega_{1,2,3}\in\C$ linearly independent over $\Z$, then $\varphi(u)=const$.
The restricted values of bases $p^n=q^m$, $n,m\in\Z$ (or, equivalently,
$r^n={\tilde q}^m$ or ${\tilde r}^n={\tilde p}^m$) may be called the
torsion points, since the Jacobi theorem fails for them.

The function
\beq
G(u;\mathbf{\omega})
= e^{-\frac{\pi \textup{i}}{3}B_{3,3}(u|\mathbf{\omega})}
\Gamma(e^{-2\pi \textup{i} \frac{u}{\omega_3}};\tilde r,\tilde p),
\lab{mod-e-gamma}\ee
where  $|\tilde p|,|\tilde r|<1$ and
\begin{eqnarray*}
&& B_{3,3}(u|\m_1,\m_2,\m_3)=\frac{1}{\m_1\m_2\m_3}
\Biggl(u^3-\frac{3u^2}{2}\sum_{k=1}^3\m_k
\\ && \makebox[2em]{}
+\frac{u}{2}\left(\sum_{k=1}^3\m_k^2+3\sum_{1\leq j<k \leq 3}\m_j\m_k\right)
-\frac{1}{4}\left(\sum_{k=1}^3\m_k\right)\sum_{1\leq j<k\leq 3}\m_j\m_k\Biggr)
\end{eqnarray*}
is the third Bernoulli polynomial,
satisfies the same three equations and normalization
as function \eqref{unit-e-gamma}. Hence they coincide, and this fact yields
one of the $SL(3;\Z)$-group modular transformation laws
for the elliptic gamma function.
From the expression \eqref{mod-e-gamma} it is easy to see that
$G(u;\mathbf{\omega})$ is a meromorphic function of $u$ for
$\omega_1/\omega_2>0$, i.e. when $|q|=1$.
The region $|q|>1$ is similar to $|q|<1$, it can be reached by
a symmetry transformation.

The theory of generalized gamma functions was built by Barnes \cite{Barnes}.
Implicitly, the function $\Gamma(z;p,q)$ appeared in the free energy
per site of Baxter's eight vertex model \cite{bax:partition}
(see also \cite{tf} and \cite{fel-var:elliptic}) -- exactly in the form
which will be used below in the superconformal indices context.
A systematic investigation of its properties was launched
by Ruijsenaars in \cite{rui:first}.
Its $SL(3,\Z)$-group transformation properties were described
in \cite{fel-var:elliptic}.
The modified (``unit circle") elliptic gamma function $G(u;\mathbf{\omega})$
was introduced in \cite{Spiridonov3} (see also \cite{unit}).
Both elliptic gamma functions are directly related to the Barnes multiple
gamma function of the third order \cite{FR,Spiridonov3}.

In terms of the $ \Gamma(z;p,q)$-function one can write
$$
\theta(x;p;q)_n=\frac{\Gamma(xq^n;p,q)}{\Gamma(x;p,q)}.
$$
The short-hand conventions
\begin{eqnarray*}
&& \Gamma(t_1,\ldots,t_k;p,q):=\Gamma(t_1;p,q)\cdots\Gamma(t_k;p,q),\quad
\\ &&
\Gamma(tz^{\pm1}; p,q):=\Gamma(tz ;p,q)\Gamma(tz^{-1}; p,q),\quad
\Gamma(z^{\pm2}; p,q):=\Gamma(z^2; p,q)\Gamma(z^{-2}; p,q)
\end{eqnarray*}
are used below. The simplest properties of $ \Gamma(z;p,q)$ are:
\begin{itemize}
  \item the symmetry $\Gamma(z; p,q)=\Gamma(z; q,p)$,
  \item the finite difference equations of the first order
$$
\Gamma(qz;p,q)=\theta(z;p)\Gamma(z ;p,q),\quad \Gamma(pz;
p,q)=\theta(z;q)\Gamma(z;p,q),
$$
  \item the reflection equation
$$
\Gamma(z;p,q)\Gamma(pq/z;p,q)=1,
$$
  \item the duplication formula
$$
\Gamma(z^2;p,q)=\Gamma(z,-z,q^{1/2}z,-q^{1/2}z, p^{1/2}z,-p^{1/2}z,
(pq)^{1/2}z,-(pq)^{1/2}z;p,q),
$$
  \item the limiting relations
$$
\lim_{p\to0}\Gamma(z;p,q)=\frac{1}{(z;q)_\infty},\qquad
\lim_{z\to1}(1-z)\Gamma(z;p,q)=\frac{1}{(p;p)_\infty(q;q)_\infty}.
$$
\end{itemize}

\begin{defn}\cite{spi:theta1,spi:cirm}
A meromorphic function \, $f(x_1,\dots,\, x_n;p)$
of $n$ variables $x_j\in\C^*$, which together with $p\in\C$ compose
all indeterminates of this function, is called totally
$p$-elliptic if
$$
f(px_1,\ldots,x_n;p)=\ldots=f(x_1,\ldots,px_n;p)=f(x_1,\ldots,x_n;p),
$$
and if its divisor forms a nontrivial manifold of the maximal possible dimension.
\end{defn}

Note that here positions of zeros and poles of elliptic functions
are considered as indeterminates (i.e., they are not fixed in advance).

Consider $n$-dimensional integrals
\[
I(y_1,\ldots, y_m)=\int_{x\in D} \Delta(x_1,\dots,x_n;y_1,\ldots,y_m)
\prod_{j=1}^n\frac{dx_j}{x_j},
\]
where $D\subset \C^n$ is some domain of integration
and $\Delta(x_1,\dots,x_n;y_1,\ldots,
y_m)$ is a meromorphic function
of $x_j$ and $y_k$, where $y_k$ denote the ``external" parameters.

\begin{defn}\cite{Spiridonov3}
The integral $I(y_1,\ldots, y_m;p,q)$ is called the elliptic hypergeometric
integral if there
are two distinguished complex parameters $p$ and $q$ such that $I$'s
kernel $\Delta(x_1,\dots,$ $x_n;y_1,\ldots,y_m;p,q)$ satisfies the following
system of linear first order $q$-difference equations in the integration
variables $x_j$:
$$
\frac{\Delta(\ldots qx_j\ldots;y_1,\ldots,y_m;p,q)}
{\Delta(x_1,\dots,x_n;y_1,\ldots,y_m;p,q)}
= h_j(x_1,\dots,x_n;y_1,\ldots,y_m;q;p),
$$
where $h_j$ are some $p$-elliptic functions of the variables $x_j$,
$$
h_j(\ldots px_i\ldots;y_1,\ldots,y_m;q;p)=h_j(x_1,\dots,x_n;y_1,\ldots,y_m;q;p).
$$
The kernel $\Delta$ is called then the elliptic hypergeometric term,
and the functions $h_j(x_1,\dots,$ $x_n;y_1,\ldots,y_m;q;p)$---the
$q$-certificates.
\end{defn}

This definition is not the most general possible one of such kind,
but it is sufficient for the purposes of the present paper.
The elliptic hypergeometric series can be introduced as sums of residues of
particular sequences of poles of the elliptic hypergeometric integral
kernels \cite{die-spi:elliptic}
and, because of the convergence difficulties, they are less general
than the integrals.
In the one-dimensional case, $n=1$, the structure of admissible
elliptic hypergeometric
terms $\Delta$ can be described explicitly. Indeed, any meromorphic
$p$-elliptic function $f(px)=f(x)$ can be written in the form
$$
f_p(x)=z\prod_{k=1}^N\frac{\theta(t_kx;p)}{\theta(w_kx;p)},\qquad
\prod_{k=1}^Nt_k=\prod_{k=1}^Nw_k,
$$
where $z,t_1,\dots,t_N,w_1,\dots,w_N$ are arbitrary complex parameters.
The positive integer $N$ is called the order of the elliptic function, and
the linear constraint on parameters -- the balancing condition.
From the identity
$$
z=\frac{\theta(zx,px;p)}{\theta(pzx,x;p)}
$$
we see that $z$ is not a distinguished parameter -- it can be obtained
from $t_k$ and $w_k$ by appropriate reduction without spoiling the
balancing condition.  Therefore we set $z=1$.

Now, for $|q|<1$, the general solution of the equation $\Delta(qx)=f_p(x)\Delta(x)$ is
$$
\Delta(x)= \varphi(x)\prod_{k=1}^N\frac{\Gamma(t_kx; p,q)}{\Gamma(w_kx; p,q)},
\qquad \varphi(x)=\prod_{k=1}^M\frac{\theta(a_kx;q)}{\theta(b_kx;q)},\quad
\prod_{k=1}^Ma_k=\prod_{k=1}^Mb_k,
$$
where $\varphi(qx)=\varphi(x)$ is an arbitrary $q$-elliptic function.
However, since
$$
\varphi(x)=\prod_{k=1}^M\frac{\Gamma(pa_kx,b_kx; p,q)}{\Gamma(a_kx,pb_kx ;p,q)},
$$
one can obtain $\varphi(x)$ from ratios of $\Gamma$-functions
after replacing $N$ by $N+2M$ and appropriate specification
of the original parameters $t_k$ and $w_k$ with the balancing condition preserved.
Therefore we can drop $\varphi(x)$ function and find that the general elliptic
hypergeometric term for $n=1$ has the form:
$$
\Delta(x;t_1,\dots,t_N,w_1,\dots,w_N;p,q) =
\prod_{k=1}^N\frac{\Gamma(t_kx ;p,q)}{\Gamma(w_kx;p,q)},
\quad \prod_{k=1}^N\frac{t_k}{w_k}=1.
$$
This function is symmetric in $p$ and $q$, i.e. we can
repeat the above considerations with these parameters permuted.
Then, for incommensurate $p$ and $q$ (i.e., when $p^j\neq q^k,\, j,k\in\Z$),
the equations
$$
\Delta(qx)=f_p(x)\Delta(x),\qquad
\Delta(px)=f_q(x)\Delta(x)
$$
determine $\Delta(x)$ uniquely up to a multiplicative constant.

For $|q|>1$,
$$
\Delta(x;t_1,\dots,t_N,w_1,\dots,w_N;p,q) =
\prod_{k=1}^N\frac{\Gamma(q^{-1}w_kx ;p,q^{-1})}{\Gamma(q^{-1}t_kx;p,q^{-1})},
\quad \prod_{k=1}^N\frac{t_k}{w_k}=1.
$$
For $|q|=1$, the requirement of meromorphicity in $x$ is too strong.
To define integrals in this case one has to use the modified elliptic
gamma function $G(u;\mathbf{\omega})$,
or modular transformations, which we skip for brevity.

In analogy with the series case considered in \cite{spi:theta1}, it is natural
to extend the notion of total ellipticity to elliptic hypergeometric terms
entering integrals \cite{Spiridonov3}.

\begin{defn}
An elliptic hypergeometric integral
\[
I(y_1,\ldots, y_m;p,q)=\int_{x\in D} \Delta(x_1,\dots,x_n;y_1,\ldots,y_m;p,q)
\prod_{j=1}^n\frac{dx_j}{x_j}
\]
is called totally elliptic if all its kernel's $q$-certificates
$h_j(x_1,\dots,x_n;y_1,\ldots,y_m;q;p)$, $j=1,\dots, n+m$,
are totally elliptic functions, i.e. they are $p$-elliptic in all variables
$x_1,\ldots, x_n, y_1,\ldots,y_m$ and $q$. In particular,
$$
h_j(x_1,\dots,x_n;y_1,\ldots,y_m;pq;p)=h_j(x_1,\dots,x_n;y_1,\ldots,y_m;q;p).
$$
\end{defn}

\begin{theorem}[Rains, Spiridonov, 2004]
Given $\Z^n\to \Z$ maps $\epsilon(m^{(a)})=\epsilon(m^{(a)}_1,\dots,m^{(a)}_n)$, $a=1,\dots,M$,
with finite support, define the meromorphic function
\begin{equation}
\Delta(x_1,\dots,x_n;p,q)
= \prod_{a=1}^M \Gamma(x_1^{m_1^{(a)}}x_2^{m_2^{(a)}}
\dots x_n^{m_n^{(a)}}; p,q)^{\epsilon(m^{(a)})}.
\label{e-term}\end{equation}
Suppose $\Delta$ is a totally elliptic hypergeometric term, i.e. its
$q$-certificates are $p$-elliptic functions of $q$ and $x_1,\ldots,x_n$.
Then these certificates are also modular invariant.
\end{theorem}
The proof is elementary. The $q$-certificates have the explicit form
$$
h_i(x;q;p)=\frac{\Delta(\dots qx_i\dots ;p,q)}{\Delta(x_1,\dots,x_n;p,q)}
=\prod_{a=1}^M\theta(x^{m^{(a)}};p;q)_{m_i^{(a)}}^{\epsilon(m^{(a)})}.
$$
The conditions for $h_i$ to be elliptic in $x_j$ yield the constraints
\begin{eqnarray} \label{con1}
&& \sum_{a=1}^M \epsilon(m^{(a)}) m_i^{(a)} m_j^{(a)} m_k^{(a)} = 0,\\
&& \sum_{a=1}^M \epsilon(m^{(a)}) m_i^{(a)} m_j^{(a)}     = 0
\label{con2}\end{eqnarray}
for $1\le i,j,k\le n$.
The conditions of ellipticity in $q$ add one more constraint
\begin{equation} \label{con3}
\sum_{a=1}^M \epsilon(m^{(a)}) m_i^{(a)} =0.
\end{equation}
The latter equation guarantees that $h_i$ has an equal number of theta functions
in its numerator and denominator. The modular invariance of $h_i$ follows
then automatically from the transformation property (\ref{mod-tr}).
Such a direct relation between total ellipticity and modularity
was conjectured to be true in general in \cite{spi:theta1}.

The simplest known nontrivial totally elliptic hypergeometric term corresponds
to $n=6,\, M=29$ and has the form \cite{spi:short}
$$
\Delta(x;t_1,\dots,t_5;p,q)=\frac{\prod_{j=1}^5
\Gamma (t_jx^{\pm1},t_j^{-1}\prod_{i=1}^5t_i; p,q)}
{\Gamma (x^{\pm2},\prod_{i=1}^5t_i\,x^{\pm1}; p,q)
\prod_{1\le i < j\le 5}\Gamma (t_it_j; p,q)}.
$$

\begin{theorem}\cite{Spiridonov2} {\bf Elliptic beta integral.}
For $|p|, |q|, |t_j|<1$, $|\prod_{j=1}^5t_j|<|pq|$,
one has
\begin{equation}
\frac{(p;p)_\infty (q;q)_\infty}{4\pi \textup{i}}
\int_{\Te}\Delta(x;t_1,\dots,t_5;p,q) \frac{dx}{x}=1,
\label{ell-beta} \end{equation}
where $\Te$ is the unit circle with positive orientation.
\end{theorem}

The Euler beta integral evaluation formula (\ref{beta-int}) lies at the bottom
of this identity. On the corresponding degeneration road
one finds many interesting integrals, including the Rahman and Askey-Wilson
$q$-beta integrals \cite{aar}. Formula (\ref{ell-beta}) served as an entry ticket
to the large class of new exactly
computable integrals discussed in \cite{die-spi:elliptic,die-spi:selberg,unit,
Rains,Spiridonov3,spi-war:inversions}, which is essentially extended by
the conjectures presented in this paper.
In \cite{Spiridonov3,spi:thesis,spi:cs} the elliptic beta integral
was generalized to an elliptic analogue of the Gauss
hypergeometric function obeying many classical properties.
For a survey of this function and its generalizations to
higher order elliptic hypergeometric functions and multiple
integrals on root systems, see \cite{Spiridonov1}.

Two totally elliptic hypergeometric terms
associated with the elliptic beta integrals of type I on root systems
$BC_n$ \cite{die-spi:selberg} and $A_n$ \cite{Spiridonov3}
were constructed in \cite{spi:short}. One more similar example for the
root system $A_n$ was built in \cite{spi-war:inversions}.
Some time ago, using the combination of techniques
introduced in \cite{spi:short} and \cite{RS}, the first author has
further generalized the former two terms to an arbitrary number of parameters
\cite{spi:cirm}. For instance, define the kernel
$$
\Delta_n(z,t;p,q)=\prod_{1\leq i<j\leq n} \frac{1}{\eg(z_i^{\pm 1}z_j^{\pm 1};p,q)}
\prod_{j=1}^n\frac{\prod_{i=1}^{2n+2m+4}\eg(t_iz_j^{\pm 1};p,q)}
{\eg(z_j^{\pm 2};p,q)}
$$
and the type I $BC_n$-elliptic hypergeometric integral:
$$
I_n^{(m)}(t_1,\ldots,t_{2n+2m+4})=
\frac{(p;p)_\infty^n(q;q)_\infty^n}{2^n n!(2\pi \textup{i})^n}
\int_{\Te^n}\Delta_n(z,t;p,q)\prod_{j=1}^n\frac{dz_j}{z_j},
$$
where $|t_j|<1$ and $\prod_{j=1}^{2n+2m+4}t_j=(pq)^{m+1}.$

\begin{theorem}\cite{Rains} For $|pq|^{1/2}<|t_j|<1$,
the integrals $I^{(m)}_n$ satisfy the relation
\begin{equation}
I_n^{(m)}(t_1,\ldots,t_{2n+2m+4})
=\prod_{1\leq r<s\leq 2n+2m+4}\eg(t_rt_s;p,q)\;
I_m^{(n)}\left(\frac{\sqrt{pq}}{t_1},\ldots,\frac{\sqrt{pq}}{t_{2n+2m+4}}\right).
\label{ra}\end{equation}
\end{theorem}
This is an elliptic analogue of the symmetry transformation for some
plain hypergeometric integrals established by Dixon in \cite{Dixon}.

\begin{theorem}\cite{spi:cirm}
The ratio
$$
\rho(z,y;t;p,q)=\prod_{1\leq r<s\leq 2n+2m+4}\eg(t_rt_s;p,q)^{-1}
\frac{\Delta_n(z;t;p,q)}{\Delta_m(y/\sqrt{pq};\sqrt{pq}/t;p,q)}
$$
is the totally elliptic hypergeometric term. That is all ratios
$\rho(\dots, qv,\dots)/\rho(\dots,v,\dots)$ for $v\in \{z_1,\dots,$
$z_n,y_1,\dots,y_m,$ $t_1,\dots,t_{2n+2m+4}\}$ are $p$-elliptic
functions of all variables $z_i,y_k,t_l$, and $q$.
\end{theorem}

The term $\rho(z,y;t;p,q)$ contains elliptic gamma functions with
non-removable integer powers of $pq$ in the argument.
Therefore the ansatz (\ref{e-term}) does not cover all interesting totally
elliptic hypergeometric terms. As we shall show below,
there are also examples of terms depending on fractional powers of $pq$.
For them the total ellipticity condition is slightly modified: it is
necessary to consider dilations of the parameter $q$ by appropriate powers
of $p$. Introducing the variable $x_0=(pq)^{1/K},\, K=1,2,\dots$
and adding to the arguments of elliptic gamma functions
in (\ref{e-term}) the multipliers $x_0^{m_0^{(a)}}$,
it is not difficult to find the general form of
constraints on integers $m_j^{(a)}$ and $\epsilon(m^{(a)})$ guaranteeing
total ellipticity (with special $p^K$-ellipticity condition for the variable $q$).
However, these constraints look essentially less beautiful than the Diophantine
equations described above. Moreover, at the moment it is not clear which
part of the modular transformation group survives because of the
presence of fractional parts of modular variables in the arguments of
respective elliptic functions-certificates.

In the present work, we have checked that {\em all} nontrivial
relations for elliptic hypergeometric integrals described below
define totally elliptic hypergeometric terms through the ratios of
the corresponding integral kernels. Namely, this property was verified
for the equalities of superconformal indices in
\begin{itemize}
  \item the initial Seiberg duality
(\ref{IE-seiberg}) and (\ref{IM-seiberg}); $SP$-groups duality (\ref{Sp-el}) and
(\ref{Sp-m}); \cite{spi:cirm}
  \item multiple dualities for
$SP(2N)$ gauge group (\ref{SP2N1}), (\ref{IM1}), (\ref{IM2}) and
(\ref{SP2N3});
  \item new duality for $SP(2N)$ group (\ref{SPn1}) and (\ref{SPn2});
  \item multiple duality for $SU(2N)$ gauge group
(\ref{SU2N_1}), (\ref{SU2N_2}), (\ref{SU2N_3}) and (\ref{SU2N_4});
  \item KS type dualities for unitary groups (\ref{intKSelE})
and (\ref{intKSelM}) (see Appendix D for a detailed consideration of this
case); (\ref{KS2adj1}) and (\ref{KS2adj2}); (\ref{KSg1_1}) and
  (\ref{KSg1_2});  (\ref{KSg2_1}) and (\ref{KSg2_2}); (\ref{KSg3_1}) and
  (\ref{KSg3_2});  (\ref{Br1_1}) and (\ref{Br1_2});  (\ref{Br2_1}) and (\ref{Br2_2});  (\ref{Br3_1}) and (\ref{Br3_2});
  \item KS type dualities for symplectic groups  (\ref{KSsp1_1}) and
(\ref{KSsp1_2}); (\ref{KSsp2_1}) and (\ref{KSsp2_2}); (\ref{KSsp3_1})
and (\ref{KSsp3_2}); (\ref{KSsp4_1}) and (\ref{KSsp4_2});
  \item confinement for $SU(N)$ group theories (\ref{s1_1}) and (\ref{s1_2});
(\ref{s2_1}) and (\ref{s2_2}); (\ref{s3_1}) and (\ref{s3_2});
(\ref{s4_1}) and (\ref{s4_2}); (\ref{s5_1}) and (\ref{s5_2});
(\ref{s6_1}) and (\ref{s6_2});
(\ref{s8_1}) and (\ref{s8_2}); (\ref{s9_1}) and (\ref{s9_2}); (\ref{s10_1}) and
  (\ref{s10_2});  (\ref{s11_1}) and (\ref{s11_2}); (\ref{s12_1}) and (\ref{s12_2});
  \item confinement for $SP(2N)$ group theories (\ref{sp1_1}) and
  (\ref{sp1_2});  (\ref{sp2_1}) and (\ref{sp2_2}); (\ref{sp3_1}) and (\ref{sp3_2});
  \item dualities for the $G_2$ gauge group (\ref{G2E}) and (\ref{G2M});
  (\ref{G2S})  and (\ref{G2SM}).
\end{itemize}

Our auxiliary file with the details of these verifications is
bigger than the present paper. During this work we have found
a number of mistakes in the description of hypercharges
of the fields in some original papers.
On the basis of this large amount of computations, we put
forward the following conjecture.

\begin{conjecture}\label{conjecture1}
The condition of total ellipticity
for the elliptic hypergeometric terms is necessary
for the existence of the exact integration formulas for elliptic beta
integrals or of the nontrivial Weyl group symmetry transformations
for the elliptic hypergeometric integrals.
\end{conjecture}

It is known that behind each elliptic hypergeometric integral there
is a terminating elliptic hypergeometric series appearing from the
residue calculus for restricted values of parameters \cite{die-spi:elliptic}.
The above conjecture has a natural meaning in terms of such series---it
simply demands that the summation or transformation identities for them
involve ratios of Jacobi forms with appropriate quasiperiodicity and modularity
properties in the sense of Eichler and Zagier \cite{eic-zag:theory}.
Already this fact is sufficient (when there are no fractional powers of
$pq$) for the confirmation of the series
identities to rather high powers of small $\log q$ expansions
\cite{die-spi:elliptic}.

For a given elliptic hypergeometric integral there may exist more than
one totally elliptic hypergeometric term.
For the terms associated with elliptic beta integrals discussed in
\cite{spi:short,spi-war:inversions}
there existed a complementary difference equation with the totally elliptic
function coefficients. During our work we have found examples of
fake terms which do not lead to identities (or fake anomaly matching
conditions without real duality).
Therefore analysis of the sufficiency condition for existence of
nontrivial identities looks much more neat -- it should address
the non-uniqueness questions and the list of additional admissible
technical tools. Sometimes the ratio of a given elliptic hypergeometric
integral kernel to
itself with different integration variables yields the totally elliptic
hypergeometric term. It may happen that for a fixed set of parameters,
it is sufficient to have totally elliptic hypergeometric terms
of a more complicated nature than the latter one, and then at least one of
them will lead to a nontrivial relation between integrals.

\section{Superconformal index}

\subsection{$\mathcal{N}=1$ superconformal algebra}

In $4$ dimensional space-time the
\textit{conformal} algebra $SO(4,2)$ is formed by the generators of
translations $P_a$, special conformal transformations $K_a$,
$SO(3,1)$ Lorentz group rotations, $M_{ab}=-M_{ba}$,  and the
dilations $H$. The commutation relations have the form  \cite{DO-ann}
 \beqa \label{Alg} &&
[M_{ab},P_c] = i (\eta_{ac}P_b-\eta_{bc}P_a), \ \ \
[M_{ab},K_c]=i(\eta_{ac}K_b-\eta_{bc}K_a), \nonumber \\ &&
[M_{ab},M_{cd}]=i(\eta_{ac}M_{bd}-\eta_{bc}M_{ad}-\eta_{ad}M_{bc}+\eta_{bd}M_{ac}),
\\ \nonumber && [H,P_a]=P_a, \ \ \ [H,K_a]=-K_a, \ \ \
[K_a,P_b]=-2iM_{ab}-2\eta_{ab}H,\eeqa where
$\eta_{ab}=diag(-1,1,1,1)$ and all indices take values $a=0,1,2,3$.
In terms of the matrix
\beq
M_{AB} =        \left(
                \begin{array}{ccc}
                M_{ab} & -\frac 12 (P_a-K_a) & -\frac 12 (P_a+K_a) \\
                \frac 12 (P_b-K_b) & 0 & iH \\
                \frac 12 (P_b+K_b) & -iH & 0
                \end{array}
                \right),
\eeq
where $A,B=0,\dots,5$, relations (\ref{Alg}) are
rewritten in the simpler form \cite{DO-ann}
\beq
[M_{AB},M_{CD}]=i(\eta_{AC}M_{BD}-\eta_{BC}M_{AD}-\eta_{AD}M_{BC}
+\eta_{BD}M_{AC})
\eeq
with $\eta_{AB}=diag(-1,1,1,1,1,-1)$.

In the spinorial basis one defines
\beqa && P_{\alpha \al} =
(\sigma^a)_{\alpha \al}P_a, \ \ \ \ \ \ \ \ \ \ \ \ \ \ K^{\al
\alpha}=(\overline{\sigma}^a)^{\al \alpha}K_a, \nonumber \\ &&
M_{\alpha}^{\beta}=-\frac i4
(\sigma^a\overline{\sigma}^b)_{\alpha}^{\beta}M_{ab}, \ \ \ \ \
\overline{M}^{\al}_{\be}=-\frac i4
(\overline{\sigma}^a\sigma^b)^{\al}_{\be}M_{ab}, \eeqa
where $\alpha,\al, \beta, \be=1,2$,
$$
\sigma^a \ = \ (I,\sigma^i), \ \ \ \ \ \overline{\sigma}^a \ = \
(I,-\sigma^i)
$$
and
$\sigma^i$ are the usual Pauli matrices
\begin{equation}
\sigma^1 \ = \ \left(
                        \begin{array}{cc}
                          0 & 1 \\
                          1 & 0 \\
                        \end{array}
                      \right), \ \ \
\sigma^2 \ = \ \left(
                        \begin{array}{cc}
                          0 & -i \\
                          i & 0 \\
                        \end{array}
                      \right), \ \ \
\sigma^3 \ = \ \left(
                        \begin{array}{cc}
                          1 & 0 \\
                          0 & -1 \\
                        \end{array}
                      \right).
\end{equation}

Using the standard angular momentum generators, we set
$$
M_{\alpha}^{\ \beta}=\left(\begin{array}{cc}
  J_3 & J_+ \\
  J_- & -J_3
\end{array}\right), \ \ \ \overline{M}_{\ \dot{\beta}}^{\dot{\alpha}}
=\left(\begin{array}{cc}
  \overline{J}_3 & \overline{J}_+ \\
  \overline{J}_- & -\overline{J}_3
\end{array}\right),
$$
with $[J_3,J_{\pm}]=\pm J_{\pm}$, $[J_+,J_-]=2J_3$ and
similar relations for $\overline{J}_{\pm}, \overline{J}_3.$
Then the tensor $M_{ab}$ is expressed through these operators as
{\tiny
$$
M_{ab}=\left(\begin{array}{cccc}
  0 & \frac i2 (\overline{J}_+ + \overline{J}_- - J_+ - J_- ) & \frac 12 (J_+ + \overline{J}_- - \overline{J}_+ - J_-) & i (\overline{J}_3 - J_3)  \\
  -\frac i2 (\overline{J}_+ + \overline{J}_- - J_+ - J_- ) & 0 & -(J_3 + \overline{J}_3) & \frac i2 (J_+ + \overline{J}_+ - J_- - \overline{J}_-) \\
  -\frac 12 (J_+ + \overline{J}_- - \overline{J}_+ - J_-) & (J_3 + \overline{J}_3) & 0 & -\frac 12 (J_+ + J_- + \overline{J}_+ + \overline{J}_- ) \\
  -i (\overline{J}_3 - J_3) & -\frac i2 (J_+ + \overline{J}_+ - J_- - \overline{J}_-)
  & \frac 12 (J_+ + J_- + \overline{J}_+ + \overline{J}_- ) & 0
\end{array}\right).
$$
}

The conformal algebra (\ref{Alg}) can be rewritten now as
\beqa\label{com1} && [M_{\alpha}^{\ \beta},M_{\gamma}^{\ \delta}] =
\delta_{\gamma}^{\beta}M_{\alpha}^{\
\delta}-\delta_{\alpha}^{\delta}M_{\gamma}^{\ \beta}, \ \ \ \ \ \
[\overline{M}^{\al}_{\ \be},\overline{M}^{\ga}_{\
\de}]=\delta_{\be}^{\ga}\overline{M}^{\al}_{\
\de}-\delta_{\de}^{\al}\overline{M}^{\ga}_{\ \be}, \nonumber \\
&& [M_{\alpha}^{\ \beta},P_{\gamma
\de}]=\delta_{\gamma}^{\beta}P_{\alpha \de}-\frac 12
\delta_{\alpha}^{\beta}P_{\gamma \de}, \ \ \ \ \ \ \ \
[\overline{M}^{\al}_{\ \be},P_{\gamma
\de}]=-\delta_{\de}^{\al}P_{\gamma \be} + \frac 12
\delta_{\be}^{\al}P_{\gamma\de},
 \nonumber \\ &&
[M_{\alpha}^{\
\beta},K^{\ga \delta}]=-\delta_{\alpha}^{\delta}K^{\ga \beta} +
\frac 12 \delta_{\alpha}^{\beta}K^{\ga \delta}, \ \ \
[\overline{M}^{\al}_{\ \be},K^{\ga \delta}]=\delta_{\be}^{\ga}K^{\al
\delta} - \frac 12 \delta_{\be}^{\al}K^{\ga \delta},
\nonumber
\\ && [M_{\alpha}^{\ \beta},H]=0, \ \ \ \ \ \ \ \ \ \ \ \ \
 \ \ \ \ \ \ \ \ \ \ \ \ \ \ \ [\overline{M}^{\al}_{\ \be},H]=0,
\nonumber \\ &&
[H,P_{\alpha \be}]=P_{\alpha \be}, \ \ \ \ \ \ \ \ \ \ \ \ \ \ \ \ \
\ \ \ \ \ \ \ \ \ [H,K^{\al \beta}]=-K^{\al \beta},
\eeqa
$$
[P_{\alpha \be},K^{\ga \delta}] = 4 \left( \delta_{\be}^{\ga} M_{\alpha}^{\delta} - \delta_{\alpha}^{\delta}
\overline{M}_{\be}^{\ga} + \delta_{\be}^{\ga} \delta_{\alpha}^{\delta} H \right).
$$

$SO(4,2)$ (or $SU(2,2)$) algebra can be extended by adding
supercharges $Q_{\alpha},\overline{Q}_{\al}$ and their
superconformal partners $S^{\alpha},\overline{S}^{\al}$.
Supercharges satisfy the anticommutator relations \cite{DO-ann,Terning}
\beq\label{com2} \{Q_{\alpha},\overline{Q}_{\al}\}=2P_{\alpha \al},
\ \ \ \ \
\{Q_{\alpha},Q_{\beta}\}=\{\overline{Q}_{\al},\overline{Q}_{\be}\}=0,
\eeq while their superconformal partners obey \beq\label{com3}
\{\overline{S}^{\al},S^{\alpha}\}=2K^{\al \alpha}, \ \ \ \ \
\{\overline{S}^{\al},\overline{S}^{\be}\}=\{S^{\alpha},S^{\beta}\}=0.
\eeq

The cross-anti-commutators of $Q_{\alpha}$ and $S_{\alpha}$ have
the form \beq\label{com4} \{Q_{\alpha},\overline{S}^{\al}\}=0, \ \ \
\ \ \ \ \{ S^{\alpha},\overline{Q}_{\al} \}=0,\eeq while \beqa \{
Q_{\alpha},S^{\beta} \}&=&4 \left( M_{\alpha}^{\ \beta}+\frac 12
\delta_{\alpha}^{\beta}H + \frac 34 \delta_{\alpha}^{\beta} R
\right),\nonumber \\ \{ \overline{S}^{\al},\overline{Q}_{\be} \}&=&4
\left( \overline{M}^{\al}_{\ \be}-\frac 12 \delta^{\al}_{\be}H +
\frac 34 \delta^{\al}_{\be} R \right),\eeqa
where $R$ is the $R$-charge generating $U(1)_R$-symmetry group.

The bosonic and fermionic generators cross-commute as
\beqa\label{com5} && [M_{\alpha}^{\
\beta},Q_{\gamma}]=\delta_{\gamma}^{\beta}Q_{\alpha}-\frac 12
\d_{\a}^{\b} Q_{\g}, \ \ \ \ \ [M_{\a}^{\ \b},\overline{Q}_{\ga}]=0,
\nonumber \\ && [M_{\alpha}^{\
\beta},S^{\gamma}]=-\delta^{\gamma}_{\a} S^{\b}+\frac 12
\d_{\a}^{\b} S^{\g}, \ \ \ \ \ [M_{\a}^{\ \b},\overline{S}^{\ga}]=0,
\nonumber \\ && [\overline{M}^{\al}_{\ \b},Q_{\g}]=0, \ \ \ \ \
[\overline{M}^{\al}_{\ \be},\overline{Q}_{\ga}]=-\delta_{\ga}^{\al}
\overline{Q}_{\be} + \frac 12 \d^{\al}_{\be} \overline{Q}_{\ga},
\nonumber \\ && [\overline{M}^{\al}_{\ \b},S^{\g}]=0, \ \ \ \ \
[\overline{M}^{\al}_{\ \be},\overline{S}^{\ga}]=\delta^{\ga}_{\be}
\overline{S}^{\al} - \frac 12 \d^{\al}_{\be} \overline{S}^{\ga},
\nonumber \\ && [P_{\a \be},S^{\g}]=\d_{\a}^{\g} \overline{Q}_{\be}, \ \ \ \ \ \ \ \ \ [P_{\a \be},\S^{\ga}]=\d_{\be}^{\ga} Q_{\a},  \nonumber \\
&& [K^{\al \b},Q_{\g}]=\d_{\g}^{\b} \overline{S}^{\al},\ \ \ \ \ \ \ \ \ [K^{\al \b},\overline{Q}_{\ga}]=\d_{\ga}^{\al} S^{\b},\nonumber \\
&& [H,Q_{\a}]=\frac 12 Q_{\a}, \ \ \ \ \ \ \ \ \ \ \ \
[H,\overline{Q}_{\al}]=\frac 12 \overline{Q}_{\al}, \nonumber \\
&& [H,S^{\a}]=-\frac 12 S^{\a}, \ \ \ \ \ \ \ \ \ \ \ \
[H,\overline{S}^{\al}]=-\frac 12 \overline{S}^{\al}. \eeqa

The $R$-charge commutes with all bosonic generators and has
non-trivial commutators only with the supercharges and their
superconformal partners
\beqa\label{com6} && [R,Q_{\a}]=-Q_{\a}, \ \
\ \ \ \ \ \ \ \ \ \
[R,\overline{Q}_{\al}]= \overline{Q}_{\al}, \nonumber \\
&& [R,S^{\a}]= S^{\a}, \ \ \ \ \ \ \ \ \ \ \ \
[R,\overline{S}^{\al}]=- \overline{S}^{\al}. \eeqa

To simplify the shape of the $\mathcal{N}=1$ superconformal algebra
one introduces the notations
\beq
\mathcal{M}_{\mathcal{A}}^{\ \mathcal{B}} = \left(
                                            \begin{array}{cc}
                                              M_{\a}^{\ \b}+\frac 12 \d_{\a}^{\b}H & \frac 12 P_{\a\be} \\
                                              \frac 12 K^{\al \b} & \overline{M}^{\al}_{\ \be}-\frac 12 \d^{\al}_{\be}H \\
                                            \end{array}
                                          \right), \ \ \
                                          \mathcal{Q}_{\mathcal{A}}=\left(
                                                                      \begin{array}{c}
                                                                        Q_{\a} \\
                                                                        \overline{S}^{\al} \\
                                                                      \end{array}
                                                                    \right),
                                                                    \
                                                                    \
                                                                    \
                                                                    \overline{\mathcal{Q}}^{\mathcal{B}}=\left(
                                                                                                           \begin{array}{cc}
                                                                                                             S^{\b} & \overline{Q}_{\be} \\
                                                                                                           \end{array}
                                                                                                         \right)
 \eeq
Then the  (anti)commutators
(\ref{com1}),(\ref{com2}),(\ref{com3}),(\ref{com4}),(\ref{com5}),(\ref{com6})
combine to \cite{Dolan}
\beqa \label{algebra} &&
[\M_{\A}^{\ \B},\M_{\mathcal{C}}^{\
\D}]=\d_{\mathcal{C}}^{\B} \M_{\A}^{\ \D}-\d_{\A}^{\D}\M_{\mathcal{C}}^{\B}, \nonumber
\\ && [\M_{\A}^{\ \B},\Q_{\mathcal{C}}]=\d_{\mathcal{C}}^{\B}\Q_{\A}-\frac 14
\d_{\A}^{\B}\Q_{\mathcal{C}}, \ \ \ \ \ [\M_{\A}^{\
\B},\Qa^{\mathcal{C}}]=-\d_{\A}^{\mathcal{C}}\Qa^{\B}+\frac 14 \d_{\A}^{\B}\Qa^{\mathcal{C}},
\nonumber \\ && [R,\Q_{\A}]=-\Q_{A}, \ \ \ \ \ \ \ \
[R,\Qa^{\B}]=\Qa^{\B}, \nonumber \\ && \{ \Q_{\A},\Qa^{\B}
\}=4\M_{\A}^{\B}+3\d_{\A}^{\B}R, \ \ \ \ \ \{\Q_{\A},\Q_{\B}\}=0, \
\ \ \ \ \{\Qa^{\A},\Qa^{\B}\}=0, \eeqa where $$\d_{\A}^{\B}=\left(
                                                              \begin{array}{cc}
                                                                \d_{\a}^{\b} & 0 \\
                                                                0 & \d^{\al}_{\be} \\
                                                              \end{array}
                                                            \right).
$$

\subsection{The index}

Suppose an operator $Q$ and its Hermitian conjugate
$Q^{\dag}$ satisfy the relations
\beqa\label{IW}
&& \{ Q,Q\} = 0, \ \ \ \ \  \{ Q^{\dag},Q^{\dag}\} = 0, \ \ \ \ \
\{Q,Q^{\dag}\} = 2H,
\eeqa
where $H$ is the Hamiltonian ($=P_0$) of a taken system. This is a universal
situation valid down to the non-relativistic quantum mechanics.
The Witten index \cite{Witten} defined as $Tr(-1)^{\rm F}$
tells (under certain conditions) whether the supersymmetry is broken
spontaneously or not. By definition the operator $(-1)^{\rm F}$ is
\beq
(-1)^{\rm F} \ = \ \exp(2 \pi \textup{i} J_z),\ \ \ \ \  \{Q,(-1)^{\rm F}\} =0,
\eeq
where in the
spinorial basis $J_z=-J_3-\overline{J}_3$. It
distinguishes bosonic states $|b\rangle$ from the fermionic ones $|f\rangle$,
$$
(-1)^{\rm F}|b\rangle=|b\rangle,
\ \ \ \ \ \ (-1)^{\rm F}|f\rangle=-|f\rangle.
$$
Because of the cancellation of contributions of states with positive energies
to $Tr(-1)^{\rm F}$, this trace formally can be evaluated using the zero-energy
states
\beq
Tr(-1)^{\rm F}=n_b^{E=0}-n_f^{E=0},
\eeq
where $n_b^{E=0}$ and $n_f^{E=0}$
are the numbers of bosonic and fermionic ground states.
Therefore, if $Tr(-1)^{\rm F}\neq 0$, supersymmetry is not broken.
However, because of the presence of infinitely many states,
one needs a regulator commuting with $Q$ (to save cancellations).
Then the regularized Witten index is defined as
\beq
I_W\ = \ Tr((-1)^{\rm F}e^{-\b H}),
\eeq
and formally it does not depend on the parameter $\b$.

As to $\mathcal{N}=1$ superconformal theories,
there are different possibilities to realize relation (\ref{IW}),
due to the superconformal operators $S^{\alpha},\overline{S}^{\al}$.
Namely, one picks a generator $\Q$ with its adjoint $\Q^{\dag}$, such that
\beq
\{ \Q,\Q^{\dag} \}=2 \mathcal{H},
\eeq
where $\mathcal{H}$ {\it does not} coincide with the Hamiltonian.
Then one can consider the subspace of the Hilbert space composed of
the BPS states $|\psi\rangle $ annihilated by $\mathcal{H}$,
 $\mathcal{H} |\psi\rangle =0$, and define the Witten index
$I_W = Tr((-1)^{\rm F}e^{-\b \mathcal{H}})$. However, the space of such states
$|\psi\rangle $ is infinite dimensional and one has to introduce other regulators,
which leads to a nontrivial generalization of the index itself.

For $SU(2,2|1)$ group, there are four non-trivial
choices for supercharges $\Q$, $\Q^{\dag}$, which can be used for constructing the
superconformal index:
\beqa\label{reg}  \{ Q_1,S^1 \} = 2  \big(
H+2J_3+\frac 32 R \big); 
\qquad && \{ Q_2,S^2 \} = 2 \big(
H-2J_3+\frac 32 R \big); \nonumber  \\  \{
\overline{Q}_{1 },-\S^{1 } \} = 2 \big(
H-2\overline{J}_3-\frac 32 R \big); \qquad 
&& \{\overline{Q}_{2 },-\S^{2 } \} = 2 \big(
H+2\overline{J}_3-\frac 32 R \big).
\eeqa

The generators commuting with the corresponding pairs of supercharges are
\begin{eqnarray*}
\overline{M}^{\al}_{\ \be},H+\frac 12 R,P_{2 \al}, K^{\al 2};
\qquad \
\overline{M}^{\al}_{\ \be},H+\frac 12 R,P_{1 \al}, K^{\al 1};
\\
M_{\a}^{\ \b},H-\frac 12 R,P_{\a 2 }, K^{2  \a};
\qquad \
M_{\a}^{\ \b},H-\frac 12 R,P_{\a 1 }, K^{1  \a},
\end{eqnarray*}
respectively, see (\ref{algebra}).
Let us stick to the choice
$$
\Q=\overline{Q}_{1 }, \ \ \   \Q^{\dag}=-\S^{1 }, \ \ \
\H=H-2\overline{J}_3-\frac 32 R.
$$
Composing the matrix \cite{Dolan}
\beq
\M_{A}^{\ B} = \left(
                                                  \begin{array}{cc}
                                     M_{\a}^{\ \b}+\frac 12 \d_{\a}^{\b}\R & \P_{\a} \\
                                          \Pa^{\b} & -\R+\frac 12 \H \\
                                                  \end{array}
                                                \right),
\eeq
where $\P_{\a}= \frac 12 P_{\a 2 }$, $\Pa^{\b}=\frac 12 K^{2  \b}$, and
$$
\R=H-\frac 12 R
$$
we come to the $SU(2,1)$ Lie algebra with the relations
\beq\label{su21}
[\M_{A}^{\ B},\M_{C}^{\ D}]=\d_{C}^{B}\M_{A}^{\ D} -
\d_{A}^{D}\M_{C}^{\ B}.
\eeq

To regularize the trace over the infinite dimensional space of zero modes of
$\mathcal{H}$, we use all operators commuting between themselves and
with the distinguished supercharges $\mathcal{Q}$ and $\mathcal{Q}^{\dag}$.
In our case one additional regulator is $t^{\mathcal{R}}$
for some arbitrary complex variable $t$ restricted as $|t|<1$ to ensure damping.
Since $M_{\a}^{\ \b}$ commute with $\overline{Q}_{1 }$ and
$\overline{S}^{1 }$, there is one more regulator $x^{2J_3}$, $|x|<1$,
resolving the degeneracy ensured by $M_{\a}^{\ \b}$. Finally, one defines
\cite{Romelsberger1,Romelsberger2}
\begin{equation}
{\rm ind}(t,x) \ = \ Tr(-1)^{\rm F}x^{2J_3}t^{\mathcal{R}}e^{-\beta{\mathcal H}}.
\end{equation}
This index explicitly depends on the chemical potentials $x$ and $t$,
in difference from the variable $\beta$.

In the presence of internal symmetries, one can introduce more
regulators to resolve the degeneracies.
For $U(1)^k$ global symmetry group, one introduces chemical potentials
$\mu_j,\, j=1,\dots,k$, and extends the superconformal index as
\begin{equation}
{\rm ind}(t,x,\mu_j) \ = \ Tr(-1)^{\rm F}x^{2J_3}t^{\mathcal{R}}
 e^{\sum_{j=1}^k \mu_j q_j},
\end{equation}
where $q_j$ is the generator of $j$-th $U(1)$-group. For a non-abelian local gauge
invariance group $G$ with the maximal torus generators $G_a,\, a=1,\ldots,$ rank $G$,
and a flavor group $F$ with the maximal torus generators $F_j,\, j=1,\ldots,$ rank $F$,
the index reads
\begin{equation}\label{ind1}
{\rm ind}(t,x,z,y) \ = \ Tr \left( (-1)^{\rm F}x^{2J_3}t^{\mathcal{R}}
e^{\sum_{a=1}^{rank\, G} g_aG^a} e^{\sum_{j=1}^{rank\, F} f_jF^j}
\right),
\end{equation}
where $g_a$ and $f_j$ are the chemical potentials for groups $G$ and
$F$ respectively. We assume that the global abelian groups enter the
flavor group contributions in (\ref{ind1}). From
the representation theory it is known that
$Tr \exp(\sum_{i=1}^{rank\, G} g_iG^i) = \chi_{G}(z)$ is the character
of the corresponding representation of the gauge group $G$, where
$z$ is the set of complex eigenvalues of matrices realizing $G$. The
same is valid for the flavor group $F$: $Tr \exp(\sum_{j=1}^{rank\,
M} f_jF^j) = \chi_{F}(y)$ is the character of the representations
forming the space of free field states, and $y$ is the set of
complex eigenvalues of matrices realizing $F$.

Since all physical observables are gauge invariant, one is interested
in the index for gauge singlet operators. Therefore formula (\ref{ind1}) is
averaged over the gauge group, which yields the matrix integral
\begin{equation}\label{Ind}
I(t,x,y) \ = \ \int_G d \mu(g)\,
Tr \left( (-1)^{\rm F}x^{2J_3}t^{\mathcal{R}}
e^{\sum_{a=1}^{rank\, G} g_aG^a} e^{\sum_{j=1}^{rank\, F} f_jF^j}\right),
\end{equation}
where $d \mu(g)$ is the $G$-invariant matrix group measure.
This is the superconformal index -- the key object for our purposes.
By construction, it has the meaning of a particular combination of
$SU(2,2|1)\times G\times F$ group characters naturally restricted
to the space of BPS states and integrated over the gauge group.

\subsection{Calculation of the index}

Explicit computation of the superconformal index for ${\mathcal N}=1$
theories  was performed by R\"omelsberger \cite{Romelsberger2}.
According to his prescription one should first compute the trace
in index (\ref{ind1}) over the single particle states, which yields
the formula
\begin{eqnarray}
i(t,x,z,y) &=& \frac{2t^2 - t(x+x^{-1})}{(1-tx)(1-tx^{-1})}
\chi_{adj}(z)\cr &+& \sum_j
\frac{t^{2r_j}\chi_{R_F,j}(y)\chi_{R_G,j}(z) - t^{2-2r_j}\chi_{{\bar
R}_F,j}(y)\chi_{{\bar R}_G,j}(z)}{(1-tx)(1-tx^{-1})},
\label{index}\end{eqnarray}
where the first term represents
contribution of the gauge fields belonging to the adjoint representation
of the group $G$, and the sum over $j$ corresponds to the chiral
matter superfields $\varphi_j$ transforming as the gauge group
representations $R_{G,j}$ and non-abelian flavor symmetry group representations
$R_{F,j}$. The functions $\chi_{adj}(z)$, $\chi_{R_F,j}(y)$ and
$\chi_{R_G,j}(z)$ are the corresponding characters -- their explicit
forms for major classical groups are described in Appendix A.

In the original Romelsberger's formula the denominators are written
as $1-t\chi_{SU(2),f}(\gamma)+t^2$, where
$\chi_{SU(2),f}(\gamma)$ is the character for the fundamental
representation of the $SU(2)$ subgroup in (\ref{su21}). Parametrizing
it by the eigenvalue $x$,  one comes to (\ref{index}).

The $U(1)_R$-group contribution to (\ref{index}) is described by the terms
$t^{2R_j}$ and $t^{2-2R_j}$ resulting from a chiral scalar field with
the $R$-charge $2R_j$ and the fermion partner of the conjugate anti-chiral
fields whose $R$-charge is $-2R_j$. In the presence of additional global
$U(1)$-groups the variables $r_j$ have the form
$$
r_j \ = \ R_j + \sum_{l=1}^{k}q_{jl} \mu_l,
$$
where $2R_j$ is the $R$-charge of the $j$-th chiral superfield,
$q_{jl}$ are the normalized hypercharges of the $j$-th matter superfield for $l$-th
$U(1)$-group and $2\mu_l$ is the chemical potential for the latter group.

To obtain the full superconformal index, this single
particle states index is inserted into the ``plethystic" exponential
with the subsequent averaging over the gauge group:
\begin{equation}\label{Ind_fin}
I(t,x,y) \ = \ \int_G d \mu(g)\,
\exp \bigg ( \sum_{n=1}^{\infty}
\frac 1n i\big(t^n ,x^n, z^n , y^ n\big ) \bigg ).
\end{equation}
Similar objects appeared in computation of partition functions of
different statistical mechanics models and quantum field theories,
see, e.g., \cite{bax:partition,ska,Sun,Nekrasov02,Kinney,
Nakayama,Plethystic,Plethystic2,Dolan0}.

Clearly there are two qualitatively different contributions
to superconformal indices -- from the matter fields and the gauge fields.
The generic form of a matter field single particle states
contribution to $i(t,x,z,y)$ in the
presence of some global $U(1)$ symmetry group is
\begin{equation}\label{1chiral}
i_S(t,x,y) = \frac{t^{2R}y -t^{2-2R}y^{-1}}{(1-tx)(1-tx^{-1})},
\end{equation}
where $t,x$ are the same variables as in (\ref{index}) and $y=t^{2\mu}$ is
the chemical potential for the $U(1)$ group. It is
convenient to introduce new parametrization
\begin{equation}
p=tx, \quad q=tx^{-1}, \quad w=t^{2R}y,
\label{pqbases}\end{equation}
where $p$ and $q$ are (in general, complex) parameters satisfying
the constraints $|q|,|p|<1$.  As a result, we can write
$$
i_S(p,q,w)= \frac{w - pqw^{-1}}{(1-p)(1-q)}.
$$
Then the described index building algorithm yields
(cf. \cite{bax:partition})
\begin{equation}
\exp \bigg ( \sum_{n=1}^\infty \frac 1n
i_S(p^n,q^n,w^n ) \bigg ) = \prod_{j,k=0}^{\infty}
\frac{1-w^{-1} p^{j+1}q^{k+1}}{1 - w\, p^j\, q^k}=:\Gamma(w;p,q).
\end{equation}
This result corresponds to formula (69) in \cite{Romelsberger2} after
the identifications $w:=t^qu, \ p:=ty, \ q:=ty^{-1}.$ However, the fact
that this index coincides with the elliptic gamma function was
recognized only by Dolan and Osborn in \cite{Dolan}.

For the gauge field part one can set
\begin{equation}
i_V(p,q,z) \ = \ \frac{2t^2 - t(x+x^{-1})}{(1-tx)(1-tx^{-1})}
\chi_{adj}(z) = \left( -\frac{p}{1-p}- \frac{q}{1-q} \right)
\chi_{adj}(z).
\label{scalar}\end{equation}
For the $SU(2)$ group one has $\chi_{adj}(z)=z^2+z^{-2}+1$.
Substituting pieces of this expression in the corresponding
places of the index, we obtain the following characteristic
building blocks
\begin{eqnarray*}
&& \exp \bigg ( \sum_{n=1}^\infty \frac 1n \left(
-\frac{p^n}{1-p^n}- \frac{q^n}{1-q^n} \right) ( z^{2n} +
z^{-2n} ) \bigg ) =\frac{\theta(z^2;p) \theta(z^2;q)}{(1-z^2)^2} \\
\nonumber && \makebox[8em]{} =
\frac{1}{(1-z^2)(1-z^{-2})\Gamma(z^{\pm 2};p,q)}
\end{eqnarray*}
and
$$
\exp \bigg ( \sum_{n=1}^\infty \frac 1n \left( -\frac{p^n}{1-p^n}-
\frac{q^n}{1-q^n} \right) \bigg ) = (p;p)_{\infty} (q;q)_{\infty}.
$$
Similar expressions are found for field contributions for
the higher rank gauge groups.

\section{Seiberg duality for unitary gauge groups}

First we consider the usual $\mathcal{N}=1$ supersymmetric quantum
chromodynamics (SQCD) as an electric theory with the
internal symmetry groups \cite{Seiberg}
$$
G=SU(N), \ \ \ \ F=SU(N_f) \times  SU(N_f) \times U(1)_B,
$$
where $U(1)_B$ is generated by the baryon number charge (the $U(1)_R$
group enters the superconformal group).
This supersymmetric version of QCD has two chiral scalar
multiplets $Q$ and ${\tilde Q}$ belonging to the fundamental $f$ and
anti-fundamental $\bar f$ representations of $SU(N)$ respectively,
each carrying a baryon number, and the vector multiplet $V$ in the
adjoint representation of $G$. The field content of the electric theory is
collected in the following table

\vbox{ \hskip1.5cm \vbox{\tabskip=0pt \offinterlineskip \hrule
\halign{&\vrule# &\strut \ \hfil#\  \cr
height2pt&\omit&&\omit&&\omit&&\omit&& \omit && \omit &\cr &\  \  Field
\hfil   && \    $SU(N)$  \  && \  $SU(N_f)$ \ && \ $SU(N_f)$ \ && \
$U(1)_B$ \ &&  \ $U(1)_R$ \  &\cr
height2pt&\omit&&\omit&&\omit&&\omit&& \omit && \omit &\cr
\noalign{\hrule} height2pt&\omit&&\omit&&\omit&&\omit&& \omit &&
\omit & \cr & \ $Q$  \ \hfil &&  $f$  \hfil     &&    $f$     \hfil
&& $1$ \hfil
 &&  $q_B=1$    \hfil    &&  $2R_Q={\tilde N}/N_f$   \hfil        & \cr
& \ ${\tilde Q}$  \ \hfil &&  ${\bar f}$  \hfil     &&    $1$ \hfil
&&  ${\bar f}$ \hfil &&  $\widetilde{q}_{B}=-1$    \hfil    &&
$2R_{\widetilde{Q}}={\tilde N}/N_f$ \hfil & \cr & \ $V$  \ \hfil &&
${\rm adj}$ \hfil && $1$ \hfil  &&  $1$ \hfil
&&  $0$    \hfil    &&  $2R_V=1$   \hfil        & \cr
height2pt&\omit&&\omit&&\omit&&\omit&& \omit &&\omit &\cr } \hrule} }
Here $q_B, \widetilde{q}_B$ denote the baryonic charge and $R_Q,
R_{\widetilde{Q}}, R_V$ are half $R$-charges of the fields.

The dual magnetic theory has the symmetry groups
$$
G=SU(\widetilde{N}), \ \ \  F=SU(N_f) \times SU(N_f) \times U(1)_B,
$$
where $\widetilde{N} \ = \ N_f-N$. Its field content is fixed in the table
below

\vbox{ \hskip1.5cm \vbox{\tabskip=0pt \offinterlineskip \hrule
\halign{&\vrule# &\strut \ \hfil#\  \cr
height2pt&\omit&&\omit&&\omit&&\omit&& \omit && \omit &\cr &\ Field
\ \hfil   && \    $SU({\tilde N})$  \  && \  $SU(N_f)$ \ && \
$SU(N_f)$ \  && \ $U(1)_B$ \ &&  \ $U(1)_R$ \  &\cr
height2pt&\omit&&\omit&&\omit&&\omit&& \omit && \omit &\cr
\noalign{\hrule} height2pt&\omit&&\omit&&\omit&&\omit&& \omit &&
\omit & \cr & \ $q$  \ \hfil &&  $f$  \hfil     &&    ${\bar f}$
\hfil
 &&  $1$ \hfil
 &&  $q_B'=N/{\tilde N}$    \hfil    &&  $2R_q=N/N_f$   \hfil        & \cr
& \ ${\tilde q}$  \ \hfil &&  ${\bar f}$  \hfil     &&    $1$ \hfil
&&  $f$ \hfil &&  $\widetilde{q}_B'=-N/{\tilde N}$    \hfil    &&
$2R_{\widetilde{q}}=N/N_f$ \hfil    & \cr & \ $M$  \ \hfil && $1$
\hfil     &&    $f$     \hfil            && ${\bar f}$ \hfil && $0$
\hfil    &&  $2R_M=2{\tilde N}/N_f$   \hfil    & \cr & \ $\tilde V$
\ \hfil && ${\rm adj}$ \hfil &&    $1$ \hfil
  &&  $1$ \hfil
&&  $0$    \hfil    &&  $2R_V=1$   \hfil      & \cr
height2pt&\omit&&\omit&&\omit&&\omit&& \omit &&\omit &\cr } \hrule}
}
This duality is supposed to work only in the conformal window
$3N/2 < N_f < 3N,$ following from the demand that both dual
theories are asymptotically free in the one-loop approximation.
The one-loop beta function for the gauge coupling
is given by
$$
\beta_g = - \frac{g^3}{16 \pi^2}
\big( \frac{11}{3} T(\textbf{adj}) - \frac 23 T(F)
 - \frac 13 T(S)\big),
$$
 where $T(F)$ is the sum of Casimir coefficients
$T(\textbf{r})$ (see Appendix C for more details) over all fermions,
$T(S)$ is the similar sum over all scalars and $T(\textbf{adj})$ is
$T(\textbf{r})$ for the adjoint representation. For a summary of this
and two loop renormalization group results, see \cite{OneLoop}.

The $r_j$-charges of fields coming from $U(1)_R$ and $U(1)_B$
currents in the electric theory are
$$
r_Q \ = \ R_Q + q_B x, \ \
\ \ \ r_{\widetilde{Q}} \ = \ R_{\widetilde{Q}} + \widetilde{q}_B x,
$$
where $x$ is the $U(1)_B$-group chemical potential.
In the magnetic theory we set
$$
r_q \ = \ R_q + q_B' x , \ \ \
\ \ \ r_{\widetilde{q}} \ = \ R_{\widetilde{q}} + \widetilde{q}_B' x,
\ \ \ \ \ \ r_M \ = \ R_M.
$$

Then the single particle states index for the electric theory has the form
\begin{eqnarray}
&& i_E(p,q,z,s,t) = - \left( \frac{p}{1-p} + \frac{q}{1-q} \right)
\chi_{SU(N),adj}(z) \\  \nonumber && \makebox[1em]{} +
\frac{1}{(1-p)(1-q)} \Big( (pq)^{r_Q} \chi_{SU(N_f),f}(s)
\chi_{SU(N),f}(z) - (pq)^{1-r_Q} \chi_{SU(N_f),\overline{f}}(s)
\chi_{SU(N),\overline{f}}(z) \\  \nonumber && \makebox[1em]{}
+ (pq)^{r_{\widetilde{Q}}}
\chi_{SU(N_f),\overline{f}}(t) \chi_{SU(N),\overline{f}}(z) -
(pq)^{1-r_{\widetilde{Q}}} \chi_{SU(N_f),f}(t)
\chi_{SU(N),f}(z)\Big).
\end{eqnarray}

For the magnetic theory we have
\begin{eqnarray}
&& i_M(p,q,z,s,t) = - \left( \frac{p}{1-p} + \frac{q}{1-q} \right)
\chi_{SU(\widetilde{N}),adj}(z) \\  \nonumber && \makebox[1em]{} +
\frac{1}{(1-p)(1-q)} \Big( (pq)^{r_q}
\chi_{SU(N_f),\overline{f}}(s) \chi_{SU(\widetilde{N}),f}(z) -
(pq)^{1-r_q} \chi_{SU(N_f),f}(s)
\chi_{SU(\widetilde{N}),\overline{f}}(z)
\\  \nonumber && \makebox[1em]{} +
(pq)^{r_{\widetilde{q}}} \chi_{SU(N_f),f}(t)
\chi_{SU(\widetilde{N}),\overline{f}}(z) -
(pq)^{1-r_{\widetilde{q}}} \chi_{SU(N_f),\overline{f}}(t)
\chi_{SU(\widetilde{N}),f}(z)
\\  \nonumber && \makebox[1em]{} +
(pq)^{r_M} \chi_{SU(N_f),f}(s) \chi_{SU(N_f),\overline{f}}(t) -
(pq)^{1-r_M} \chi_{SU(N_f),\overline{f}}(s)
\chi_{SU(N_f),f}(t)\Big).
\end{eqnarray}

The superconformal indices take the form (see the invariant measures
in Appendix B)
\begin{eqnarray}\label{SUE}
 I_E &=&   \frac{(p;p)_{\infty}^{N-1} (q;q)_{\infty}^{N-1}}{N!} \\
\nonumber && \times \int_{\mathbb{T}^{N-1}}
 \frac{\prod_{i=1}^{N_f} \prod_{j=1}^{N}
    \Gamma((pq)^{r_Q} s_i z_j,(pq)^{\widetilde{r}_Q}t^{-1}_i z^{-1}_j;p,q)}{\prod_{1 \leq i < j \leq N} \Gamma(z_i z^{-1}_j,z_i^{-1} z_j;p,q)}
    \prod_{j=1}^{N-1}
\frac{d z_j}{2 \pi \textup{i} z_j},
\end{eqnarray}
where $\prod_{j=1}^Nz_j=\prod_{i=1}^{N_f}s_i= \prod_{i=1}^{N_f}t_i=1$,
and
\begin{eqnarray}\label{SUM}
    I_M &=& \frac{(p;p)_{\infty}^{\widetilde{N}-1} (q;q)_{\infty}^{\widetilde{N}-1}}{\widetilde{N}!} \prod_{1 \leq i,j \leq N_f} \Gamma((pq)^{r_M} s_i t^{-1}_j;p,q)
 \\ \nonumber && \times \int_{\mathbb{T}^{\widetilde{N}-1}}
     \frac{\prod_{i=1}^{N_f} \prod_{j=1}^{\widetilde{N}} \Gamma((pq)^{r_q} s_i^{-1} z_j,(pq)^{r_{\widetilde{q}}} t_i z_j^{-1};p,q)}{\prod_{1 \leq i < j \leq \widetilde{N}} \Gamma(z_i z_j^{-1},z_i^{-1}
    z_j;p,q)} \prod_{j=1}^{\widetilde{N}-1}  \frac{d z_j}{2 \pi \textup{i} z_j},
\end{eqnarray}
where $\prod_{j=1}^{\widetilde{N}}z_j=1$. Let us renormalize the variables
\begin{eqnarray}
s_i  \rightarrow  (pq)^{-r_Q} s_i, \ \ \
t_i^{-1}  \rightarrow  (pq)^{-r_{\widetilde{Q}}} t_i^{-1}, \ \ \
i=1,\ldots,N_f.
\end{eqnarray}

Then the superconformal indices are rewritten as the following elliptic
hypergeometric integrals
\begin{eqnarray}\label{IE-seiberg}
&& I_E =   \frac{(p;p)_{\infty}^{N-1} (q;q)_{\infty}^{N-1}}{N!} \\
\nonumber && \makebox[3em]{} \times
 \int_{\mathbb{T}^{N-1}}
 \frac{\prod_{i=1}^{N_f} \prod_{j=1}^{N}
    \Gamma(s_i z_j,t^{-1}_i z^{-1}_j;p,q)}
{\prod_{1 \leq i < j \leq N} \Gamma(z_i z^{-1}_j,z_i^{-1} z_j;p,q)}
\prod_{j=1}^{N-1} \frac{d z_j}{2 \pi \textup{i} z_j}
\end{eqnarray}
and
\begin{eqnarray}\label{IM-seiberg}
&&    I_M=\frac{(p;p)_{\infty}^{\widetilde{N}-1} (q;q)_{\infty}^{\widetilde{N}-1}}
{\widetilde{N}!} \prod_{1 \leq i,j \leq N_f}
    \Gamma(s_i t^{-1}_j;p,q)
 \\ \nonumber && \makebox[3em]{} \times \int_{\mathbb{T}^{\widetilde{N}-1}}
     \frac{\prod_{i=1}^{N_f} \prod_{j=1}^{\widetilde{N}}
     \Gamma(S^{1/\widetilde{N}} s_i^{-1} z_j,T^{-1/\widetilde{N}} t_i z_j^{-1};p,q)}{\prod_{1 \leq i < j \leq \widetilde{N}}
     \Gamma(z_i z_j^{-1},z_i^{-1}z_j;p,q)}\prod_{j=1}^{\widetilde{N}-1}  \frac{d z_j}{2 \pi \textup{i} z_j},
\end{eqnarray}
where $S =\prod_{i=1}^{N_f}s_i, \  T=\prod_{i=1}^{N_f}t_i$, and the
balancing condition reads $ST^{-1} = (pq)^{N_f-N}.$

As shown by Dolan and Osborn \cite{Dolan}, the equality $I_E =I_M$
coincides with the $A_n \leftrightarrow A_m$ root systems
symmetry transformation
established by Rains \cite{Rains}. For $N = \widetilde{N}= 2$
this identity is a simple consequence of the symmetry
transformation for an elliptic analogue of the Gauss hypergeometric
function discovered earlier by the first author in
\cite{Spiridonov3}. Note that this equality of integrals is valid
for any $N_f$, while the Seiberg duality is expected to exist only in the
conformal window, where we have appropriate $R-$charges yielding
an anomaly free theory. One cannot extrapolate the duality outside
this window except of the boundary points $N_f=3N/2$ and $N_f=3N$
(we thank A. Schwimmer and S. Theisen for a discussion on this point).
However, this does not mean that for the electric theory outside
the conformal window there cannot be {\it different} magnetic duals.
We present a number of such examples in a separate paper \cite{SV_CW}.

The needed equality between elliptic hypergeometric integrals
is rigorous only under certain constraints on the parameters.
The kernels of the integrals are meromorphic functions of integration
variables $z_j\in \C^*$.
There are two qualitatively different geometric sequences of poles of these
kernels---some of them converge to zero $z_j=0$ and others go to infinity.
So, the equality $I_E =I_M$ with the integration contours $\mathbb{T}$
on both sides is true provided $\mathbb{T}$ separates these two types of
pole sequences. In the present situation this is guaranteed for
$|S|^{1/\tilde N}<|s_i|<1$ and $1<|t_i|<|T|^{1/\tilde N}$.
All the relations for superconformal indices described below have
similar constraints on the parameters, but we shall not describe them
for brevity, assuming that the separability conditions
for pole sequences are satisfied by the contour $\mathbb{T}$.

\section{Intriligator-Pouliot duality for symplectic gauge groups}

The electric theory  has the overall symmetry group
$$
G=SP(2N), \ \ \ \ \ F=SU(2N_f),
$$
and the following matter field content

\vbox{  \hskip2.5cm \vbox{\tabskip=0pt \offinterlineskip \hrule
\halign{&\vrule# &\strut \ \hfil#\  \cr
height2pt&\omit&&\omit&&\omit&&\omit&\cr &\     \ \hfil   && \
$SP(2N)$  \  && \  $SU(2N_f)$ \ &&  \ $U(1)_R$ \  &\cr
height2pt&\omit&&\omit&&\omit&&\omit&\cr \noalign{\hrule}
height2pt&\omit&&\omit&&\omit&&\omit& \cr & \ $Q$  \ \hfil &&  $f$
\hfil     &&    $f$     \hfil &&  $1 - (N+1)/ N_f$ \hfil  & \cr }
\hrule} }
In this and all other tables below we drop the vector superfields
$V$ (or $\widetilde V$, except for the confining theories where this
field is absent), since they are always described by the adjoint
representation of $G$ and singlets of $F$.

The dual magnetic theory constructed by Intriligator and Pouliot
\cite{Intriligator1} has the same flavor group and the gauge group
$G=SP(2\widetilde{N}),$ where $\widetilde{N} = N_f-N-2$, with the field
content described in the table below

\vbox{ \hskip2.5cm \vbox{\tabskip=0pt \offinterlineskip \hrule
\halign{&\vrule# &\strut \ \hfil#\  \cr
height2pt&\omit&&\omit&&\omit&&\omit&\cr &\     \ \hfil   && \
$SP(2 \widetilde{N})$  \  && \  $SU(2N_f)$ \ &&  \ $U(1)_R$ \  &\cr
height2pt&\omit&&\omit&&\omit&&\omit&\cr \noalign{\hrule}
height2pt&\omit&&\omit&&\omit&&\omit& \cr & \ $q$  \ \hfil &&  $f$
\hfil     &&    ${\bar f}$ \hfil            &&  $(N +1)/ N_f$ \hfil
& \cr & \ $M$  \ \hfil &&  $1$  \hfil     && $T_A$ \hfil
  &&  $2(\widetilde{N}+1)/N_f$ \hfil
& \cr } \hrule} }

The conformal window for this duality is $3(N+1)/2 < N_f < 3(N+1).$

For these theories we have the following indices (in the renormalized
variables) \cite{Dolan}
\begin{eqnarray}\label{Sp-el}
 I_E &=&   \frac{(p;p)_{\infty}^{N} (q;q)_{\infty}^{N}}{2^N N!}
\int_{\mathbb{T}^N}   \frac{\prod_{i=1}^{2N_f} \prod_{j=1}^{N}
    \Gamma(t_i z_j^{\pm 1};p,q)}{\prod_{1 \leq i < j \leq N} \Gamma(z_i^{\pm 1}z^{\pm 1}_j;p,q)\prod_{j=1}^N \Gamma(z_j^{\pm 2};p,q)}
\prod_{j=1}^{N} \frac{d z_j}{2 \pi \textup{i} z_j}
\end{eqnarray}
and
\begin{eqnarray}\label{Sp-m}
    I_M &=& \frac{(p;p)_{\infty}^{\widetilde{N}} (q;q)_{\infty}^{\widetilde{N}}}{2^{\widetilde{N}} \widetilde{N}!}
     \prod_{1 \leq i < j \leq 2N_f} \Gamma(t_it_j;p,q)    \\ \nonumber && \times \int_{\mathbb{T}^{\widetilde{N}}}
   \frac{\prod_{i=1}^{2N_f} \prod_{j=1}^{\widetilde{N}}
    \Gamma((pq)^{1/2}t_i^{-1}z_j^{\pm 1};p,q)}{\prod_{1 \leq i < j \leq \widetilde{N}} \Gamma(z_i^{\pm 1}z^{\pm 1}_j;p,q)
    \prod_{j=1}^{\widetilde{N}} \Gamma(z_j^{\pm
2};p,q)}     \prod_{j=1}^{\widetilde{N}} \frac{d z_j}{2 \pi \textup{i} z_j},
\end{eqnarray}
with the balancing condition $\prod_{i=1}^{2N_f}t_i \ = \
(pq)^{N_f-N-1}.$
For $N =  \widetilde{N} = 1$, the equality $I_E=I_M$ is a
consequence of the symmetry transformation established in
\cite{Spiridonov3}. For arbitrary ranks $N, \ \widetilde{N}$, the
needed identity (\ref{ra}) was proven by Rains in \cite{Rains}.
After the degeneration to the rational integrals
level, it reduces to the Dixon transformation formula \cite{Dixon}.

\section{Multiple duality for $SP(2N)$ gauge group}\label{SPDual}

There exists a multiple duality phenomenon,
when one electric theory has many magnetic duals. In this section we describe
theories with $SP(2N)$ gauge group, where multiple duality
is ensured by $W(E_7)$, the Weyl group for the exceptional
root system $E_7$. However, we skip the description of this group
referring for details to \cite{SV}.

We take $\mathcal{N}=1$ SQCD electric theory with the
symmetry groups ${G} = SP(2N)$ and ${F} = SU(8) \times U(1).$
This model has  one chiral scalar multiplet $Q$ belonging to the
fundamental representations of $G$ and $F$, a vector multiplet $V$
in the adjoint representation, and the antisymmetric $SP(2N)$-tensor
field $X$, see the table below
\begin{center}
\begin{tabular}{|c|c|c|c|c|}
  \hline
    & $SP(2N)$ & $SU(8)$ & $U(1)$ & $U(1)_R$ \\  \hline
  $Q$ & $f$ & $f$ & $- \frac{N-1}{4}$ & $\frac{1}{2}$ \\
  $X$ & $T_A$ & 1 & 1 & $0$ \\
\hline
\end{tabular}
\end{center}
For $N=1$, the field $X$ is absent and $U(1)$-group is completely
decoupled. In \cite{SV} we were giving in tables halves of the $R$-charges.

This electric theory and its particular magnetic dual (with $N>1$) were
considered in \cite{Csaki1}. However, as described in \cite{SV},
there are other dual theories. In a special section below we show that
the 't~Hooft anomaly matching conditions are fulfilled for all these
new dualities.

The electric superconformal index is
\begin{eqnarray}\label{SP2N1}\nonumber
    && I_E = \frac{(p;p)_{\infty}^N (q;q)_{\infty}^N }{2^N N!}
\Gamma((pq)^s;p,q)^{N-1} \int_{{\mathbb T}^N}\prod_{1 \leq i < j
\leq N} \frac{\Gamma((pq)^s z_i^{\pm 1} z_j^{\pm 1};p,q)}
{\Gamma(z_i^{\pm 1} z_j^{\pm 1};p,q) }
    \\     &&   \makebox[4em]{} \times
\prod_{j=1}^N \frac{\prod_{i=1}^8  \Gamma((pq)^{r_Q} y_i z_j^{\pm
1};p,q)} {\Gamma(z_j^{\pm 2};p,q)} \frac{d z_j}{2 \pi \textup{i} z_j},
\end{eqnarray}
where $r_Q= R_Q +e_Q s,$  $r_X = e_X s,$
and $2R_Q=1/2$ is the $R$-charge of the $Q$-field, $e_Q= - (N-1)/4$
and $e_X=1$ are the $U(1)$-group hypercharges with $s$ being
its chemical potential.

The first (new) class of magnetic theories has the symmetry groups
$$
{G} \ = \ SP(2N), \qquad {F} \ = \ SU(4) \times SU(4) \times U(1)_B
\times U(1).
$$
Its field content is fixed in the table below
\begin{center}
\begin{tabular}{|c|c|c|c|c|c|c|}
  \hline
                & $SP(2N)$            & $SU(4)$    & $SU(4)$    & $U(1)_B$ & $U(1)$ & $U(1)_R$
\\  \hline
  $q$             & $f$                & $f$            & 1            & $-1$   &   $-\frac{N-1}{4}$   &  $\frac{1}{2}$
  \\
 $\widetilde{q}$& $f$   & 1            &$f$&  $1$   &  $-\frac{N-1}{4}$    &  $\frac{1}{2}$
 \\
  $Y$             & $T_A$              & 1            &    1         & 0    &  1    & 0
  \\
  $M_J  $             & 1         & $T_A$            &    1       & $2$     &  $\frac{2J-N+1}{2}$   & 1
  \\
  $\widetilde{M}_J $    & 1      & 1            &    $T_A$       & $-2$    &  $\frac{2J-N+1}{2}$    & 1     \\
\hline
\end{tabular}
\end{center}
In the tables of this section the capital index $J$ takes
values $0,\ldots,N-1$, which is not indicated for brevity.
The superconformal index in this magnetic theory is
\begin{eqnarray}  \nonumber &&
    I_M^{(1)} =
\prod_{J=0}^{N-1} \prod_{1 \leq i < j \leq 4} \Gamma((pq)^{r_{M_J}}
y_i y_j;p,q) \prod_{5 \leq i < j \leq 8}
\Gamma((pq)^{r_{\widetilde{M}_J}} y_i y_j;p,q)
    \\  \nonumber   &&  \makebox[2em]{} \times
\Gamma((pq)^{s};p,q)^{N -1}\frac{(p;p)_{\infty}^{N}
(q;q)_{\infty}^{N}}{2^N N!} \int_{{\mathbb T}^N} \prod_{1 \leq i < j
\leq N} \frac{\Gamma((pq)^{s} z_i^{\pm 1} z_j^{\pm 1};p,q)} {
\Gamma(z_i^{\pm 1} z_j^{\pm 1};p,q) }
    \\    && \makebox[2em]{}   \times
\prod_{j=1}^N \frac{\prod_{i=1}^4 \Gamma((pq)^{r_q}v^{-2} y_i z^{\pm
1}_j;p,q) \prod_{i=5}^8 \Gamma((pq)^{r_{\widetilde q}}v^2 y_i z^{\pm
1}_j;p,q)} {\Gamma(z_j^{\pm 2};p,q)}\frac{d z_j}{2 \pi \textup{i} z_j},
\label{IM1}\end{eqnarray}
where $v=\sqrt[4]{y_1y_2y_3y_4}$ and
\begin{eqnarray*}
r_q \ = \ R_q - \frac{N-1}{4} s, \ \ \ r_{\widetilde{q}} \ = \
R_{\widetilde{q}} - \frac{N-1}{4} s, \ \ \ r_Y \ = \ s,
\\
r_{M_J} \ = \ R_{M_J} - \frac12 (N-1-2J) s, \ \ \
r_{\widetilde{M}_J} \ = \ R_{\widetilde{M}_J} - \frac12 (N-1-2J)s.
\end{eqnarray*}

The second (new) class of dual magnetic theories has the same
symmetries  as in the previous case, but different
representation content as described in the
following table
\begin{center}
\begin{tabular}{|c|c|c|c|c|c|c|}
  \hline
                & $SP(2N)$            & $SU(4)$    & $SU(4)$    & $U(1)_B$ & $U(1)$ & $U(1)_R$            \\  \hline
  $q$           & $f$                & $\overline{f}$   & 1   & $1$     &   $-\frac{N-1}{4}$    & $\frac{1}{2}$  \\
 $\widetilde{q}$& $f$   & 1            &$\overline{f}$&
$-1$   &   $-\frac{N-1}{4}$    & $\frac{1}{2}$ \\
  $Y$             & $T_A$       & 1            &    1         & 0    & 1    & 0                          \\
  $M_J $          & 1              & $f$            &    $f$         & 0    & $\frac{2J-N+1}{2}$    & 1       \\
\hline
\end{tabular}
\end{center}

The index for this magnetic theory is given by
\begin{eqnarray}\nonumber
 &&   I_M^{(2)} =\Gamma((pq)^{ s };p,q)^{N-1} \prod_{J=0}^{N-1}
\prod_{i=1}^4 \prod_{j=5}^8 \Gamma((pq)^{r_{M_J}} y_i y_j;p,q)
    \nonumber \\ \nonumber   && \makebox[2em]{} \times
\frac{(p;p)_{\infty}^{N} (q;q)_{\infty}^{N}}{2^N N!}\int_{{\mathbb
T}^N} \prod_{1 \leq i < j \leq N} \frac{\Gamma((pq)^{ s } z_i^{\pm
1} z_j^{\pm 1};p,q)} {\Gamma(z_i^{\pm 1} z_j^{\pm 1};p,q)}
    \\     &&  \makebox[2em]{} \times
 \prod_{j=1}^N \frac{\prod_{i=1}^4
\Gamma((pq)^{r_q} v^2y_i^{-1} z^{\pm 1}_j;p,q) \prod_{i=5}^8
 \Gamma((pq)^{r_{\widetilde q}}v^{-2} y^{-1}_i z^{\pm 1}_j;p,q)}
{\Gamma(z_j^{\pm 2};p,q)}\frac{d z_j}{2 \pi \textup{i} z_j},
\label{IM2}\end{eqnarray} where
$$
r_q =r_{\widetilde{q}}= \frac14 - \frac{N-1}{4} s, \quad
 r_Y = s, \quad
r_{M_J} = \frac12 - \frac12 (N-1-2J) s.
$$

Finally, the third type of magnetic theories, which was constructed
originally by Cs\'aki, Skiba and Schmaltz in \cite{Csaki1},
has the symmetry groups $G =SP(2N)$ and ${F}=SU(8) \times U(1),$
and its fields content is
\begin{center}
\begin{tabular}{|c|c|c|c|c|}
  \hline
   & $SP(2N)$ & $SU(8)$ & $U(1)$ & $U(1)_R$ \\  \hline
  $q$ & $f$ & $\overline{f}$ & $-\frac{N-1}{4}$ & $\frac{1}{2}$ \\
  $Y$ & $T_A$ & 1 & 1 & 0  \\
  $M_J $ & 1 & $T_A$ & $\frac{2J-N+1}{2}$ & 1 \\
\hline
\end{tabular}
\end{center}

Corresponding magnetic superconformal index has the form
\begin{eqnarray}\label{SP2N3}
 &&   I_M^{(3)}= \Gamma((pq)^{r_Y};p,q)^{N -1}
\prod_{J=0}^{N-1} \prod_{1 \leq i < j \leq 8} \Gamma((pq)^{r_{M_J}}
y_i y_j;p,q)\ \frac{(p;p)_{\infty}^{N} (q;q)_{\infty}^{N}}{2^N N!}
    \\ \nonumber     && \makebox[2em]{} \times
\int_{{\mathbb
T}^N} \prod_{1 \leq i < j \leq N} \frac{\Gamma((pq)^{r_Y} z_i^{\pm
1} z_j^{\pm 1};p,q)} {\Gamma(z_i^{\pm 1} z_j^{\pm 1};p,q)}
\prod_{j=1}^N  \frac{\prod_{i=1}^8 \Gamma((pq)^{r_q} y_i^{-1}
z_j^{\pm 1};p,q)} { \Gamma(z_j^{\pm 2};p,q)} \frac{d z_j}{2 \pi \textup{i}
z_j},
\end{eqnarray}
where
$$
r_q \ = \ \frac{1-s(N-1)}{4},\quad r_Y=s,\quad r_{M_J} \ = \
sJ+\frac{1-s(N-1)}{2}.
$$

The $SP(2)$ gauge group case can be obtained from the
tables above by substituting $N=1$ and deleting fields $X$ in the
electric theory and $Y$ in the magnetic theories, which decouple
completely. The number of mesons in dual theories is reduced as well.
Equality of superconformal indices for $N=1$ follows from the results
of \cite{Spiridonov3}, and the needed identities for elliptic hypergeometric
integrals for $N>1$ were established in \cite{Rains}.
As argued in \cite{SV}, there should be in total 72 theories dual to each
other -- this number equals to the dimension of the coset group $W(E_7)/S_8$
responsible for the dualities (in this respect, see also \cite{Leigh:1996ds}).

\section{A new $SP(2N) \leftrightarrow SP(2M)$ groups duality}\label{SpSpD}

We take as the electric theory SQCD based on the symmetry groups
$$
G=SP(2M), \ \ \  F=SU(4)\times SP(2l_1) \times SP(2l_2) \times
\ldots \times SP(2l_K) \times U(1)
$$
with the fields content fixed in the table below

\begin{center}
\begin{tabular}{|c|c|c|c|c|c|c|c|c|}
  \hline
   & $SP(2M)$ & $SU(4)$ & $SP(2l_1)$ & $SP(2l_2)$ & $\ldots$ & $SP(2l_K)$ & $U(1)$ & $U(1)_R$ \\  \hline
  $W_1$ & $f$ & $\overline{f}$ & 1 & 1 & $\ldots$ & 1 & $-\frac{M-N-2}{4}$ & 0
  \\
  $Q_1$ & $f$ & 1 & $f$ & 1 & $\ldots$ & 1 & $-\frac{n_1}{2}$ & 1 \\

  $Q_1$ & $f$ & 1 & 1 & $f$ & $\ldots$ & 1 & $-\frac{n_2}{2}$ & 1 \\

  $\ldots$ & & & & & & & & \\
  $Q_K$ & $f$ & 1 & 1 & 1 & $\ldots$ & $f$ & $-\frac{n_K}{2}$ & 1 \\

  $X$ & $T_A$ & 1 & 1 & 1 & $\ldots$ & 1 & 1 & 0  \\
\hline
\end{tabular}
\end{center}
where $n_1 \neq n_2 \neq \ldots \neq n_K$ and $\sum_{i=1}^{K} l_i
n_i= M+N.$

The dual magnetic theory has $G=SP(2N)$ and the same flavor group;
the fields content is described below

\begin{center}
\begin{tabular}{|c|c|c|c|c|c|c|c|c|}
  \hline
   & $SP(2N)$ & $SU(4)$ & $SP(2l_1)$ & $SP(2l_2)$ & $\ldots$ & $SP(2l_K)$ & $U(1)$ & $U(1)_R$ \\  \hline
  $w_1$ & $f$ & $f$ & 1 & 1 & $\ldots$ & 1 & $\frac{M-N+2}{4}$ & 0
  \\
  $q_1$ & $f$ & 1 & $f$ & 1 & $\ldots$ & 1 & $-\frac{n_1}{2}$ & 1 \\

  $q_1$ & $f$ & 1 & 1 & $f$ & $\ldots$ & 1 & $-\frac{n_2}{2}$ & 1 \\

  $\ldots$ & & & & & & & & \\
  $q_K$ & $f$ & 1 & 1 & 1 & $\ldots$ & $f$ & $-\frac{n_K}{2}$ & 1 \\

  $N_j$ & 1 & $\overline{T}_A$ & $1$ & 1 & $\ldots$ & 1 & $j-\frac{M-N-2}{2}$ & 0
  \\
  $M_{1,k_1}$ & 1 & $\overline{f}$ & $f$ & 1 & $\ldots$ & 1 & $-\frac{M-N-2}{4}-\frac{n_1}{2}+k_1$ & 1
  \\
  $M_{2,k_2}$ & 1 & $\overline{f}$ & 1 & $f$ & $\ldots$ & 1 & $-\frac{M-N-2}{4}-\frac{n_2}{2}+k_2$ & 1
  \\
  $\ldots$ & & & & & & & & \\
  $M_{K,k_K}$ & 1 & $\overline{f}$ & 1 & 1 & $\ldots$ & $f$ & $-\frac{M-N-2}{4}-\frac{n_K}{2}+k_K$ & 1
  \\
  $Y$ & $T_A$ & 1 & 1 & 1 & $\ldots$ & 1 & 1 & 0  \\
\hline
\end{tabular}
\end{center}
where $j=0,\ldots,M-N-1$ and $k_i = 0,\ldots,n_i-1$ for any
$i=1,\ldots,K$. Here we assume that $M\geq N$ (for $M=N$ the fields $N_j$ are
are absent).

The superconformal indices have the form
\begin{eqnarray}\label{SPn1}&&
I_E =\frac{(p;p)_\infty^M (q;q)_\infty^M} {2^M M!} \Gamma(t;p,q)^{M-1}
\int_{\mathbb{T}^M} \prod_{1\le
i<j\le M} \frac{\Gamma(tz_i^{\pm1}z_j^{\pm1};p,q)}{
\Gamma(z_i^{\pm1}z_j^{\pm1};p,q)}
\\ \nonumber && \makebox[2em]{} \times
\prod_{j=1}^M\frac{\prod_{k=1}^4\Gamma(tt^{-1}_kz_j^{\pm1};p,q)
\prod_{r=1}^{K} \prod_{j=1}^{l_r} \Gamma(s_{r,j}z_j^{\pm1};p,q)}
{\Gamma(z_j^{\pm2};p,q) \prod_{r=1}^{K} \prod_{j=1}^{l_r}
\Gamma(t^{n_r}s_{r,j}z_j^{\pm1};p,q)} \frac{dz_j}{2\pi \textup{i} z_j}
\end{eqnarray}
and
\begin{eqnarray}\label{SPn2} &&
I_M =\frac{(p;p)_\infty^N (q;q)_\infty^N} {2^N N!} \Gamma(t;p,q)^{N-1}
\prod_{i=0}^{M-N-1}\prod_{1 \le k < r \le 4}
\Gamma(t^{i+2}t_k^{-1}t_r^{-1};p,q)
\\ \nonumber && \makebox[2em]{} \times
\prod_{r=1}^4\prod_{m=1}^{K}
\prod_{i=1}^{l_m}\prod_{k_m=0}^{n_m-1}
\frac{\Gamma(t^{k_m+1}t_r^{-1}s_{m,i};p,q)}
{\Gamma(t^{k_m}t_rs_{m,i};p,q)}
\int_{\mathbb{T}^N} \prod_{1\le
i<j\le N} \frac{\Gamma(tz_i^{\pm1}z_j^{\pm1};p,q)}{
\Gamma(z_i^{\pm1}z_j^{\pm1};p,q)}
\\ \nonumber && \makebox[3em]{} \times
\prod_{j=1}^N\frac{\prod_{k=1}^4\Gamma(t_kz_j^{\pm1};p,q)
\prod_{r=1}^{K} \prod_{j=1}^{l_r} \Gamma(s_{r,j}z_j^{\pm1};p,q)}
{\Gamma(z_j^{\pm2};p,q) \prod_{r=1}^{K}
\prod_{j=1}^{l_r}\Gamma(t^{n_r}s_{r,j}z_j^{\pm1};p,q)}
\frac{dz_j}{2\pi \textup{i} z_j},
\end{eqnarray}
with the balancing condition $\prod_{r=1}^4t_r=t^{2+M-N}.$

We have checked that the anomalies of these two theories match
(see below), which is a very strong indication that the theories
are dual to each other. This is another new duality that we
have found. It has rather complicated structure with
the flavor group composed of an arbitrary number of simple group
components. The renormalization group analysis shows that the asymptotic
freedom is present on the electric side for $M > \sum_{i=1}^K l_i/2 - 1$
and on the magnetic side for $N > \sum_{i=1}^K l_i/2 - 1.$

The equality of elliptic hypergeometric integrals $I_E=I_M$, which gives another
argument supporting this duality, coincides with the Rains Conjecture 1
from \cite{rai:lit}
(it was used  as  a starting point for the derivation of the described duality).
As we have known after the completion of this work, this
conjecture is proven recently by van de Bult \cite{bult}.

\section{Multiple duality for $SU(2N)$ gauge group}\label{Mult2N}

We describe now the  multiple duality phenomenon for $SU(2N)$ gauge
group. The overall flavor symmetry group of the theories is rather
unusual. For $N=1$, this multiple duality coincides with that for
$SP(2)$ group, see \cite{SV}. For $N>2$, one has $F=SU(4) \times
SU(4) \times U(1)_1 \times U(1)_2 \times U(1)_B. $ For $N=2$, the
flavor subgroup $U(1)_1$ is replaced by $SU(2)$. The field content
of the electric theory for $N>2$ is shown in the table below

\begin{center}
\begin{tabular}{|c|c|c|c|c|c|c|c|}
  \hline
   & $SU(2N)$ & $SU(4)$ & $SU(4)$ & $U(1)_1$ & $U(1)_2$ & $U(1)_B$ & $U(1)_R$ \\
\hline
  $Q$ & $f$ & $f$ & $1$ & 0 & $2N-2$ & 1 & $\frac 12$ \\
  $\widetilde{Q}$ & $\overline{f}$ & 1 & $f$ & 0 & $2N-2$ & -1 & $\frac 12$
  \\
  $A$ & $T_A$ & 1 & 1 & 1 & -4 & 0 & 0 \\
  $\overline{A}$ & $\overline{T}_A$ & 1 & 1 & -1 & -4 & 0 & 0 \\
\hline
\end{tabular}
\end{center}
Corresponding superconformal index has the form
\begin{eqnarray}\label{SU2N_1}
&&I_E= \frac{(p;p)_{\infty}^{2N-1} (q;q)_{\infty}^{2N-1} }{(2N)!}
\int_{\mathbb{T}^{2N-1}} \prod_{1 \leq j < k \leq 2N}
\frac{\Gamma(Uz_jz_k,Vz_j^{-1}z_k^{-1};p,q)}{\Gamma(z_j^{-1}z_k,z_jz_k^{-1};p,q)}
\nonumber\\  && \makebox[3em]{} \times \prod_{j=1}^{2N}
\prod_{k=1}^{4} \Gamma(s_kz_j,t_kz_j^{-1};p,q)\prod_{j=1}^{2N-1}
\frac{dz_j}{2\pi \textup{i} z_j},
\end{eqnarray}
where $\prod_{j=1}^{2N}z_j=1$ and the balancing condition reads
$(UV)^{2N-2} ST =(pq)^2$ with $S=\prod_{k=1}^4 s_k $ and $T =\prod_{k=1}^4 t_k.$
This is the two-parameter (higher order) extension of the type II
elliptic beta integral for the root system $A_{2N-1}$
introduced by Spiridonov in \cite{Spiridonov3}.

For $N\geq2$, magnetic dual theories have the same gauge and global
symmetry groups. The first (new) dual theory
has the field content described for $N>2$ in the table below
\begin{center}
\begin{tabular}{|c|c|c|c|c|c|c|c|}
  \hline
   & $SU(2N)$ & $SU(4)$ & $SU(4)$ & $U(1)_1$ & $U(1)_2$ & $U(1)_B$ & $U(1)_R$ \\  \hline
  $q$ & $f$ & $f$ & 1 & 0 & $2N-2$ & -1 & $\frac 12$ \\
  $\overline{q}$ & $\overline{f}$ & 1 & $f$ & 0 & $2N-2$ & 1 & $\frac 12$
  \\
  $a$ & $T_A$ & 1 & 1 & 1 & -4 & 0 & 0 \\
  $\overline{a}$ & $\overline{T}_A$ & 1 & 1 & -1 & -4 & 0 & 0 \\
  $H_m$ & 1 & $T_A$ & 1 & -1 & $4N-8-8m$ & 2 & 1 \\
  $G$ & 1 & $T_A$ & 1 & $N-1$ & 0 & 2 & 1 \\
  $\overline{H}_m$ & 1 & 1 & $T_A$ & 1 & $4N-8-8m$ & -2 & 1 \\
  $\overline{G}$ & 1 & 1 & $T_A$ & $-N+1$ & 0 & -2 & 1 \\
\hline
\end{tabular}
\end{center}
where $m=0,\ldots,N-2.$ This leads to the magnetic index
\begin{eqnarray}\label{SU2N_2}
&& I_M^{(1)}= \prod_{1 \leq i < j \leq 4}\Big[
\Gamma(U^{N-1}s_is_j,V^{N-1}t_it_j;p,q)
\prod_{m=0}^{N-2} \Gamma(V (UV)^{m}s_is_j,U (UV)^{m}t_it_j;p,q)\Big]
 \nonumber \\ && \makebox[2em]{} \times
\frac{(p;p)_{\infty}^{2N-1} (q;q)_{\infty}^{2N-1}}{(2N)!}
\int_{\mathbb{T}^{2N-1}}  \prod_{1 \leq j < k \leq 2N}
\frac{\Gamma(Vz_jz_k,Uz_j^{-1}z_k^{-1};p,q)}{\Gamma(z_j^{-1}z_k,z_jz_k^{-1};p,q)}
  \\ \nonumber && \makebox[4em]{} \times \prod_{j=1}^{2N} \prod_{k=1}^{4} \Gamma(\sqrt[4]{T/S}
s_kz_j,\sqrt[4]{S/T}t_kz_j^{-1};p,q) \prod_{j=1}^{2N-1}
\frac{dz_j}{2\pi \textup{i} z_j}.
\end{eqnarray}

Our second dual theory was found by Cs\'aki et al in \cite{Csaki2}.
Its field content for $N>2$ is described in the table below
\begin{center}
\begin{tabular}{|c|c|c|c|c|c|c|c|}
  \hline
   & $SU(2N)$ & $SU(4)$ & $SU(4)$ & $U(1)_1$ & $U(1)_2$ & $U(1)_B$ & $U(1)_R$ \\  \hline
  $q$ & $f$ & $\overline{f}$ & 1 & 0 & $2N-2$ & 1 & $\frac 12$ \\
  $\overline{q}$ & $\overline{f}$ & 1 & $\overline{f}$ & 0 & $2N-2$ & -1 & $\frac 12$
  \\
  $a$ & $T_A$ & 1 & 1 & 1 & -4 & 0 & 0 \\
  $\overline{a}$ & $\overline{T}_A$ & 1 & 1 & -1 & -4 & 0 & 0 \\
  $M_k$ & 1 & $f$ & $f$ & 0 & $4N-4-8k$ & 0 & 1 \\
\hline
\end{tabular}
\end{center}
where $k=0,\ldots,N-1$. Its superconformal index has the form
\begin{eqnarray}\label{SU2N_3}
&& I_M^{(2)}= \frac{(p;p)_{\infty}^{2N-1} (q;q)_{\infty}^{2N-1}
}{(2N)!} \prod_{m=0}^{N-1} \prod_{k, l=1}^4 \Gamma((UV)^ms_kt_l;p,q)
\\ \nonumber && \makebox[-1em]{} \times \int_{\mathbb{T}^{2N-1}}  \prod_{1 \leq j < k \leq 2N}
\frac{\Gamma(Uz_jz_k,Vz_j^{-1}z_k^{-1};p,q)}{\Gamma(z_j^{-1}z_k,z_jz_k^{-1};p,q)}
\nonumber
\prod_{j=1}^{2N}\prod_{k=1}^{4} \Gamma(\sqrt{S}
s_k^{-1}z_j,\sqrt{T}t_k^{-1}z_j^{-1};p,q)\prod_{j=1}^{2N-1}
\frac{dz_j}{2\pi \textup{i} z_j}. \nonumber
\end{eqnarray}

Our third, again new, duality corresponds to the theory described below for $N>2$
\begin{center}
\begin{tabular}{|c|c|c|c|c|c|c|c|}
  \hline
   & $SU(2N)$ & $SU(4)$ & $SU(4)$ & $U(1)_1$ & $U(1)_2$ & $U(1)_B$ & $U(1)_R$ \\  \hline
  $q$ & $f$ & $\overline{f}$ & 1 & 0 & $2N-2$ & -1 & $\frac 12$ \\
  $\overline{q}$ & $\overline{f}$ & 1 & $\overline{f}$ & 0 & $2N-2$ & 1 & $\frac 12$
  \\
  $a$ & $T_A$ & 1 & 1 & 1 & -4 & 0 & 0 \\
  $\overline{a}$ & $\overline{T}_A$ & 1 & 1 & -1 & -4 & 0 & 0 \\
  $M_k$ & 1 & $f$ & $f$ & 0 & $4N-4-8k$ & 0 & 1 \\
  $H_m$ & 1 & $T_A$ & 1 & -1 & $4N-8-8m$ & 2 & 1 \\
  $G$ & 1 & $T_A$ & 1 & $N-1$ & 0 & 2 & 1 \\
  $\overline{H}_m$ & 1 & 1 & $T_A$ & 1 & $4N-8-8m$ & -2 & 1 \\
  $\overline{G}$ & 1 & 1 & $T_A$ & $-N+1$ & 0 & -2 & 1 \\
\hline
\end{tabular}
\end{center}
where $k=0,\ldots,N-1,\ m=0,\ldots,N-2.$ Its superconformal index reads
\begin{eqnarray}\label{SU2N_4}
&& I_M^{(3)}= \frac{(p;p)_{\infty}^{2N-1} (q;q)_{\infty}^{2N-1}
}{(2N)!} \prod_{m=0}^{N-1} \prod_{k, l=1}^4 \Gamma((UV)^ms_kt_l;p,q)
\\ \nonumber && \makebox[1em]{} \times
 \prod_{1 \leq i < j\leq 4}\Big[ \Gamma(U^{N-1}s_is_j,V^{N-1}t_it_j;p,q)
\prod_{m=0}^{N-2} \Gamma(V (UV)^{m}s_is_j,U (UV)^{m}t_it_j;p,q) \Big]  \nonumber
\\ && \makebox[-1em]{}  \times \int_{\mathbb{T}^{2N-1}}
\prod_{1 \leq j < k \leq 2N}
\frac{\Gamma(Vz_jz_k,Uz_j^{-1}z_k^{-1};p,q)}
{\Gamma(z_j^{-1}z_k,z_jz_k^{-1};p,q)}
\prod_{j=1}^{2N} \prod_{k=1}^{4} \Gamma(\sqrt[4]{ST}
s_k^{-1}z_j,\sqrt[4]{ST}t_k^{-1}z_j^{-1};p,q)\prod_{j=1}^{2N-1}
\frac{dz_j}{2\pi \textup{i} z_j}.
\nonumber \end{eqnarray}

From the duality arguments for these field theories, we
conjecture that $I_E=I_M^{(1)}=I_M^{(2)}=I_M^{(3)}$ under certain
constraints on the integral parameters, which yield
new powerful elliptic hypergeometric integral identities.
Instead of the $W(E_7)$ Weyl group symmetry
in parameters, existing for $N=1$, only its subgroup
of reflection transformations consistent with the permutational
$S_4\times S_4$ symmetry group survives. Nevertheless, preliminary
considerations indicate that these relations should be provable
by an appropriate analog of the method used in \cite{Rains} for
proving $W(E_7)$-identities for $BC_N$-integrals of type II.
We have checked that the reduction from $N_f=4$ to $N_f=3$ realized
by the constraint $s_4t_4=pq$ reduces superconformal indices
to Spiridonov's $A_{2N-1}$-elliptic beta integral \cite{Spiridonov3},
i.e. equality of indices in this case is proven rigorously.

For $N=2$, superconformal indices  are given by the same
integrals. However, in this case $\prod_{1 \leq j < k \leq
4}f(z_iz_j) =\prod_{1 \leq j < k \leq 4}f(z_i^{-1}z_j^{-1})$ for
arbitrary function $f(x)$, and the parameters $U$  and $V$ unify to a
doublet, meaning that the fields $A$  and $\bar A$,  $a$  and $\bar
a$ unify to fundamentals of the $SU(2)$ group, which replaces the
$U(1)_1$ flavor subgroup.

An interesting situation occurs in the limit $V\to 1$ (or $U\to 1$).
Some of the poles coming from the integrand factor $\Gamma(Vz_i^{-1}z_j^{-1};p,q)$
approach the integration contour and it is necessary to deform $\mathbb{T}$
before taking this limit. A careful residue calculus shows that
in this limit the leading asymptotic contribution to all four superconformal
indices are given by the residues of the poles at $z_jz_k=V\to 1,\ j\neq k.$
As a result, $N-1$ integrations are taken away, there remain only
$N$-dimensional integrals over, say, $z_{2j-1},\ j=1,\ldots, N$, variables.
The latter integrals coincide exactly with the indices of four theories
appearing in multiple $SP(2N)$-duality described above.
Thus we have shown, that our multiple $SU(2N)$-dualities
contain $SP(2N)$ dualities as special subcases.
The first mathematical observation that the type II hypergeometric
identities for $BC_N$-root system can be obtained from type II
relations for both $A_{2N-1}$ and $A_{2N}$ root systems has been
done in \cite{spi-war:psi}, where various new multiple ${}_6\psi_6$
summation formulas on root systems have been suggested. Here we extend
this observation to the (expected) relations between type II elliptic
hypergeometric integrals. On the physical ground, such a
relation between the particular $SU(4)$ and $SP(4)$ gauge group dualities
was observed in \cite{Csaki2}. Note that the further limit $U=1$
leads to the $SU(2)^N$-gauge group theories whose indices are given
by $N$-th power of the indices of $SP(2)$-group models with $N_f=4$
constructed in \cite{SV}.

The attempts in \cite{Csaki2} to construct an analogous duality for
even rank gauge groups $SU(2N+1)$ have failed.
We have succeeded in solving this problem; corresponding results
together with the residue calculus details will be presented in a separate paper.

\section{Kutasov-Schwimmer type dualities for the unitary gauge group}

Now we pass to generalizations of the Seiberg
dualities for unitary and symplectic gauge groups $G$ discovered by
Kutasov and Schwimmer (KS) \cite{Kutasov,KS} and studied in detail
in \cite{Kutasov:1995ss} and other papers. For brevity, we skip separate
global symmetry group descriptions since they can be read off
easily from the field contents of the theories given in the
tables. The first column in the tables describes gauge
group representations for fields, while
other columns, except of the very last one, describe representations and
hypercharges for subgroups of the flavor group $F$.
Also, we skip the detailed description of single particle
state indices  and write out directly the integrals for the
superconformal indices together with
the balancing condition, if there is any.
In this section we describe such dualities for $G=SU(N)$.

\subsection{$SU(N)$ gauge group with the adjoint matter field}\label{KSel}

The following electric-magnetic duality is described in \cite{KS}.
The field content of the electric theory is

\vspace*{2mm}

\begin{tabular}{|c|c|c|c|c|c|}
  \hline
                & $SU(N)$            & $SU(N_f)$    & $SU(N_f)$    & $U(1)_B$ &
                $U(1)_R$
                \\  \hline
  $Q$             & $f$                & $f$            & 1            & 1        & $2r=1-\frac{2N}{(K+1)N_f}$
  \\
 $\widetilde{Q}$& $\overline{f}$   & 1            &$\overline{f}$&
-1       &   2$\widetilde{r}$=$1-\frac{2N}{(K+1)N_f}$ \\
  $X$             & $adj$              & 1            &    1         & 0        & $2s=\frac{2}{K+1}$
  \\
\hline
\end{tabular}

\vspace*{2mm}

The magnetic theory ingredients are collected in the following table

\vspace*{2mm}

\begin{tabular}{|c|c|c|c|c|c|}
  \hline
                & $SU(\widetilde{N})$            & $SU(N_f)$    & $SU(N_f)$    & $U(1)_B$ & $U(1)_R$                                    \\  \hline
  $q$             & $f$                & $\overline{f}$            & 1            & $N/\widetilde{N}$        & $2r'=1-\frac{2\widetilde{N}}{(K+1)N_f}$
  \\
 $\widetilde{q}$& $\overline{f}$   & 1            &$f$&
$-N/\widetilde{N}$       &
2$\widetilde{r}'=1-\frac{2\widetilde{N}}{(K+1)N_f}$
\\
  $Y$             & $adj$              & 1            &    1         & 0        & $2s=\frac{2}{K+1}$
  \\
  $M_j$             & 1              & $f$            &    $\overline{f}$         & 0        & $2r_{M_j}$=$2-\frac{4N}{(K+1)N_f}+\frac{2(j-1)}{K+1}$
  \\
\hline
\end{tabular}

\vspace*{2mm}

Here $j=1, \ldots, K$  and the dual gauge group dimension is
\begin{equation}\label{KSSU}
\widetilde{N} \ = \ K N_f - N, \ \ \ K=1,2,\ldots,
\end{equation}
with the constraint $N_f > N/K$.

Defining $U=(pq)^s=(pq)^{\frac{1}{K+1}}$, we find the
following indices for these theories
\begin{eqnarray}\label{intKSelE}
&&    I_E =\frac{(p;p)_{\infty}^{N-1} (q;q)_{\infty}^{N-1}}{N!} \Gamma(U;p,q)^{N-1}
     \int_{\mathbb{T}^{N-1}}  \prod_{1 \leq i < j \leq N}
 \frac{\Gamma(U z_i z_j^{-1},U z_i^{-1} z_j;p,q) }{ \Gamma(z_i z^{-1}_j,z_i^{-1}
    z_j;p,q)} \nonumber \\  &&  \makebox[3em]{}  \times
   \prod_{i=1}^{N_f} \prod_{j=1}^{N}
    \Gamma(s_i z_j,t^{-1}_i z^{-1}_j;p,q)\prod_{j=1}^{N-1}
\frac{d z_j}{2 \pi \textup{i} z_j},
\end{eqnarray}
where $\prod_{j=1}^Nz_j=1$, the balancing condition reads
$U^{2N}ST^{-1} = (pq)^{N_f}$ with $S=\prod_{i=1}^{N_f}s_i,\ T=
\prod_{i=1}^{N_f}t_i$, and
\begin{eqnarray}\label{intKSelM}
&&    I_M = \frac{ (p;p)_{\infty}^{\widetilde{N}-1}
(q;q)_{\infty}^{\widetilde{N}-1} }{\widetilde{N}!}
\Gamma(U;p,q)^{\widetilde{N}-1}  \prod_{l=1}^K \prod_{1 \leq i,j
\leq N_f} \Gamma(U^{l-1} s_i    t^{-1}_j;p,q) \\ \nonumber &&
\makebox[3em]{} \times \int_{T^{\widetilde{N}-1}} \prod_{1 \leq i <
j \leq \widetilde{N}} \frac{\Gamma(U z_i  z_j^{-1},U  z_i^{-1}
z_j;p,q)}{ \Gamma(z_i z_j^{-1},z_i^{-1} z_j;p,q)} \\ \nonumber &&
\makebox[3em]{} \times  \prod_{i=1}^{N_f}
\prod_{j=1}^{\widetilde{N}} \Gamma(U(ST)^{\frac{K}{2\widetilde{N}}}
s_i^{-1} z_j, U(ST)^{-\frac{K}{2\widetilde{N}}} t_i z_j^{-1};p,q)
\prod_{j=1}^{\widetilde{N}-1} \frac{d z_j}{2 \pi \textup{i} z_j},
\end{eqnarray}
where $\prod_{j=1}^{\widetilde{N}}z_j=1$.

An important fact is that these theories contain matter fields in the
adjoint representation of the gauge group. The conjecture that $I_E=I_M$
(under appropriate contour separability constraints mentioned earlier)
represents a new type of elliptic hypergeometric identities,
which was not met earlier \cite{Spiridonov1}. Therefore we describe
in Appendix D the total ellipticity property
hidden behind this identity. In the large $N, N_f$ limit (with fixed $N/N_f$)
the equality of $I_E$ and $I_M$ was confirmed up to a few
terms of the corresponding expansion in \cite{Dolan} using the method of
\cite{Dolan0}.

\subsection{Two adjoint matter fields case}

This duality was considered by Brodie and Strassler in \cite{Brodie1,Brodie2}.
The electric theory is

\vspace*{2mm}

\begin{center}
\begin{tabular}{|c|c|c|c|c|c|}
  \hline
   & $SU(N)$ & $SU(N_f)$ & $SU(N_f)$ & $U(1)_B$ & $U(1)_R$ \\  \hline
  $Q$ & $f$ & $f$ & 1 & 1 & $1-\frac{N}{N_f(K+1)}$ \\
  $\widetilde{Q}$ & $\overline{f}$ & 1 & $\overline{f}$ & $-1$ & $1-\frac{N}{N_f(K+1)}$
  \\
  $X$ & $adj$ & 1 & 1 & 0 & $\frac{2}{K+1}$  \\
  $Y$ & $adj$ & 1 & 1 & 0 & $\frac{K}{K+1}$  \\
\hline
\end{tabular}
\end{center}

\vspace*{2mm}

The magnetic theory has the following matter field content

\vspace*{2mm}

\begin{center}
\begin{tabular}{|c|c|c|c|c|c|}
  \hline
   & $SU(\widetilde{N})$ & $SU(N_f)$ & $SU(N_f)$ & $U(1)_B$ & $U(1)_R$ \\  \hline
  $q$ & $f$ & $\overline{f}$ & 1 & $\frac{N}{\widetilde{N}}$ & $1-\frac{\widetilde{N}}{N_f(K+1)}$
  \\
  $\widetilde{q}$ & $\overline{f}$ & 1 & $f$ & $-\frac{N}{\widetilde{N}}$ & $1-\frac{\widetilde{N}}{N_f(K+1)}$ \\
  $X$ & $adj$ & 1 & 1 & 0 & $\frac{2}{K+1}$  \\
  $Y$ & $adj$ & 1 & 1 & 0 & $\frac{K}{K+1}$  \\
  $M_{LJ}$ & 1 & $f$ & $\overline{f}$ & 0 &
                           $2-\frac{2N}{N_f(K+1)}+\frac{2L+KJ}{K+1}$ \\
\hline
\end{tabular}
\end{center}

\vspace*{2mm}
Here $K$ is odd, $0 \leq L \leq K-1$, $J=0,1,2$, and
\beq\label{KS2}
\widetilde{N}=3KN_f-N.
\eeq

Corresponding electric superconformal index has the form
\begin{eqnarray}\label{KS2adj1}
&&I_E= \frac{(p;p)_{\infty}^{N-1} (q;q)_{\infty}^{N-1}}{N!}\
\Gamma(U,U^{\frac K2};p,q)^{N-1}
\\  \nonumber && \makebox[4em]{} \times
\int_{\mathbb{T}^{N-1}}   \prod_{1 \leq i < j \leq N}
\frac{\Gamma(Uz_iz_j^{-1},Uz_i^{-1}z_j,
U^{K/2}z_iz_j^{-1},U^{K/2}z_i^{-1}z_j;p,q)}{\Gamma(z_iz_j^{-1},z_i^{-1}z_j;p,q)}
\\ \nonumber && \makebox[6em]{} \times \prod_{i=1}^{N_f}
\prod_{j=1}^N \Gamma(s_iz_j,t_i^{-1}z_j^{-1};p,q) \prod_{j=1}^{N-1}
\frac{d z_j}{2 \pi \textup{i} z_j}, \eeqa
where $\prod_{j=1}^Nz_j=1$,  $U=(pq)^{\frac{1}{K+1}}$,
and the balancing condition reads $U^NST^{-1}=(pq)^{N_f}$
with $S = \prod_{i=1}^{N_f}s_i, \ T=\prod_{i=1}^{N_f}t_i.$
The magnetic index looks like
\beqa\label{KS2adj2} &&
I_M=\frac{(p;p)_{\infty}^{N-1} (q;q)_{\infty}^{N-1}}{N!}\
\Gamma(U,U^{\frac K2};p,q)^{\widetilde{N}-1}
\prod_{L=0}^{K-1} \prod_{J=0}^2 \prod_{i,j=1}^{N_f}
\Gamma(U^{L+KJ/2}s_it_j^{-1};p,q) \nonumber
\\ && \makebox[4em]{} \times \int_{\mathbb{T}^{\widetilde{N}-1}}
\prod_{1 \leq i < j \leq \widetilde{N}}
\frac{\Gamma(Uz_iz_j^{-1},Uz_i^{-1}z_j;p,q)
\Gamma(U^{K/2}z_iz_j^{-1},U^{K/2}z_i^{-1}z_j;p,q)}
{\Gamma(z_iz_j^{-1},z_i^{-1}z_j;p,q)}
\\ \nonumber && \makebox[4em]{} \times \prod_{i=1}^{N_f}
\prod_{j=1}^{\widetilde{N}} \Gamma(U^{\frac{2-K}{2}}
(ST)^{\frac{3K}{2\widetilde{N}}} s_i^{-1}z_j , U^{\frac{2-K}{2}}
(ST)^{-\frac{3K}{2\widetilde{N}}} t_iz_j^{-1};p,q)
\prod_{j=1}^{\widetilde{N}-1} \frac{d z_j}{2 \pi \textup{i} z_j},
\end{eqnarray}
where $\prod_{j=1}^{\widetilde{N}}z_j=1.$
Again, the conjectured equality $I_E=I_M$ is a new type of identities
requiring a rigorous proof.

\subsection{Generalized KS type dualities}

These dualities were considered in \cite{ILS}.

\subsubsection{First pair of dual theories}

Electric theory: \\

\begin{tabular}{|c|c|c|c|c|c|c|}
  \hline
    & $SU(N)$ & $SU(N_f)$ & $SU(N_f)$ & $U(1)$ & $U(1)_B$ & $U(1)_R$ \\  \hline
  $Q$ & $f$ & $f$ & 1 & 0 & $\frac 1N$ & $2r=1-\frac{N+2K}{(K+1)N_f}$ \\
  $\widetilde{Q}$ & $\overline{f}$ & 1 & $f$ & 0 & $-\frac 1N$ & $2r=1-\frac{N+2K}{(K+1)N_f}$ \\
  $X$ & $T_A$ & 1 & 1 & 1 & $\frac 2N$ & $2s=\frac{1}{K+1}$ \\
  $\widetilde{X}$ & $\overline{T}_A$ & 1 & 1 & -1 & $-\frac 2N$ & $2s=\frac{1}{K+1}$
  \\
\hline
\end{tabular}

$\hspace*{1mm}$

Magnetic theory: \\

 \begin{tabular}{|c|c|c|c|c|c|c|}
  \hline
    & $SU(\widetilde{N})$ & $SU(N_f)$ & $SU(N_f)$ & $U(1)$ & $U(1)_B$ & $U(1)_R$ \\  \hline
  $q$ & $f$ & $\overline{f}$ & 1 & $\frac{K(N_f-2)}{\widetilde{N}}$ & $\frac{1}{\widetilde{N}}$ & $2r'=1-\frac{\widetilde{N}+2K}{(K+1)N_f}$
  \\
  $\widetilde{q}$ & $\overline{f}$ & 1 & $\overline{f}$ & $-\frac{K(N_f-2)}{\widetilde{N}}$ & $-\frac{1}{\widetilde{N}}$ & $2r'=1-\frac{\widetilde{N}+2K}{(K+1)N_f}$
  \\
  $Y$ & $T_A$ & 1 & 1 & $\frac{N-N_f}{\widetilde{N}}$ & $\frac{2}{\widetilde{N}}$ & $2s=\frac{1}{K+1}$
  \\
  $\widetilde{Y}$ & $\overline{T}_A$ & 1 & 1 & $-\frac{N-N_f}{\widetilde{N}}$ & $-\frac{2}{\widetilde{N}}$ & $2s=\frac{1}{K+1}$
  \\
  $M_j$ & 1 & $f$ & $f$ & 0 & 0 &
  $\frac{\widetilde{N}-N+(2j+1)N_f}{N_f(K+1)}$ \\
  $P_r$ & 1 & $T_A$ & 1 & -1 & 0 &
  $\frac{\widetilde{N}-N+(2r+2)N_f}{N_f(K+1)}$ \\
  $\widetilde{P}_r$ & 1 & 1 & $T_A$ & 1 & 0 &
  $\frac{\widetilde{N}-N+(2r+2)N_f}{N_f(K+1)}$ \\
\hline
\end{tabular}

$\hspace*{1mm}$

Here $j=0,\ldots,K$, $r=0,\ldots,K-1$, and
\begin{equation}\label{KS3}
\widetilde{N} \ = \ (2K+1)N_f-4K-N, \ \ \ K=0,1,2,\ldots.
\end{equation}

The electric index is
\begin{eqnarray}\label{KSg1_1}
I_E &=& \frac{(p;p)_{\infty}^{N-1} (q;q)_{\infty}^{N-1}}{N!}
\int_{\mathbb{T}^{N-1}} \prod_{1 \leq i < j \leq N}
\frac{\Gamma(Uz_iz_j,U^{-1}(pq)^{\frac{1}{K+1}}z_i^{-1}z_j^{-1};p,q)}{\Gamma(z_i^{-1}z_j,z_iz_j^{-1};p,q)}
\nonumber\\  && \times \prod_{j=1}^{N} \prod_{k=1}^{N_f}
\Gamma(s_kz_j,t_kz_j^{-1};p,q)\prod_{j=1}^{N-1} \frac{dz_j}{2 \pi \textup{i}
z_j},
\end{eqnarray}
where $\prod_{j=1}^Nz_j=1$  and $U$ is  an  arbitrary parameter.
The magnetic index is
\begin{eqnarray}\label{KSg1_2}
&& \makebox[-1em]{}
I_M = \prod_{j=0}^K \prod_{k, l =1}^{N_f} \Gamma((pq)^{\frac{j}{K+1}}s_kt_l;p,q)
\prod_{r=0}^{K-1} \prod_{1 \leq k < l \leq N_f}
\Gamma(U^{-1}(pq)^{\frac{r+1}{K+1}}s_ks_l,U(pq)^{\frac{r}{K+1}}t_kt_l;p,q)
\nonumber\\
\nonumber && \makebox[1em]{} \times
 \frac{(p;p)_{\infty}^{\widetilde{N}-1}
(q;q)_{\infty}^{\widetilde{N}-1}}{\widetilde{N}!}
 \int_{\mathbb{T}^{\widetilde{N}-1}}  \prod_{1
\leq i < j \leq \widetilde{N}}
\frac{\Gamma(\widetilde{U}z_iz_j,\widetilde{U}^{-1}(pq)^{\frac{1}{K+1}}z_i^{-1}z_j^{-1};p,q)}{\Gamma(z_i^{-1}z_j,z_iz_j^{-1};p,q)}
 \nonumber
\\ && \makebox[1em]{} \times \prod_{j=1}^{\widetilde{N}}
\prod_{k=1}^{N_f} \Gamma((U\widetilde{U})^{\frac 12}
s_k^{-1}z_j,(U\widetilde{U})^{-\frac 12} (pq)^{\frac{1}{K+1}}
t_k^{-1}z_j^{-1};p,q) \prod_{j=1}^{\widetilde{N}-1} \frac{dz_j}{2
\pi \textup{i} z_j},
 \nonumber\end{eqnarray}
where  $\prod_{j=1}^{\widetilde{N}}z_j=1$, the balancing condition looks as
$ST=(pq)^{N_f-\frac{N+2K}{K+1}}$ with  $S =\prod_{j=1}^{N_f} s_j,
\  T=\prod_{j=1}^{N_f}t_j$, and $\widetilde{U}=U^{\frac{N-N_f }{\widetilde{N}}}
(ST^{-1})^{\frac{1}{\widetilde{N}}}
(pq)^{\frac{\widetilde{N}-N+N_f }{2\widetilde{N}(K+1)}}$.

\subsubsection{Second pair of dual theories}

Electric theory: \\

\begin{tabular}{|c|c|c|c|c|c|c|}
  \hline
    & $SU(N)$ & $SU(N_f)$ & $SU(N_f)$ & $U(1)$ & $U(1)_B$ & $U(1)_R$ \\  \hline
  $Q$ & $f$ & $f$ & 1 & 0 & $\frac 1N$ & $2r=1-\frac{N-2K}{(K+1)N_f}$ \\
  $\widetilde{Q}$ & $\overline{f}$ & 1 & $f$ & 0 & $-\frac 1N$ & $2r=1-\frac{N-2K}{(K+1)N_f}$
  \\
  $X$ & $T_S$ & 1 & 1 & 1 & $\frac 2N$ & $2s=\frac{1}{K+1}$ \\
  $\widetilde{X}$ & $\overline{T}_S$ & 1 & 1 & -1 & $-\frac 2N$ & $2s=\frac{1}{K+1}$
  \\
\hline
\end{tabular}

$\hspace*{1mm}$

Magnetic theory: \\

\begin{tabular}{|c|c|c|c|c|c|c|}
  \hline
    & $SU(\widetilde{N})$ & $SU(N_f)$ & $SU(N_f)$ & $U(1)$ & $U(1)_B$ & $U(1)_R$ \\  \hline
  $q$ & $f$ & $\overline{f}$ & 1 & $\frac{K(N_f+2)}{\widetilde{N}}$ & $\frac{1}{\widetilde{N}}$ & $2r'=1-\frac{\widetilde{N}-2K}{(K+1)N_f}$ \\
  $\widetilde{q}$ & $\overline{f}$ & 1 & $\overline{f}$ & $-\frac{K(N_f+2)}{\widetilde{N}}$ & $-\frac{1}{\widetilde{N}}$ & $2r'=1-\frac{\widetilde{N}-2K}{(K+1)N_f}$ \\
  $Y$ & $T_S$ & 1 & 1 & $\frac{N-N_f}{\widetilde{N}}$ & $\frac{2}{\widetilde{N}}$ & $2s=\frac{1}{K+1}$
  \\
  $\widetilde{Y}$ & $\overline{T}_S$ & 1 & 1 & $-\frac{N-N_f}{\widetilde{N}}$ & $-\frac{2}{\widetilde{N}}$ & $2s=\frac{1}{K+1}$
  \\
  $M_j$ & 1 & $f$ & $f$ & 0 & 0 &
  $\frac{\widetilde{N}-N+(2j+1)N_f}{N_f(K+1)}$ \\
  $P_r$ & 1 & $T_S$ & 1 & -1 & 0 &
  $\frac{\widetilde{N}-N+(2r+2)N_f}{N_f(K+1)}$ \\
  $\widetilde{P}_r$ & 1 & 1 & $T_S$ & 1 & 0 &
  $\frac{\widetilde{N}-N+(2r+2)N_f}{N_f(K+1)}$ \\
\hline
\end{tabular}

$\hspace*{1mm}$

Here $j=0,\ldots,K$, $r=0,\ldots,K-1$, and
\begin{equation}\label{KS4}
\widetilde{N} \ = \ (2K+1)N_f+4K-N, \ \ \ K=0,1,2,\ldots.
\end{equation}

The electric index is given by the integral
\begin{eqnarray}\label{KSg2_1}
I_E &=& \frac{(p;p)_{\infty}^{N-1} (q;q)_{\infty}^{N-1}}{N!}
\int_{\mathbb{T}^{N-1}} \prod_{1 \leq i < j \leq N}
\frac{\Gamma(Uz_iz_j,U^{-1} (pq)^{\frac{1}{K+1}}
z_i^{-1}z_j^{-1};p,q)}{\Gamma(z_i^{-1}z_j,z_iz_j^{-1};p,q)}
\\ \nonumber &&  \times \prod_{j=1}^{N} \Gamma(Uz_j^2,U^{-1} (pq)^{\frac{1}{K+1}}z_j^{-2};p,q)
\prod_{k=1}^{N_f} \Gamma(s_kz_j,t_kz_j^{-1};p,q)\prod_{j=1}^{N-1}
\frac{dz_j}{2 \pi \textup{i} z_j},
\end{eqnarray}
where $\prod_{j=1}^Nz_j=1$.
The magnetic index is
\begin{eqnarray}\nonumber
&& \makebox[-1em]{}
I_M = \prod_{j=0}^K \ \ \prod_{k,l =1}^{N_f}
\Gamma( (pq)^{\frac{j}{K+1}}s_kt_l;p,q)
\prod_{r=0}^{K-1}  \ \ \prod_{1 \leq k < l \leq
N_f} \Gamma(U^{-1} (pq)^{\frac{r+1}{K+1}}s_ks_l,U
(pq)^{\frac{r}{K+1}}t_kt_l;p,q)
\\  \label{KSg2_2} && \makebox[1em]{} \times
\prod_{r=0}^{K-1} \prod_{k=1}^{N_f}
\Gamma(U^{-1}
(pq)^{\frac{r+1}{K+1}} s_k^2,U (pq)^{\frac{r}{K+1}} t_k^2;p,q)
\frac{(p;p)_{\infty}^{\widetilde{N}-1}
(q;q)_{\infty}^{\widetilde{N}-1}}{\widetilde{N}!}
 \\  \nonumber && \makebox[1em]{} \times
\int_{\mathbb{T}^{\widetilde{N}-1}} \prod_{1 \leq i < j \leq
\widetilde{N}} \frac{\Gamma(\widetilde{U}z_iz_j,\widetilde{U}^{-1}
(pq)^{\frac{1}{K+1}}
z_i^{-1}z_j^{-1};p,q)}{\Gamma(z_i^{-1}z_j,z_iz_j^{-1};p,q)}
\prod_{j=1}^{\widetilde{N}}
\Gamma(\widetilde{U}z_j^2,\widetilde{U}^{-1} (pq)^{\frac{1}{K+1}}
z_j^{-2};p,q)
 \nonumber
\\ &&  \makebox[1em]{} \times \prod_{j=1}^{\widetilde{N}}
\prod_{k=1}^{N_f} \Gamma((U\widetilde{U})^{\frac 12}
s_k^{-1}z_j,(U\widetilde{U})^{-\frac 12} (pq)^{\frac{1}{K+1}}
t_k^{-1}z_j^{-1};p,q)\prod_{j=1}^{\widetilde{N}-1} \frac{dz_j}{2 \pi
i z_j},
 \nonumber\end{eqnarray}
where $\prod_{j=1}^{\widetilde{N}}z_j=1$,
the balancing condition reads $ST=(pq)^{N_f-\frac{N-2K}{K+1}}$ with
$S = \prod_{j=1}^{N_f} s_j, \ T= \prod_{j=1}^{N_f}t_j$,
and $\widetilde{U}=U^{\frac{N-N_f }{\widetilde{N}}}
(ST^{-1})^{\frac{1}{\widetilde{N}}}
(pq)^{\frac{\widetilde{N}-N+N_f }{2\widetilde{N}(K+1)}}$.

\subsubsection{Third pair of dual theories}
In comparison with the dualities described in previous two subsections,
this case involves non-abelian flavor subgroups of different ranks.

The electric theory: \\

\begin{tabular}{|c|c|c|c|c|c|c|}
  \hline
    & $SU(N)$ & $SU(N_f)$ & $SU(N_f-8)$ & $U(1)$ & $U(1)_B$ & $U(1)_R$ \\  \hline
  $Q$ & $f$ & $f$ & 1 & $-(2K+1)+\frac{2(4K+3)}{N_f}$ & $\frac 1N$ & $2r=1-\frac{N+2(4K+3)}{2(K+1)N_f}$ \\
  $\widetilde{Q}$ & $\overline{f}$ & 1 & $f$ & $2K+1+\frac{2(4K+3)}{N_f-8}$ & $-\frac 1N$ & $2\widetilde{r}=1-\frac{N-2(4K+3)}{2(K+1)(N_f-8)}$ \\
  $X$ & $T_A$ & 1 & 1 & 1 & $\frac 2N$ & $2s=\frac{1}{2(K+1)}$ \\
  $\widetilde{X}$ & $\overline{T}_S$ & 1 & 1 & -1 & $-\frac 2N$ & $2s=\frac{1}{2(K+1)}$
  \\
\hline
\end{tabular}

$\hspace*{1mm}$

The magnetic theory: \\

{\makebox[-1em]{}
\begin{tabular}{|c|c|c|c|c|c|c|}
  \hline
    & $SU(\widetilde{N})$ & $SU(N_f)$ & $SU(N_f-8)$ & $U(1)$ & $U(1)_B$ & $U(1)_R$ \\  \hline
  $q$ & $f$ & $\overline{f}$ & 1 & $2K+1-\frac{2(4K+3)}{N_f}$ & $\frac{1}{\widetilde{N}}$ & $2r'=1-\frac{\widetilde{N}+2(4K+3)}{2(K+1)N_f}$ \\
  $\widetilde{q}$ & $\overline{f}$ & 1 & $\overline{f}$ & $-2K-1-\frac{2(4K+3)}{N_f-8}$ & $-\frac{1}{\widetilde{N}}$ & $2\widetilde{r}'=1-\frac{\widetilde{N}-2(4K+3)}{2(K+1)(N_f-8)}$ \\
  $Y$ & $T_A$ & 1 & 1 & -1 & $\frac{2}{\widetilde{N}}$ & $2s=\frac{1}{2(K+1)}$ \\
  $\widetilde{Y}$ & $\overline{T}_S$ & 1 & 1 & 1 & $-\frac{2}{\widetilde{N}}$ & $2s=\frac{1}{2(K+1)}$ \\
  $M_J$ & 1 & $f$ & $f$ & $\frac{2(4K+3)(2N_f-8)}{N_f(N_f-8)}$ & 0 &
  $2(r+\widetilde{r})+\frac{J}{K+1}$  \\
  $P_{2L}$ & 1 & $T_S$ & 1 & $-4K-3+\frac{4(4K+3)}{N_f}$ & 0 & $4r+\frac{4L+1}{2(K+1)}$  \\
  $P_{2M+1}$ & 1 & $T_A$ & 1 & $-4K-3+\frac{4(4K+3)}{N_f}$ & 0 & $4r+\frac{4M+3}{2(K+1)}$  \\
  $\widetilde{P}_{2L}$ & 1  & 1 & $T_A$ & $4K+3+\frac{2(4K+3)}{N_f-8}$ & 0 & $4\widetilde{r}+\frac{4L+1}{2(K+1)}$  \\
  $\widetilde{P}_{2M+1}$ & 1  & 1 & $T_S$ & $4K+3+\frac{2(4K+3)}{N_f-8}$ & 0 & $4\widetilde{r}+\frac{4M+3}{2(K+1)}$  \\
\hline
\end{tabular}
$\hspace*{1mm}$
}

Here $J=0,\ldots,2K+1,\, L=0,\ldots,K$, $M=0,\ldots,K-1$, and
\begin{equation}\label{KS5}
\widetilde{N} \ = \ (4K+3)(N_f-4)-N, \ \ \ K=0,1,2,\ldots.
\end{equation}

The electric index is
\begin{eqnarray}\label{KSg3_1}
I_E &=& \frac{(p;p)_{\infty}^{N-1} (q;q)_{\infty}^{N-1}}{N!}
\int_{\mathbb{T}^{N-1}} \prod_{1 \leq i < j \leq N}
\frac{\Gamma(Uz_iz_j,U^{-1} (pq)^{\frac{1}{2(K+1)}}
z_i^{-1}z_j^{-1};p,q)}{\Gamma(z_i^{-1}z_j,z_iz_j^{-1};p,q)}
\\ \nonumber &&  \times \prod_{j=1}^{N} \Gamma(U^{-1} (pq)^{\frac{1}{2(K+1)}} z_j^{-2};p,q)
\prod_{k=1}^{N_f} \Gamma(s_kz_j;p,q) \prod_{l=1}^{N_f-8}
\Gamma(t_lz_j^{-1};p,q)\prod_{j=1}^{N-1} \frac{dz_j}{2 \pi \textup{i} z_j} .
\end{eqnarray}
with $\prod_{j=1}^Nz_j=1$ and the balancing condition
$ STU^{-4} = (pq)^{N_f-4-\frac{N+2}{2(K+1)}}$, where
$S = \prod_{j=1}^{N_f} s_j,$ $T =\prod_{j=1}^{N_f-8} t_j.$
The magnetic index is
\begin{eqnarray}\label{KSg3_2}
&& I_M = \frac{(p;p)_{\infty}^{\widetilde{N}-1}
(q;q)_{\infty}^{\widetilde{N}-1}}{\widetilde{N}!} \prod_{J=0}^{2K+1}
\prod_{i=1}^{N_f} \prod_{j=1}^{N_f-8}
\Gamma((pq)^{\frac{J}{2(K+1)}}s_it_j;p,q)
\end{eqnarray}
\begin{eqnarray}\nonumber
&& \times \prod_{l=0}^{2K}
\prod_{1 \leq i < j \leq N_f}
\Gamma((pq)^{\frac{l+1}{2(K+1)}}U^{-1}s_is_j;p,q) \prod_{l=0}^K
\prod_{i=1}^{N_f} \Gamma((pq)^{\frac{2l+1}{2(K+1)}}U^{-1}s_i^2;p,q)
\\ \nonumber &&
\makebox[2em]{} \times \prod_{m=0}^{2K} \prod_{1 \leq i < j \leq
N_f-8} \Gamma((pq)^{\frac{m}{2(K+1)}}Ut_it_j;p,q) \prod_{m=0}^{K-1}
\prod_{i=1}^{N_f-8} \Gamma((pq)^{\frac{2m+1}{2(K+1)}}Ut_i^2;p,q)
 \\
\nonumber && \makebox[-1em]{} \times
\int_{\mathbb{T}^{\widetilde{N}-1}} \prod_{1 \leq i < j \leq
\widetilde{N}} \frac{\Gamma(\widetilde{U}z_iz_j,\widetilde{U}^{-1}
(pq)^{\frac{1}{2(K+1)}}
z_i^{-1}z_j^{-1};p,q)}{\Gamma(z_i^{-1}z_j,z_iz_j^{-1};p,q)}
 \prod_{j=1}^{\widetilde{N}}\Big[
\Gamma(\widetilde{U}^{-1} (pq)^{\frac{1}{2(K+1)}} z_j^{-2};p,q)
 \\ \nonumber && \makebox[0em]{} \times
\prod_{k=1}^{N_f} \Gamma((U\widetilde{U})^{\frac 12}s_k^{-1}z_j;p,q)
\prod_{l=1}^{N_f-8} \Gamma((U\widetilde{U})^{-\frac 12} (pq)^{\frac{1}{2(K+1)}}
t_l^{-1}z_j^{-1};p,q)\Big]
\prod_{j=1}^{\widetilde{N}-1} \frac{dz_j}{2 \pi {\textup i} z_j} ,
\end{eqnarray}
where
$\prod_{j=1}^{\widetilde{N}}z_j=1$ and
$ \widetilde{U}=\big( S^2 U^{N-N_f} \big)^{\frac{1}{\widetilde{N}}}.$

\subsection{Adjoint, symmetric and
conjugate symmetric tensor matter fields}

This duality was constructed by Brodie and Strassler \cite{Brodie2}.
The electric theory is

$\hspace*{1mm}$

\begin{center}
\begin{tabular}{|c|c|c|c|c|c|c|}
  \hline
   & $SU(N)$ & $SU(N_f)$ & $SU(N_f)$ & $U(1)$ & $U(1)_B$ & $U(1)_R$ \\  \hline
  $Q$ & $f$ & $f$ & 1 & 0 & $\frac{1}{N}$ & $1-\frac{N-2}{N_f(K+1)}$ \\
  $\widetilde{Q}$ & $\overline{f}$ & 1 & $\overline{f}$ & 0 & $-\frac 1N$ & $1-\frac{N-2}{N_f(K+1)}$ \\
  $X$ & $adj$ & 1 & 1 & 0 & 0 & $\frac{2}{K+1}$  \\
  $Y$ & $T_S$ & 1 & 1 & 1 & $\frac 2N$ & $\frac{K}{K+1}$  \\
  $\widetilde{Y}$ & $\overline{T}_S$ & 1 & 1 & -1 & $-\frac 2N$ & $\frac{K}{K+1}$  \\
\hline
\end{tabular}
\end{center}

$\hspace*{1mm}$

The magnetic theory is

$\hspace*{1mm}$

\begin{center}
\begin{tabular}{|c|c|c|c|c|c|c|}
  \hline
   & $SU(\widetilde{N})$ & $SU(N_f)$ & $SU(N_f)$ & $U(1)$ & $U(1)_B$ & $U(1)_R$ \\  \hline
  $q$ & $f$ & $\overline{f}$ & 1 & $\frac{KN_f+2}{\widetilde{N}}$ & $\frac{1}{\widetilde{N}}$ & $1-\frac{\widetilde{N}-2}{N_f(K+1)}$ \\
  $\widetilde{q}$ & $\overline{f}$ & 1 & $f$ & $-\frac{KN_f+2}{\widetilde{N}}$ & $-\frac{1}{\widetilde{N}}$ & $1-\frac{\widetilde{N}-2}{N_f(K+1)}$ \\
  $X$ & $adj$ & 1 & 1 & 0 & 0 & $\frac{2}{K+1}$  \\
  $Y$ & $T_S$ & 1 & 1 & $\frac{N-KN_f}{\widetilde{N}}$ & $\frac{2}{\widetilde{N}}$ & $\frac{K}{K+1}$  \\
  $\overline{Y}$ & $\overline{T}_S$ & 1 & 1 & $-\frac{N-KN_f}{\widetilde{N}}$ & $-\frac{2}{\widetilde{N}}$ & $\frac{K}{K+1}$  \\
  $N_I$ & 1 & $f$ & $\overline{f}$ & 0 & 0 &
  $\frac{2I}{K+1}+\frac{2K}{K+1}+2-2\frac{N-2}{N_f(K+1)}$ \\
  $M_I$ & 1 & $f$ & $\overline{f}$ & 0 & 0 &
  $\frac{2I}{K+1}+2-2\frac{N-2}{N_f(K+1)}$ \\
  $P_{2J+1}$ & 1 & $T_A$ & 1 & -1 & 0 &
  $2\frac{2J+1}{K+1}+\frac{K}{K+1}+2-2\frac{N-2}{N_f(K+1)}$ \\
  $P_{2J}$ & 1 & $T_S$ & 1 & -1 & 0 &
  $2\frac{2J}{K+1}+\frac{K}{K+1}+2-2\frac{N-2}{N_f(K+1)}$ \\
  $\widetilde{P}_{2J+1}$ & 1 & 1 & $\overline{T}_A$ & 1 & 0 &
  $2\frac{2J+1}{K+1}+\frac{K}{K+1}+2-2\frac{N-2}{N_f(K+1)}$ \\
  $\widetilde{P}_{2J}$ & 1 & 1 & $\overline{T}_S$ & 1 & 0 &
  $2\frac{2J}{K+1}+\frac{K}{K+1}+2-2\frac{N-2}{N_f(K+1)}$ \\
\hline
\end{tabular}
\end{center}

$\hspace*{1mm}$

Here $K$ is odd, $I=0,1,\ldots,K-1$, $J=0,1,\ldots,\frac{K-1}{2}$,
but there  are no fields $P_K, \tilde P_K$, and
\beq\label{B1}
\widetilde{N}=3KN_f+4-N.
\eeq

The indices are
\begin{eqnarray}\label{Br1_1}\makebox[-2em]{}
&&I_E= \frac{(p;p)_{\infty}^{N-1} (q;q)_{\infty}^{N-1}}{N!}
\Gamma(U;p,q)^{N-1}
\int_{\mathbb{T}^{N-1}}   \prod_{1 \leq i < j \leq N}
\frac{\Gamma(Uz_iz_j^{-1},Uz_i^{-1}z_j;p,q)}{\Gamma(z_iz_j^{-1},z_i^{-1}z_j;p,q)}
\nonumber \\ && \makebox[4em]{} \times \prod_{1 \leq i < j \leq N}
\Gamma(U^{K/2}XYz_iz_j,U^{K/2}(XY)^{-1}z_i^{-1}z_j^{-1};p,q)
\\ && \makebox[1em]{} \times
\prod_{j=1}^{N}\Big[ \Gamma(U^{K/2}XYz_j^2,U^{K/2}(XY)^{-1}z_j^{-2};p,q)
\prod_{i=1}^{N_f} \Gamma(s_iz_j,t_i^{-1}z_j^{-1};p,q)\Big]
\prod_{j=1}^{N-1} \frac{d z_j}{2 \pi \textup{i} z_j},
\nonumber \eeqa
where $U=(pq)^{\frac{1}{K+1}}$, $\prod_{j=1}^Nz_j=1$, and
\beqa\label{Br1_2} && I_M =
\frac{(p;p)_{\infty}^{\widetilde{N}-1}
(q;q)_{\infty}^{\widetilde{N}-1}}{\widetilde{N}!}
\Gamma(U;p,q)^{\widetilde{N}-1}
\prod_{L=0}^{K-1}
\prod_{i,j=1}^{N_f} \Gamma(U^{L+K}s_it_j^{-1},U^{L}s_it_j^{-1};p,q)
\nonumber \\ && \makebox[2em]{} \times \prod_{J=0}^{K-1} \prod_{1
\leq i < j \leq N_f} \Gamma((XY)^{-1}U^{J+K/2}s_is_j,XYU^{J+K/2}t_i^{-1}t_j^{-1};p,q) \nonumber \\
&& \makebox[-1em]{} \times \prod_{J=0}^{\frac{K-1}{2}}
\prod_{i=1}^{N_f}
\Gamma((XY)^{-1}U^{2J+\frac{K}{2}}s_i^2,XYU^{2J+\frac{K}{2}}t_i^{-2};p,q)
\int_{\mathbb{T}^{\widetilde{N}-1}}
\prod_{1 \leq i < j \leq \widetilde{N}}
\frac{\Gamma(Uz_iz_j^{-1},Uz_i^{-1}z_j;p,q)}{\Gamma(z_iz_j^{-1},z_i^{-1}z_j;p,q)}
\nonumber
\\ && \makebox[2em]{} \times \prod_{1 \leq
i < j \leq \widetilde{N}} \Gamma(U^{K/2}
X^{\frac{N-KN_f}{\widetilde{N}}}
Y^{\frac{N}{\widetilde{N}}}z_iz_j,U^{K/2}(X^{\frac{N-KN_f}{\widetilde{N}}}
Y^{\frac{N}{\widetilde{N}}})^{-1}z_i^{-1}z_j^{-1};p,q) \nonumber
\\ && \makebox[2em]{} \times \prod_{j=1}^{\widetilde{N}} \Big[
\Gamma(U^{K/2} X^{\frac{N-KN_f}{\widetilde{N}}}Y^{\frac{N}{\widetilde{N}}}
z_j^2,U^{K/2}(X^{\frac{N-KN_f}{\widetilde{N}}}
Y^{\frac{N}{\widetilde{N}}})^{-1}z_j^{-2};p,q)
\\ \nonumber && \makebox[-1em]{} \times \prod_{i=1}^{N_f}
\Gamma( U^{\frac{2-K}{2}}
X^{\frac{KN_f+2}{\widetilde{N}}} Y^{\frac{3KN_f+4}{2\widetilde{N}}}
s_i^{-1}z_j,U^{\frac{2-K}{2}} X^{-\frac{KN_f+2}{\widetilde{N}}}
Y^{-\frac{3KN_f+4}{2\widetilde{N}}} t_iz_j^{-1};p,q)\Big]
\prod_{j=1}^{\widetilde{N}-1} \frac{d z_j}{2 \pi \textup{i} z_j},
\nonumber \end{eqnarray}
where  $Y = (ST)^{1/N_f},\ S=\prod_{i=1}^{N_f} s_i,\
T=\prod_{i=1}^{N_f} t_i$, $X$ is an arbitrary
chemical potential associated with the $U(1)$-group,
and the balancing condition reads $U^{N-2}ST^{-1}=(pq)^{N_f}.$

\subsection{Adjoint, anti-symmetric and
conjugate anti-symmetric tensor matter fields}

This duality was considered in \cite{Brodie2}.
The electric theory is

\begin{center}
\begin{tabular}{|c|c|c|c|c|c|c|}
  \hline
   & $SU(N)$ & $SU(N_f)$ & $SU(N_f)$ & $U(1)$ & $U(1)_B$ & $U(1)_R$ \\  \hline
  $Q$ & $f$ & $f$ & 1 & 0 & $\frac{1}{N}$ & $1-\frac{N+2}{N_f(K+1)}$ \\
  $\widetilde{Q}$ & $\overline{f}$ & 1 & $\overline{f}$ & 0 & $-\frac 1N$ & $1-\frac{N+2}{N_f(K+1)}$ \\
  $X$ & $adj$ & 1 & 1 & 0 & 0 & $\frac{2}{K+1}$  \\
  $Y$ & $T_A$ & 1 & 1 & 1 & $\frac 2N$ & $\frac{K}{K+1}$  \\
  $\widetilde{Y}$ & $\overline{T}_A$ & 1 & 1 & -1 & $-\frac 2N$ & $\frac{K}{K+1}$  \\
\hline
\end{tabular}
\end{center}

The magnetic theory is

\begin{center}
\begin{tabular}{|c|c|c|c|c|c|c|}
  \hline
   & $SU(\widetilde{N})$ & $SU(N_f)$ & $SU(N_f)$ & $U(1)$ & $U(1)_B$ & $U(1)_R$ \\  \hline
  $q$ & $f$ & $\overline{f}$ & 1 & $\frac{KN_f-2}{\widetilde{N}}$ & $\frac{1}{\widetilde{N}}$ & $1-\frac{\widetilde{N}+2}{N_f(K+1)}$ \\
  $\widetilde{q}$ & $\overline{f}$ & 1 & $f$ & $-\frac{KN_f-2}{\widetilde{N}}$ & $-\frac{1}{\widetilde{N}}$ & $1-\frac{\widetilde{N}+2}{N_f(K+1)}$ \\
  $X$ & $adj$ & 1 & 1 & 0 & 0 & $\frac{2}{K+1}$  \\
  $Y$ & $T_A$ & 1 & 1 & $\frac{N-KN_f}{\widetilde{N}}$ & $\frac{2}{\widetilde{N}}$ & $\frac{K}{K+1}$  \\
  $\widetilde{Y}$ & $\overline{T}_A$ & 1 & 1 & $-\frac{N-KN_f}{\widetilde{N}}$ & $-\frac{2}{\widetilde{N}}$ & $\frac{K}{K+1}$  \\
  $N_I$ & 1 & $f$ & $\overline{f}$ & 0 & 0 &
  $\frac{2I}{K+1}+\frac{2K}{K+1}+2-2\frac{N+2}{N_f(K+1)}$ \\
  $M_I$ & 1 & $f$ & $\overline{f}$ & 0 & 0 &
  $\frac{2I}{K+1}+2-2\frac{N+2}{N_f(K+1)}$ \\
  $P_{2J+1}$ & 1 & $T_S$ & 1 & -1 & 0 &
  $2\frac{2J+1}{K+1}+\frac{K}{K+1}+2-2\frac{N+2}{N_f(K+1)}$ \\
  $P_{2J}$ & 1 & $T_A$ & 1 & -1 & 0 &
  $2\frac{2J}{K+1}+\frac{K}{K+1}+2-2\frac{N+2}{N_f(K+1)}$ \\
  $\widetilde{P}_{2J+1}$ & 1 & 1 & $\overline{T}_S$ & 1 & 0 &
  $2\frac{2J+1}{K+1}+\frac{K}{K+1}+2-2\frac{N+2}{N_f(K+1)}$ \\
  $\widetilde{P}_{2J}$ & 1 & 1 & $\overline{T}_A$ & 1 & 0 &
  $2\frac{2J}{K+1}+\frac{K}{K+1}+2-2\frac{N+2}{N_f(K+1)}$ \\
\hline
\end{tabular}
\end{center}
$\hspace*{1mm}$

Here $K$ is odd, $I=0,\ldots,K-1$, $J=0,\ldots,\frac{K-1}{2}$,
but there  are no fields $P_K, \ \tilde P_K$, and
\beq\label{B2}
\widetilde{N}=3KN_f-4-N.
\eeq

The superconformal indices are
\begin{eqnarray}\label{Br2_1}\makebox[-2em]{}
&&I_E= \frac{(p;p)_{\infty}^{N-1} (q;q)_{\infty}^{N-1}}{N!}
\Gamma(U;p,q)^{N-1}
\int_{\mathbb{T}^{N-1}}   \prod_{1 \leq i < j \leq N}
\frac{\Gamma(Uz_iz_j^{-1},Uz_i^{-1}z_j;p,q)}{\Gamma(z_iz_j^{-1},z_i^{-1}z_j;p,q)}
\\   && \makebox[-1em]{} \times \prod_{1 \leq i < j \leq N}
\Gamma(U^{K/2}XYz_iz_j,U^{K/2}(XY)^{-1}z_i^{-1}z_j^{-1};p,q)
\prod_{i=1}^{N_f} \prod_{j=1}^N
\Gamma(s_iz_j,t_i^{-1}z_j^{-1};p,q) \prod_{j=1}^{N-1} \frac{d z_j}{2
\pi \textup{i} z_j},
\nonumber \end{eqnarray}
for $U=(pq)^{\frac{1}{K+1}},$ $\prod_{j=1}^Nz_j=1$, and
\begin{eqnarray}\label{Br2_2} &&
I_M= \frac{(p;p)_{\infty}^{\widetilde{N}-1}
(q;q)_{\infty}^{\widetilde{N}-1}}{\widetilde{N}!}
\Gamma(U;p,q)^{\widetilde{N}-1}
\prod_{L=0}^{K-1}
\prod_{i,j=1}^{N_f} \Gamma(U^{L+K}s_it_j^{-1},U^{L}s_it_j^{-1};p,q)
\\ && \makebox[4em]{} \times \prod_{J=0}^{K-1} \prod_{1
\leq i < j \leq N_f} \Gamma((XY)^{-1}U^{J+K/2}s_is_j,XYU^{J+K/2}t_i^{-1}t_j^{-1};p,q) \nonumber \\
&& \makebox[-2em]{} \times \prod_{J=0}^{\frac{K-3}{2}}
\prod_{i=1}^{N_f}
\Gamma((XY)^{-1}U^{2J+1+\frac{K}{2}}s_i^2,XYU^{2J+1+\frac{K}{2}}t_i^{-2};p,q)
\int_{\mathbb{T}^{\widetilde{N}-1}}
\prod_{1 \leq i < j \leq \widetilde{N}}
\frac{\Gamma(Uz_iz_j^{-1},Uz_i^{-1}z_j;p,q)}{\Gamma(z_iz_j^{-1},z_i^{-1}z_j;p,q)}
\nonumber  \\ && \makebox[4em]{} \times
\prod_{1 \leq   i < j \leq \widetilde{N}} \Gamma(U^{K/2}
X^{\frac{N-KN_f}{\widetilde{N}}} Y^{\frac{N}{\widetilde{N}}}
z_iz_j,U^{K/2}( X^{\frac{N-KN_f}{\widetilde{N}}}
Y^{\frac{N}{\widetilde{N}}})^{-1}z_i^{-1}z_j^{-1};p,q)
\nonumber \\ \nonumber && \makebox[0em]{} \times \prod_{i=1}^{N_f}
\prod_{j=1}^{\widetilde{N}} \Gamma( U^{\frac{2-K}{2}}
X^{\frac{KN_f-2}{\widetilde{N}}} Y^{\frac{3KN_f-4}{2\widetilde{N}}}
s_i^{-1}z_j,U^{\frac{2-K}{2}}
X^{-\frac{KN_f-2}{\widetilde{N}}}
Y^{-\frac{3KN_f-4}{2\widetilde{N}}} t_iz_j^{-1};p,q)
\prod_{j=1}^{\widetilde{N}-1} \frac{d z_j}{2 \pi \textup{i} z_j} ,
 \nonumber \end{eqnarray}
where  $\prod_{j=1}^{\widetilde{N}}z_j=1$, $Y = (ST)^{1/N_f}, \
S =  \prod_{i=1}^{N_f} s_i, \
T =  \prod_{i=1}^{N_f} t_i$, $X$ is an arbitrary
parameter and the balancing condition reads $U^{N+2}ST^{-1}=(pq)^{N_f}.$

\subsection{Adjoint, anti-symmetric and
conjugate symmetric tensor matter fields}

This duality was discussed by Brodie in \cite{Brodie2}.
The electric theory is

\begin{center}
\begin{tabular}{|c|c|c|c|c|c|c|}
  \hline
   & $SU(N)$ & $SU(N_f)$ & $SU(N_f-8)$ & $U(1)$ & $U(1)_B$ & $U(1)_R$ \\  \hline
  $Q$ & $f$ & $f$ & 1 & $x_1=\frac{6}{N_f}-1$ & $\frac{1}{N}$ & $2r_1=1-\frac{N+6K}{N_f(K+1)}$ \\
  $\widetilde{Q}$ & $\overline{f}$ & 1 & $f$ & $x_2=\frac{6}{N_f-8}+1$ & $-\frac 1N$ & $2r_2=1-\frac{N-6K}{(N_f-8)(K+1)}$ \\
  $X$ & $adj$ & 1 & 1 & 0 & 0 & $\frac{2}{K+1}$  \\
  $Y$ & $T_A$ & 1 & 1 & 1 & $\frac 2N$ & $\frac{K}{K+1}$  \\
  $\widetilde{Y}$ & $\overline{T}_S$ & 1 & 1 & -1 & $-\frac 2N$ & $\frac{K}{K+1}$  \\
\hline
\end{tabular}
\end{center}
In the original paper \cite{Brodie2} there were misprints for the values
of $U(1)$-group hypercharges which were corrected in \cite{Klein}.
The magnetic theory is
\begin{center}
\begin{tabular}{|c|c|c|c|c|c|c|}
\hline & $SU(\widetilde{N})$ & $SU(N_f)$ & $SU(N_f-8)$ & $U(1)$ &
$U(1)_B$ & $U(1)_R$ \\ \hline $q$ & $f$ & $\overline{f}$ & 1 &
$1-\frac{6}{N_f}$ & $\frac{1}{\widetilde{N}}$ &
$1-\frac{\widetilde{N}+6K}{N_f(K+1)}$ \\ $\widetilde{q}$ &
$\overline{f}$ & 1 & $\overline{f}$ & $-1-\frac{6}{N_f-8}$ &
$-\frac{1}{\widetilde{N}}$ &
$1-\frac{\widetilde{N}-6K}{(N_f-8)(K+1)}$ \\   $X$ & $adj$ & 1 & 1 &
0 & 0 & $\frac{2}{K+1}$ \\   $Y$ & $T_A$ & 1 & 1 & $-1$ &
$\frac{2}{\widetilde{N}}$ & $\frac{K}{K+1}$ \\ $\widetilde{Y}$ &
$\overline{T}_S$ & 1 & 1 & 1 & $-\frac{2}{\widetilde{N}}$ &
$\frac{K}{K+1}$ \\   $N_J$ & 1 & $f$ & $f$ & $x_1+x_2$ & 0 &
$\frac{2J}{K+1}+\frac{2K}{K+1}+2r_1+2r_2$ \\   $M_J$ & 1 & $f$ & $f$
& $x_1+x_2$ & 0 & $\frac{2J}{K+1}+2r_1+2r_2$ \\   $P_{J}$ & 1 &
$T_S$ & 1 & $2x_1-1$ & 0 &
$\frac{2J}{K+1}+\frac{K}{K+1}+2-2\frac{\widetilde{N}+6K}{N_f(K+1)}$
\\   $\widetilde{P}_{J}$ & 1 & 1 & $T_A$ & $2x_2+1$ & 0 &
$\frac{2J}{K+1}+\frac{K}{K+1}+2-2\frac{\widetilde{N}-6K}{(N_f-8)(K+1)}$
\\
\hline
\end{tabular}
\end{center}

$\hspace*{1mm}$

Here $J=0,1,\ldots,K-1$ and
\beq\label{B3} \widetilde{N}=3K(N_f-4)-N.\eeq

The superconformal indices are
\begin{eqnarray}\makebox[-2em]{}
&&I_E= \frac{(p;p)_{\infty}^{N-1} (q;q)_{\infty}^{N-1}}{N!}
\Gamma(U;p,q)^{N-1}
\int_{\mathbb{T}^{N-1}} \prod_{1 \leq i < j \leq N}
\frac{\Gamma(Uz_iz_j^{-1},Uz_i^{-1}z_j;p,q)}
{\Gamma(z_iz_j^{-1},z_i^{-1}z_j;p,q)}\nonumber
\\  \label{Br3_1}
&& \makebox[4em]{} \prod_{1 \leq i < j \leq N}
\Gamma(U^{K/2}XYz_iz_j,U^{K/2}(XY)^{-1}z_i^{-1}z_j^{-1};p,q)
\\ && \makebox[0em]{} \times \prod_{i=1}^{N}
\Gamma(U^{K/2}(XY)^{-1}z_i^{-2};p,q)
 \prod_{j=1}^N \prod_{i=1}^{N_f}
\Gamma(s_iz_j;p,q) \prod_{k=1}^{N_f-8} \Gamma(t_kz_j^{-1};p,q)
\prod_{j=1}^{N-1} \frac{d z_j}{2 \pi \textup{i} z_j}
\nonumber\end{eqnarray}
for $U=(pq)^{\frac{1}{K+1}},$ $\prod_{j=1}^Nz_j=1$, and
\begin{eqnarray}\label{Br3_2}  &&\makebox[-2em]{}
I_M = \frac{(p;p)_{\infty}^{\widetilde{N}-1}
(q;q)_{\infty}^{\widetilde{N}-1}}{\widetilde{N}!}
\Gamma(U;p,q)^{\widetilde{N}-1}
\prod_{L=0}^{K-1}
\prod_{i=1}^{N_f} \prod_{j=1}^{N_f-8}
\Gamma(U^{L+K}s_it_j,U^{L}s_it_j;p,q) \nonumber \\ &&
\makebox[0em]{} \times \prod_{J=0}^{K-1} \prod_{1 \leq i < j \leq
N_f} \Gamma((XY)^{-1}U^{J+K/2}s_is_j;p,q)
\prod_{1 \leq i < j \leq N_f-8} \Gamma(XYU^{J+K/2}t_it_j;p,q) \nonumber \\
&& \makebox[0em]{} \times \prod_{J=0}^{K-1} \prod_{i=1}^{N_f}
\Gamma((XY)^{-1}U^{J+K/2}s_i^2;p,q)
\int_{\mathbb{T}^{\widetilde{N}-1}}
\prod_{1 \leq i < j \leq \widetilde{N}}
\frac{\Gamma(Uz_iz_j^{-1},Uz_i^{-1}z_j;p,q)}{\Gamma(z_iz_j^{-1},z_i^{-1}z_j;p,q)}\nonumber
\\ && \makebox[2em]{} \times \prod_{1 \leq
i < j \leq \widetilde{N}} \Gamma(U^{K/2} X^{-1}
Y^{\frac{N}{\widetilde{N}}} z_iz_j,U^{K/2}
XY^{-\frac{N}{\widetilde{N}}} z_i^{-1}z_j^{-1};p,q) \nonumber
\\ && \makebox[0em]{} \times \prod_{i=1}^{\widetilde{N}}
\Gamma(U^{K/2} X Y^{-\frac{N}{\widetilde{N}}} z_i^{-2};p,q)
\prod_{j=1}^{\widetilde{N}} \prod_{i=1}^{N_f}
\Gamma(U^{\frac{2-K}{2}}Y^{\frac{3K(N_f-4)}{2\widetilde{N}}}
s_i^{-1}z_j;p,q) \nonumber
\\ && \makebox[2em]{} \times \prod_{j=1}^{\widetilde{N}} \prod_{k=1}^{N_f-8}
\Gamma(U^{\frac{2-K}{2}} Y^{-\frac{3K(N_f-4)}{2\widetilde{N}}}
t_k^{-1}z_j^{-1};p,q)
\prod_{j=1}^{\widetilde{N}-1} \frac{d z_j}{2 \pi \textup{i} z_j},\nonumber
\end{eqnarray}
where $\prod_{j=1}^{\widetilde{N}}z_j=1$,
$Y =\big(ST^{-1}X^{2N_f-8} (pq)^{\frac{2(K-2)}{K+1}}
\big)^{\frac{1}{N_f-4}},$
and the balancing condition reads $U^{N}X^{-4}Y^{-4}ST=(pq)^{N_f-4}$
with $S = \prod_{i=1}^{N_f} s_i, \  T = \prod_{i=1}^{N_f-8} t_i$.

$\hspace*{1mm}$

The equalities $I_E=I_M$ for all the dualities described in this
section require a rigorous mathematical confirmation. For the
moment we have only one justifying argument coming from
the total ellipticity condition associated with the kernels of
the corresponding pairs of integrals.

\section{KS type dualities for symplectic gauge groups}

\subsection{The anti-symmetric tensor matter field}

For $SP(2N)$ group the following electric-magnetic duality
was discovered by Intriligator in \cite{Intriligator2}.
The electric theory: \\

\begin{tabular}{|c|c|c|c|}
  \hline
    & $SP(2N)$ & $SU(2N_f)$ & $U(1)_R$ \\  \hline
  $Q$ & $f$ & $f$ & $2r=1-\frac{2(N+K)}{(K+1)N_f}$ \\
  $X$ & $T_A$ & 1 & $2s=\frac{2}{K+1}$ \\
\hline
\end{tabular}

$\hspace*{1mm}$

The magnetic theory: \\

\begin{tabular}{|c|c|c|c|}
  \hline
   & $SP(2\widetilde{N})$ & $SU(2N_f)$ & $U(1)_R$ \\  \hline
  $q$ & f & $\overline{f}$ & $2\widetilde{r}=1-\frac{2(\widetilde{N} +K)}{(K+1)N_f}$
  \\
  $Y$ & $T_A$ & 1 & $2s=\frac{2}{K+1}$ \\
  $M_j$  & 1 & $T_A$ & $2r_j=2\frac{K+j}{K+1}- 4 \frac{\widetilde{N} +K}{(K+1)N_f}$ \\
\hline
\end{tabular}
\\

where $j=1, \ldots, K,$ and
\begin{equation}\label{KSeq}
\widetilde{N} \ = \ K(N_f-2)-N, \ \ \ K=1,2,\ldots.
\end{equation}

Defining $U=(pq)^s=(pq)^{\frac{1}{K+1}}$, we find
the following indices for these theories

\begin{eqnarray}\label{KSsp1_1}
    I_E &=& \frac{(p;p)_{\infty}^{N} (q;q)_{\infty}^{N} }{2^N N!} \Gamma(U;p,q)^{N-1} \int_{\mathbb{T}^N}
   \prod_{1 \leq i < j \leq N} \frac{\Gamma(U z_i^{\pm 1} z_j^{\pm
   1};p,q)}{\Gamma(z_i^{\pm 1} z_j^{\pm 1};p,q)}
  \nonumber  \\
    && \times \prod_{j=1}^{N} \frac{\prod_{i=1}^{2N_f}  \Gamma(s_i z_j^{\pm 1};p,q)}
     {\Gamma(z_j^{\pm 2};p,q)}\prod_{j=1}^{N} \frac{d z_j}{2 \pi \textup{i} z_j}
\end{eqnarray}
and
\begin{eqnarray}\label{KSsp1_2}
    I_M &=&  \frac{(p;p)_{\infty}^{\widetilde{N}} (q;q)_{\infty}^{\widetilde{N}} }{2^{\widetilde{N}}
    \widetilde{N}!} \Gamma(U;p,q)^{\widetilde{N}-1} \prod_{l=1}^K \prod_{1 \leq i < j \leq 2N_f} \Gamma(U^{l-1} s_i s_j;p,q)
    \\ \nonumber
    &&  \times \int_{\mathbb{T}^{\widetilde{N}}}  \prod_{1 \leq i < j \leq \widetilde{N}} \frac{\Gamma(U z_i^{\pm 1} z_j^{\pm
    1};p,q)}{\Gamma(z_i^{\pm 1} z_j^{\pm 1};p,q)} \prod_{j=1}^{\widetilde{N}} \frac{\prod_{i=1}^{2N_f} \Gamma(U s_i^{-1} z_j^{ \pm 1};p,q)}
    {\Gamma(z_j^{\pm 2};p,q)}\prod_{j=1}^{\widetilde{N}}\frac{d z_j}{2 \pi \textup{i} z_j},
\end{eqnarray}
where the balancing condition reads
$U^{2(N+K)}\prod_{i=1}^{2N_f}s_i= (pq)^{N_f}.$

\subsection{Symmetric tensor matter field}

Another  electric-magnetic duality is described by Leigh and Strassler
in \cite{LSS}. The electric theory: \\

\begin{tabular}{|c|c|c|c|}
  \hline
    & $SP(2N)$ & $SU(2N_f)$ & $U(1)_R$ \\  \hline
  $Q$ & $f$ & $f$ & $2r=1-\frac{N+1}{(K+1)N_f}$ \\
  $X$ & $adj=T_S$ & 1 & $2s=\frac{1}{K+1}$ \\
\hline
\end{tabular}

$\hspace*{1mm}$

The magnetic theory: \\

\begin{tabular}{|c|c|c|c|}
  \hline
   & $SP(2\widetilde{N})$ & $SU(2N_f)$ & $U(1)_R$ \\  \hline
  $q$ & $f$ & $\overline{f}$ & $2\widetilde{r}=1-\frac{\widetilde{N}+1}{(K+1)N_f}$
  \\
  $Y$ & $adj$ & 1 & $2s=\frac{1}{K+1}$ \\
  $M_{2j}$, $j=0, \ldots ,K$ & 1 & $T_A$ & $2r_{2j}=2- \frac{2(N+1)-2 j N_f}{(K+1)N_f}$ \\
  $M_{2j+1}$, $j=0, \ldots ,K-1$ & 1 & $T_S$ & $2r_{2j+1}=2- \frac{2(N+1)-(2j+1) N_f}{(K+1)N_f}$ \\
\hline
\end{tabular}

$\hspace*{1mm}$

Here
\begin{equation}\label{KS2eq}
\widetilde{N} \ = \ (2K + 1) N_f - N-2, \ \ \ K=0,1,2,\ldots.
\end{equation}

Defining $U=(pq)^s=(pq)^{\frac{1}{2(K+1)}}$,  we
find the following superconformal indices
\begin{eqnarray}\label{KSsp2_1}
    I_E &=& \frac{(p;p)_{\infty}^{N} (q;q)_{\infty}^{N} }{2^N N!} \Gamma(U;p,q)^{N} \int_{\mathbb{T}^N}\prod_{1 \leq i < j \leq N} \frac{\Gamma(U z_i^{\pm 1} z_j^{\pm 1};p,q)}
    {\Gamma(z_i^{\pm 1} z_j^{\pm 1};p,q)}
    \\ \nonumber
    && \times \prod_{j=1}^N \frac{\Gamma(Uz_j^{\pm
    2};p,q)}{\Gamma(z_j^{\pm 2};p,q)} \prod_{i=1}^{2N_f} \prod_{j=1}^N \Gamma(s_i z_j^{\pm
    1};p,q)\prod_{j=1}^N
     \frac{d z_j}{2 \pi \textup{i} z_j}
\end{eqnarray}
and
\begin{eqnarray}\label{KSsp2_2}
    I_M &=& \frac{(p;p)_{\infty}^{\widetilde{N}} (q;q)_{\infty}^{\widetilde{N}} }
    {2^{\widetilde{N}}  \widetilde{N}!} \Gamma(U;p,q)^{\widetilde{N}}
    \prod_{l=0}^{2K} \prod_{1 \leq i < j \leq 2N_f} \Gamma(U^l s_i s_j;p,q)
     \\ \nonumber  && \times \prod_{l=0}^{K-1}
     \prod_{i=1}^{2N_f} \Gamma(U^{2l+1} s_i^{2};p,q) \int_{\mathbb{T}^{\widetilde{N}}}
     \prod_{1 \leq i < j \leq \widetilde{N}} \frac{\Gamma(U z_i^{\pm 1} z_j^{\pm 1};p,q)
     }{\Gamma(z_i^{\pm 1} z_j^{\pm 1};p,q)}
    \\ \nonumber   && \makebox[2em]{} \times
\prod_{j=1}^{\widetilde{N}} \frac{\Gamma(U z_j^{\pm
    2};p,q)}{\Gamma(z_j^{\pm 2};p,q)} \prod_{i=1}^{2N_f} \prod_{j=1}^{\widetilde{N}}
    \Gamma(U s_i^{-1} z_j^{\pm 1};p,q)\prod_{j=1}^{\widetilde{N}} \frac{d z_j}{2 \pi \textup{i} z_j},
\end{eqnarray}
where the balancing condition
reads $U^{2(N+1)}\prod_{i=1}^{2N_f}s_i= (pq)^{N_f}.$

\subsection{Two anti-symmetric tensor matter fields}

This duality was investigated by Brodie and Strassler in
\cite{Brodie2}.
The electric theory:  \\

\begin{tabular}{|c|c|c|c|}
  \hline
    & $SP(2N)$ & $SU(2N_f)$ & $U(1)_R$ \\  \hline
  $Q$ & $f$ & $f$ & $1-\frac{N+2K+1}{(K+1)N_f}$ \\
  $X$ & $T_A$ & 1 & $\frac{2}{K+1}$ \\
  $Y$ & $T_A$ & 1 & $\frac{K}{K+1}$ \\
\hline
\end{tabular}

$\hspace*{1mm}$

The magnetic theory: \\

\begin{tabular}{|c|c|c|c|}
  \hline
   & $SP(2\widetilde{N})$ & $SU(2N_f)$ & $U(1)_R$ \\  \hline
  $q$ & $f$ & $\overline{f}$ & $1-\frac{\widetilde{N}+2K+1}{(K+1)N_f}$
  \\
  $\widetilde{X}$ & $T_A$ & 1 & $\frac{2}{K+1}$ \\
  $\widetilde{Y}$ & $T_A$ & 1 & $\frac{K}{K+1}$ \\
  $M_{J0}$, $J=0, \ldots ,K-1$ & 1 & $T_A$ & $2- \frac{N+2K+1}{ (K+1)N_f}+\frac{2J}{K+1}$ \\
  $M_{2J \ 1}$, $J=0, \ldots ,\frac{K-1}{2}$ & 1 & $T_A$ & $2- \frac{N+2K+1}{ (K+1)N_f}+\frac{2(2J)}{K+1}+\frac{K}{K+1}$ \\
  $M_{2J+1 \ 1}$, $J=0, \ldots ,\frac{K-3}{2}$ & 1 & $T_S$ & $2- \frac{N+2K+1}{ (K+1)N_f}+\frac{2(2J+1)}{K+1}+\frac{K}{K+1}$ \\
  $M_{J2}$, $J=0, \ldots ,K-1$ & 1 & $T_A$ & $2- \frac{N+2K+1}{ (K+1)N_f}+\frac{2J}{K+1}+\frac{2K}{K+1}$ \\
\hline
\end{tabular}

$\hspace*{1mm}$

Here $K$ is odd and
\begin{equation}\label{BP1}
\widetilde{N} \ = \ 3K N_f - 4K - 2 - N. \ \ \
\end{equation}

For these theories we have the following superconformal indices

\begin{eqnarray}\label{KSsp3_1}
    I_E &=& \frac{(p;p)_{\infty}^{N} (q;q)_{\infty}^{N} }{2^N N!} \Gamma(U,U^{\frac K2};p,q)^{N-1}
    \\ \nonumber
    && \times \int_{\mathbb{T}^N}\prod_{1 \leq i < j \leq N} \frac{\Gamma(U z_i^{\pm 1} z_j^{\pm 1},U^{\frac K2} z_i^{\pm 1} z_j^{\pm 1};p,q)}
    {\Gamma(z_i^{\pm 1} z_j^{\pm 1};p,q)}\prod_{i=1}^{2N_f} \prod_{j=1}^N \frac{\Gamma(s_i z_j^{\pm
    1};p,q)}{\Gamma(z_j^{\pm 2};p,q)}\prod_{j=1}^N
     \frac{d z_j}{2 \pi \textup{i} z_j}
\end{eqnarray}
where $U=(pq)^{\frac{1}{K+1}}$,
the balancing condition reads $U^{N+2K+1}\prod_{i=1}^{2N_f}s_i = (pq)^{N_f},$ and
\begin{eqnarray}\label{KSsp3_2}
    I_M &=& \frac{(p;p)_{\infty}^{\widetilde{N}} (q;q)_{\infty}^{\widetilde{N}} }
    {2^{\widetilde{N}}  \widetilde{N}!}
    \Gamma(U,U^{\frac K2};p,q)^{\widetilde{N}-1}
     \\ \nonumber  && \times
    \prod_{J=0}^{K-1} \prod_{L=0}^2 \prod_{1 \leq i < j \leq 2N_f} \Gamma(U^{J+\frac{KL}{2}} s_i
    s_j;p,q)\prod_{J=0}^{\frac{K-3}{2}} \prod_{j=1}^{2N_f} \Gamma(U^{2J+1+\frac K2} s_j^2;p,q)
    \\ \nonumber   && \times \int_{\mathbb{T}^{\widetilde{N}}}
     \prod_{1 \leq i < j \leq \widetilde{N}} \frac{\Gamma(U z_i^{\pm 1} z_j^{\pm 1},U^{\frac K2} z_i^{\pm 1} z_j^{\pm 1};p,q)
     }{\Gamma(z_i^{\pm 1} z_j^{\pm 1};p,q)} \prod_{i=1}^{2N_f} \prod_{j=1}^{\widetilde{N}}
    \frac{\Gamma(U^{1-\frac K2} s_i^{-1} z_j^{\pm 1};p,q)}{\Gamma(z_j^{\pm 2};p,q)}\prod_{j=1}^{\widetilde{N}} \frac{d z_j}{2 \pi \textup{i} z_j}.
\end{eqnarray}

\subsection{Symmetric and anti-symmetric tensor matter fields}

This duality was found in \cite{Brodie2}. The electric theory:  \\

\begin{tabular}{|c|c|c|c|}
  \hline
    & $SP(2N)$ & $SU(2N_f)$ & $U(1)_R$ \\  \hline
  $Q$ & $f$ & $f$ & $1-\frac{N+2K-1}{(K+1)N_f}$ \\
  $X$ & $T_A$ & 1 & $\frac{2}{K+1}$ \\
  $Y$ & $T_S$ & 1 & $\frac{K}{K+1}$ \\
\hline
\end{tabular}

$\hspace*{1mm}$

The magnetic theory: \\

\begin{tabular}{|c|c|c|c|}
  \hline
   & $SP(2\widetilde{N})$ & $SU(2N_f)$ & $U(1)_R$ \\  \hline
  $q$ & $f$ & $\overline{f}$ & $1-\frac{\widetilde{N}+2K-1}{(K+1)N_f}$
  \\
  $\widetilde{X}$ & $T_A$ & 1 & $\frac{2}{K+1}$ \\
  $\widetilde{Y}$ & $T_S$ & 1 & $\frac{K}{K+1}$ \\
  $M_{J0}$, $J=0, \ldots ,K-1$ & 1 & $T_A$ & $2- \frac{N+2K+1}{ (K+1)N_f}+\frac{2J}{K+1}$ \\
  $M_{2J \ 1}$, $J=0, \ldots ,\frac{K-1}{2}$ & 1 & $T_S$ & $2- \frac{N+2K+1}{ (K+1)N_f}+\frac{2(2J)}{K+1}+\frac{K}{K+1}$ \\
  $M_{2J+1 \ 1}$, $J=0, \ldots ,\frac{K-3}{2}$ & 1 & $T_A$ & $2- \frac{N+2K+1}{ (K+1)N_f}+\frac{2(2J+1)}{K+1}+\frac{K}{K+1}$ \\
  $M_{J2}$, $J=0, \ldots ,K-1$ & 1 & $T_A$ & $2- \frac{N+2K+1}{ (K+1)N_f}+\frac{2J}{K+1}+\frac{2K}{K+1}$ \\
\hline
\end{tabular}

$\hspace*{1mm}$

Here $K$ is odd and
\begin{equation}\label{BP2}
\widetilde{N} \ = \ 3K N_f - 4K + 2 - N.
\end{equation}

For these theories we have the following superconformal indices
\begin{eqnarray}\label{KSsp4_1}
&&I_E = \frac{(p;p)_{\infty}^{N} (q;q)_{\infty}^{N} }{2^N N!}
\Gamma(U;p,q)^{N-1} \Gamma(U^{\frac K2};p,q)^{N}    \\ \nonumber &&
\makebox[-1em]{} \times \int_{\mathbb{T}^N}\prod_{1 \leq i < j \leq
N} \frac{\Gamma(U z_i^{\pm 1} z_j^{\pm 1},U^{\frac K2} z_i^{\pm 1}
z_j^{\pm 1};p,q)} {\Gamma(z_i^{\pm 1} z_j^{\pm 1};p,q)}\prod_{j=1}^N
\frac{\Gamma(U^{\frac K2} z_j^{\pm 2};p,q)\prod_{i=1}^{2N_f}
\Gamma(s_i z_j^{\pm1};p,q) }{\Gamma(z_j^{\pm 2};p,q)}\prod_{j=1}^N
\frac{d z_j}{2 \pi \textup{i} z_j}
\end{eqnarray}
where $U=(pq)^{\frac{1}{K+1}}$ and the balancing condition reads
$U^{N+2K-1}\prod_{i=1}^{2N_f}s_i= (pq)^{N_f}$, and
\begin{eqnarray}\label{KSsp4_2}
&& \makebox[-1em]{}  I_M = \frac{(p;p)_{\infty}^{\widetilde{N}}
(q;q)_{\infty}^{\widetilde{N}} } {2^{\widetilde{N}}  \widetilde{N}!}
\Gamma(U;p,q)^{\widetilde{N}-1} \Gamma(U^{\frac
K2};p,q)^{\widetilde{N}} \prod_{J=0}^{K-1} \prod_{L=0}^2 \prod_{1
\leq i < j \leq 2N_f} \Gamma(U^{J+\frac{KL}{2}} s_i s_j;p,q)
\nonumber \\  && \times \prod_{J=0}^{\frac{K-1}{2}}
\prod_{j=1}^{2N_f} \Gamma(U^{2J+\frac K2} s_j^2;p,q)
\int_{\mathbb{T}^{\widetilde{N}}} \prod_{1 \leq i < j \leq
\widetilde{N}} \frac{\Gamma(U z_i^{\pm 1} z_j^{\pm 1},U^{\frac K2}
z_i^{\pm 1} z_j^{\pm 1};p,q)}{\Gamma(z_i^{\pm 1} z_j^{\pm 1};p,q)}
\\ \nonumber   && \makebox[2em]{} \times \prod_{j=1}^{\widetilde{N}}
\frac{\Gamma(U^{\frac K2} z_j^{\pm 2};p,q)\prod_{i=1}^{2N_f}
\Gamma(U^{1-\frac K2} s_i^{-1} z_j^{\pm 1};p,q) }{\Gamma(z_j^{\pm
2};p,q)}\prod_{j=1}^{\widetilde{N}} \frac{d z_j}{2 \pi \textup{i} z_j}.
\end{eqnarray}

The equalities $I_E=I_M$ for all the dualities described
in this section represent new elliptic hypergeometric identities
requiring a rigorous mathematical confirmation.

\section{Some other new dualities} \label{NewDual}

Let us denote
\begin{eqnarray}
&& I_{A_N}(\underline{t},\underline{u};p,q) =
\frac{(p;p)^N_{\infty}(q;q)^N_{\infty}}{(N+1)!}
\int_{\mathbb{T}^{N}}
\frac{\prod_{i=1}^{N+1}\prod_{r=1}^{N+3}\Gamma(t_rz_i,u_rz_i^{-1};p,q)}{\prod_{1
\leq i < j \leq
N+1}\Gamma(z_iz_j^{-1},z_i^{-1}z_j;p,q)}\prod_{j=1}^N \frac{dz_j}{2
\pi \textup{i} z_j}
\end{eqnarray}
with $\prod_{j=1}^{N+1}z_j=1$  and
the balancing condition $\prod_{i=1}^{N+3} t_iu_i \ = \ (pq)^2,$
and
\begin{eqnarray}
&& \makebox[-3em]{}
I_{BC_N}(\underline{t};p,q) =
\frac{(p;p)^N_{\infty}(q;q)^N_{\infty}}{2^NN!}
\int_{\mathbb{T}^{N}}
\frac{\prod_{i=1}^{N}\prod_{r=1}^{2N+6}\Gamma(t_rz_i^{\pm
1};p,q)}{\prod_{1 \leq i < j \leq N}\Gamma(z_i^{\pm 1} z_j^{\pm
1};p,q)\prod_{j=1}^N \Gamma(z_j^{\pm 2};p,q)}\prod_{j=1}^N
\frac{dz_j}{2 \pi \textup{i} z_j}
\end{eqnarray}
with the balancing condition $\prod_{r=1}^{2N+6}t_r \ = \ (pq)^2.$

\subsection{$SU \leftrightarrow SP$ groups mixing duality}\label{Rains1D}
The first case electric gauge group is $G=SU(N+1)$,
but the dual gauge group is of a different type $G=SP(2N)$.
The flavor symmetry group in both cases is
$F=SU(N+3) \times SU(N+3) \times U(1)_B.$
The field content of dual theories is described in the tables below

\begin{center}
\begin{tabular}{|c|c|c|c|c|c|}
  \hline
   & $SU(N+1)$ & $SU(N+3)$ & $SU(N+3)$ & $U(1)_B$ & $U(1)_R$ \\  \hline
  $Q_1$ & $f$ & $f$ & 1 & 2 & $\frac{2}{N+3}$ \\
  $Q_2$ & $\overline{f}$ & 1 & $f$ & -2 & $\frac{2}{N+3}$ \\
\hline
\end{tabular}
\end{center}

\begin{center}
\begin{tabular}{|c|c|c|c|c|c|}
  \hline
   & $SP(2N)$ & $SU(N+3)$ & $SU(N+3)$ & $U(1)_B$ & $U(1)_R$ \\  \hline
  $q_1$ & $f$ & $f$ & 1 & $-(N+1)$ & $\frac{2}{N+3}$ \\
  $q_2$ & $\overline{f}$ & 1 & $f$ & $N+1$ & $\frac{2}{N+3}$ \\
  $X_1$ & 1 & $\overline{T}_A$ & 1 & $2(N+1)$ & $2\frac{N+1}{N+3}$
  \\
  $X_2$ & 1 & 1 & $\overline{T}_A$ & $-2(N+1)$ & $2\frac{N+1}{N+3}$
  \\
\hline
\end{tabular}
\end{center}

The superconformal indices are
\begin{eqnarray}
&&I_E= I_{A_N}(t_1, \ldots, t_{N+3},u_1, \ldots, u_{N+3};p,q),
\\ \nonumber &&I_M= \prod_{1 \leq i < j \leq N+3}
\Gamma(T/t_it_j,U/u_iu_j;p,q) I_{BC_N}(\ldots (U/T)^{1/4}t_i \ldots,
\ldots (T/U)^{1/4}u_i \ldots;p,q),
\end{eqnarray}
where $T= \prod_{1 \leq i \leq N+3} t_i$ and $U = \prod_{1 \leq i \leq N+3}u_i.$

The equality $I_E = I_M$ represents the mixed elliptic
hypergeometric integrals transformation proven in \cite{Rains}.
We used this identity as a starting
point for finding the described new Seiberg-type pair of field theories.

\subsection{$SU \leftrightarrow SU$ groups mixing duality}\label{Rains2D}
Again, we use consequences of the mixed transformations derived in
\cite{Rains}. Corresponding dualities have the flavor
symmetry groups
$$
F=SU(K) \times SU(N+2-K) \times U(1)_1 \times SU(K) \times SU(N+2-K)
\times U(1)_2 \times U(1)_B,
$$
for arbitrary $0 < K < N+2.$ The matter field content of the initial
electric field  theory is given in the table

\begin{center}
\begin{tabular}{|c|c|c|c|c|c|}
  \hline
   & $SU(N)$ & $SU(N+2)$ & $SU(N+2)$ & $U(1)_B$ & $U(1)_R$ \\  \hline
  $Q_1$ & $f$ & $f$ & 1 & 1 & $\frac{2}{N+2}$ \\
  $Q_2$ & $\overline{f}$ & 1 & $f$ & -1 & $\frac{2}{N+2}$ \\
\hline
\end{tabular}
\end{center}
In order to verify the 't Hooft anomalies matching conditions for
relevant flavor symmetry subgroups, it is useful to rewrite the
latter table as

{\small
\begin{center}
\begin{tabular}{|c|c|c|c|c|c|c|c|c|c|}
  \hline
   & $SU(N)$ & $SU(K)$ & $SU(M)$ & $U(1)_1$ & $SU(K)$ & $SU(M)$ & $U(1)_2$ & $U(1)_B$ & $U(1)_R$ \\  \hline
  $q_1$ & $f$ & $f$ & 1 & $M$ & 1 & 1 & 0 & 1 & $\frac{2}{N+2}$ \\
  $q_2$ & $f$ & 1 & $f$ & $-K$ & 1 & 1 & 0 & 1 & $\frac{2}{N+2}$ \\
  $q_3$ & $\overline{f}$ & 1 & 1 & 0 & $f$ & 1 & $M$ & -1 & $\frac{2}{N+2}$ \\
  $q_4$ & $\overline{f}$ & 1 & 1 & 0 & 1 & $f$ & $-K$ & -1 & $\frac{2}{N+2}$ \\
\hline
\end{tabular}
\end{center}
} where $M=N+2-K.$ The dual theory content is described in the
following table {\tiny
\begin{center}
\begin{tabular}{|c|c|c|c|c|c|c|c|c|c|}
  \hline
   & $SU(N)$ & $SU(K)$ & $SU(M)$ & $U(1)_1$ & $SU(K)$ & $SU(M)$ & $U(1)_2$ & $U(1)_B$ & $U(1)_R$ \\  \hline
  $q_1$ & $\overline{f}$ & $f$ & 1 & $\frac{K(K-2)}{N}-K+M$ & 1 & 1 & $\frac{MK}{N}$ & $1-M$ & $\frac{2}{N+2}$ \\
  $q_2$ & $f$ & 1 & $f$ & $-\frac{K(K-2)}{N}$ & 1 & 1 & $\frac{-MK}{N}$ & $1-K$ & $\frac{2}{N+2}$ \\
  $q_3$ & $f$ & 1 & 1 & $\frac{MK}{N}$ & $f$ & 1 & $\frac{K(K-2)}{N}-K+M$ & $M-1$ & $\frac{2}{N+2}$ \\
  $q_4$ & $\overline{f}$ & 1 & 1 & $\frac{-MK}{N}$ & 1 & $f$ & $-\frac{K(K-2)}{N}$ & $K-1$ & $\frac{2}{N+2}$ \\
  $X_1$ & 1 & $f$ & 1 & $M$ & 1 & $f$ & $-K$ & 0 & $\frac{4}{N+2}$ \\
  $X_2$ & 1 & 1 & $f$ & $-K$ & $f$ & 1 & $M$ & 0 & $\frac{4}{N+2}$ \\
  $Y_1$ & 1 & $\overline{f}$ & $\overline{f}$ & $K-M$ & 1 & 1 & 0 & $N$ & $\frac{2N}{N+2}$ \\
  $Y_2$ & 1 & 1 & 1 & 0 & $\overline{f}$ & $\overline{f}$ & $K-M$ & $-N$ & $\frac{2N}{N+2}$ \\
\hline
\end{tabular}
\end{center}
}

The superconformal indices have the form
\begin{eqnarray}
&& I_E=I_{A_{N-1}}(t_1, \ldots, t_{N+2},u_1, \ldots, u_{N+2};p,q), \\
\nonumber && I_M= \prod_{1 \leq r< K, K \leq s \leq N+2}
\Gamma(t_ru_s,t_su_r,T/t_st_r,U/u_ru_s) I_{A_{N-1}}(t_1', \ldots,
t_{N+2}', u_1', \ldots, u_{N+2}';p,q),
\end{eqnarray}
where $T=\prod_{r=1}^{N+2}t_r,$ $U=\prod_{r=1}^{N+2}u_r,$ $T_K
=\prod_{r=1}^{K}t_r,$ $ U_K =\prod_{r=1}^{K}u_r$, and
$$
t_r' \ = \ (T/U)^{\frac{N-K}{2N}}(T_K/U_K)^{1/N}u_r,
 \ \ \ \ 1 \leq r < K+1,
$$
$$
t_r' \ = \ (U/T)^{\frac{K}{2N}}(T_K/U_K)^{1/N}t_r, \ \ \ \ K+1 \leq
r \leq N+2,
$$
$$
u_r' \ = \ (U/T)^{\frac{N-K}{2N}}(U_K/T_K)^{1/N}t_r, \ \ \ \ 1 \leq
r < K+1,
$$
$$
u_r' \ = \ (T/U)^{\frac{K}{2N}}(U_K/T_K)^{1/N}u_r, \ \ \ \ K+1 \leq
r \leq N+2.
$$
The equality $I_E \ = \ I_M$ for $K=1$ was suggested in
\cite{Spiridonov3} and the general relation with the complete proof
for arbitrary $K$ is given in \cite{Rains}.

\section{$S$-confinement}

Following \cite{Csaki4,Csaki3,Seiberg}, by $s$-confinement we mean
smooth confinement without chiral symmetry breaking and with a
non-vanishing confining superpotential. The theory is confined
when its infrared physics can be described completely in terms of gauge
invariant composite fields and their interactions. This description has to
be valid everywhere in the moduli space of vacua.
$s$-confinement requires also that the theory dynamically generates
a confining superpotential. Furthermore, the phase without chiral
symmetry breaking implies that the origin of the classical moduli
space serves also as a vacuum in the quantum theory. In this vacuum
all the global symmetries present in the ultraviolet regime remain unbroken.
Finally, the confining superpotential is a holomorphic function of
the confined degrees of freedom and couplings, which describe all
interactions in the extreme infrared. From the point of view of
elliptic hypergeometric functions the $s$-confinement means that the
dual theory gauge group is trivial $G=1$ (i.e., there is no
vector superfield $\widetilde V$) and the integrals
describing superconformal indices are computable exactly, defining
highly non-trivial elliptic beta integrals \cite{Spiridonov2}.

\subsection{$SU(N)$ gauge group}

In this section we present known examples of the confining theories with
the unitary gauge group.
For brevity we combine the electric and magnetic theories in a
single table separating them by the double line. The magnetic theory
fields are denoted using the conventions of \cite{Csaki4}.

\subsubsection{$SU(N)$ with $(N+1)(f + \overline{f})$}\label{TypeA1} \cite{Seiberg}
\begin{center}
\begin{tabular}{|c|c|c|c|c|c|}
  \hline
   & $SU(N)$ & $SU(N+1)$ & $SU(N+1)$ & $U(1)$ & $U(1)_R$ \\  \hline
  $Q$ & $f$ & $f$ & 1 & 1 & $ \frac{1}{N+1}$ \\
  $\widetilde{Q}$ & $\overline{f}$ & 1 & $f$ & -1 & $ \frac{1}{N+1}$ \\
\hline \hline
  $Q\widetilde{Q}$ && $f$ & $f$ & 0 & $ \frac{2}{N+1}$ \\
  $Q^N$ && $\overline{f}$ & 1 & $N$ & $ \frac{N}{N+1}$ \\
  $\widetilde{Q}^N$ && 1& $\overline{f}$ & $-N$ & $ \frac{N}{N+1}$ \\ \hline
\end{tabular}
\end{center}

The superconformal indices for these theories are equal to (after
appropriate renormalization of the parameters)
\begin{eqnarray}\label{s1_1}
&& I_E=\frac{(p;p)^{N-1}_{\infty} (q;q)^{N-1}_{\infty}}{N!}
\int_{\mathbb{T}^{N-1}} \prod_{1 \leq j < k \leq N}
\frac{1}{\Gamma(z_iz_j^{-1},z_i^{-1}z_j;p,q)} \nonumber \\
 && \makebox[3em]{}  \times \prod_{j=1}^{N} \prod_{m=1}^{N+1} \Gamma(s_m z_j,t_m
z_j^{-1};p,q)\prod_{j=1}^{N-1} \frac{d z_j}{2 \pi \textup{i}  z_j},
\end{eqnarray}
where $\prod_{j=1}^Nz_j=1$, and
\begin{eqnarray}\label{s1_2} && I_M=
\prod_{m=1}^{N+1} \Gamma(Ss_m^{-1},Tt_m^{-1};p,q)
\prod_{k,m=1}^{N+1} \Gamma(s_k t_m;p,q),
\end{eqnarray}
where $S=\prod_{m=1}^{N+1} s_m,$ $T=\prod_{m=1}^{N+1} t_m,$
with the balancing condition $ST = pq.$

The exact evaluation formula for the integral $I_E=I_M$ was conjectured and
partially confirmed in \cite{Spiridonov3}. Its complete proofs are given in
\cite{Rains,spi:short}. In the simplest $p \rightarrow 0$ limit it
is reduced to one of the Gustafson integrals \cite{gus1}.

\subsubsection{$SU(2N)$ with $T_A + 2 N \overline{f} + 4
f$} The theory with $G=SU(2N)$ gauge group and flavor group
$F=SU(2N) \times SU(4) \times U(1)_1 \times U(1)_2$ was found to be
confining in \cite{P3,PT}. The field content of both theories is
described in the table below
\begin{center}
\begin{tabular}{|c|c|c|c|c|c|c|}
  \hline
   & $SU(2N)$ & $SU(2N)$ & $SU(4)$ & $U(1)_1$ & $U(1)_2$ & $U(1)_R$ \\  \hline
  $Q$ & $f$ & $1$ & $f$ & $-2N$ & $-2N+2$ & $\frac 12$ \\
  $\widetilde{Q}$ & $\overline{f}$ & $f$ & 1 & 4 & $-2N+2$ & 0 \\
  $A$ & $T_A$ & 1 & 1 & 0 & $2N+4$ & 0 \\
\hline \hline
  $Q\widetilde{Q}$ && $f$ & $f$ & $4-2 N$ & $-4N+4$ & $\frac 12$ \\
  $A\widetilde{Q}^2$ && $T_A$ & 1 & 8 & $-2N+8$ & 0 \\
  $A^N$ && 1& 1 & 0 &  $2N^2+4N$ & 0 \\
  $A^{N-1}Q^2$ && 1& $T_A$ & $-4N$ &  $2N^2-2N$ & 1 \\
  $A^{N-1}Q^4$ && 1& 1 & $-8N$ &  $2N^2-8N$ & 2 \\
  $\widetilde{Q}^{2N}$ && 1& 1 & $8N$ &  $-4N^2+4N$ & 0 \\ \hline
\end{tabular}
\end{center}

We come to the following integrals describing the superconformal indices
\begin{eqnarray}\label{s2_1}
&&I_E= \frac{(p;p)^{2N-1}_{\infty} (q;q)^{2N-1}_{\infty}}{(2N)!}
\int_{\mathbb{T}^{2N-1}}   \prod_{1 \leq j < k \leq 2N}
\frac{\Gamma(tz_iz_j;p,q)}{\Gamma(z_iz_j^{-1},z_i^{-1}z_j;p,q)} \\
&& \makebox[3em]{} \times \prod_{j=1}^{2N} \prod_{k=1}^{2N}
\Gamma(t_kz_j^{-1};p,q) \prod_{i=1}^4 \Gamma(s_i
z_j;p,q)\prod_{j=1}^{2N-1} \frac{d z_j}{2 \pi \textup{i} z_j},
\nonumber \eeqa
with $\prod_{j=1}^{2N}z_j=1$, and
\beqa\label{s2_2} &&\makebox[-3em]{}
I_M= \prod_{1 \leq j < k \leq 2N} \Gamma(t t_j
t_k;p,q) \prod_{k=1}^{2N} \prod_{i=1}^4 \Gamma(t_k s_i;p,q)
\frac{\Gamma(t^{N},T;p,q)}{\Gamma(t^{N}T;p,q)} \prod_{1 \leq i < m
\leq 4} \Gamma(t^{\frac{2N-2}{2}}s_i s_m;p,q),
\end{eqnarray}
with the balancing condition $t^{2N-2} ST = pq,$
where $S=\prod_{i=1}^4s_i$, $T=\prod_{j=1}^{2N}t_j$.

Equality $I_E=I_M$ defines the elliptic beta integral introduced in
\cite{Spiridonov3}. It represents an elliptic extension of the
Gustafson-Rakha $q$-beta integral for odd number of integration
variables \cite{gr}.

\subsubsection{$SU(2N+1)$ with $T_A + (2N +1)\overline{f} + 4
f$} These dual models were considered in \cite{P3,PT}:

\begin{center}
\begin{tabular}{|c|c|c|c|c|c|c|}
  \hline
   & $SU(2N+1)$ & $SU(2N+1)$ & $SU(4)$ & $U(1)_1$ & $U(1)_2$ & $U(1)_R$ \\  \hline
  $Q$ & $f$ & $1$ & $f$ & $-2N-1$ & $-2N+1$ & $\frac 12$ \\
  $\widetilde{Q}$ & $\overline{f}$ & $f$ & 1 & 4 & $-2N+1$ & 0 \\
  $A$ & $T_A$ & 1 & 1 & 0 & $2N+5$ & 0 \\
\hline \hline
  $Q\widetilde{Q}$ && $f$ & $f$ & $3-2 N$ & $-4N+2$ & $\frac 12$ \\
  $A\widetilde{Q}^2$ && $T_A$ & 1 & 8 & $-2N+7$ & 0 \\
  $A^{N}Q$ && 1& $f$ & $-2N-1$ &  $2N^2+3N+1$ & $\frac 12$ \\
  $A^{N-1}Q^3$ && 1& $\overline{f}$ & $-6N-3$ &  $2N^2-3N-2$ & $\frac 32$ \\
  $\widetilde{Q}^{2N+1}$ && 1& 1 & $8N+4$ &  $-4N^2+1$ & 0 \\ \hline
\end{tabular}
\end{center}

The indices have the form
\begin{eqnarray}\label{s3_1}
&& I_E=\frac{(p;p)^{2N}_{\infty} (q;q)^{2N}_{\infty}}{(2N+1)!}
\int_{\mathbb{T}^{2N}} \prod_{1 \leq j < k \leq 2N+1}
\frac{\Gamma(tz_iz_j;p,q)}{\Gamma(z_iz_j^{-1},z_i^{-1}z_j;p,q)} \\
&& \makebox[3em]{} \times \prod_{j=1}^{2N+1} \prod_{k=1}^{2N+1}
\Gamma(t_kz_j^{-1};p,q) \prod_{i=1}^4 \Gamma(s_i
z_j;p,q)\prod_{j=1}^{2N} \frac{d z_j}{2 \pi \textup{i} z_j},
\nonumber \eeqa
with $\prod_{j=1}^{2N+1}z_j=1$, and
\beqa\label{s3_2} &&
I_M= \prod_{1 \leq j < k \leq 2N+1} \Gamma(t
t_j t_k;p,q) \prod_{k=1}^{2N+1} \prod_{i=1}^4 \Gamma(t_k s_i;p,q)
 \Gamma(T;p,q)  \prod_{i=1}^4
\frac{\Gamma(t^{N}s_i;p,q)}{\Gamma(t^{N} T s_i;p,q)},
\end{eqnarray}
where the balancing condition reads $t^{2N-1} ST = pq$
and $T=\prod_{k=1}^{2N+1} t_k,$ $S=\prod_{k=1}^4 s_k$.

The equality $I_E=I_M$  was also suggested in \cite{Spiridonov3} as
an elliptic extension of the Gustafson-Rakha $q$-beta integral with an even
number of integrations  \cite{gr}.

\subsubsection{$SU(2N+1)$ with $T_A + \overline{T}_A + 3 \overline{f} + 3
f$} Models \cite{Csaki3}:

\begin{center}
\begin{tabular}{|c|c|c|c|c|c|c|c|}
  \hline
   & $SU(2N+1)$ & $SU(3)$ & $SU(3)$ & $U(1)_1$ & $U(1)_2$ & $U(1)_B$ & $U(1)_R$ \\  \hline
  $Q$ & $f$ & $f$ & $1$ & 0 & $2N-1$ & 1 & $\frac 13$ \\
  $\widetilde{Q}$ & $\overline{f}$ & 1 & $f$ & 0 & $2N-1$ & -1 & $\frac 13$ \\
  $A$ & $T_A$ & 1 & 1 & 1 & -3 & 0 & 0 \\
  $\widetilde{A}$ & $\overline{T}_A$ & 1 & 1 & -1 & -3 & 0 & 0 \\
\hline \hline
  $Q(A\widetilde{A})^k\widetilde{Q}$ && $f$ & $f$ & 0 & $4N-2-6k$ & 0 & $\frac 23$ \\
  $\widetilde{A}(A\widetilde{A})^kQ^2$ && $T_A$ & 1 & -1 & $4N-5-6k$ & 2 & $\frac 23$ \\
  $A(A\widetilde{A})^k\widetilde{Q}^2$ && 1 & $T_A$ & 1 & $4N-5-6k$ & -2 & $\frac 23$ \\
  $A^NQ$ && $f$ & 1 & $N$ & $-N-1$ & 1 & $\frac 13$ \\
  $\widetilde{A}^N\widetilde{Q}$ && 1 & $f$ & $-N$ & $-N-1$ & -1 & $\frac 13$ \\
  $A^{N-1}Q^3$ && 1 & 1 & $N-1$ & $3N$ & 3 & 1 \\
  $\widetilde{A}^{N-1}\widetilde{Q}^3$ && 1 & 1 & $-N+1$ & $3N$ & -3 & 1 \\
  $(A\widetilde{A})^m$ && 1& 1 & 0 & $-6m$ & 0 & 0 \\ \hline
\end{tabular}
\end{center}
where $k=0,\ldots,N-1$ and $m=1,\ldots,N$.

The superconformal indices are written as
\begin{eqnarray}\label{s4_1}
&& I_E=\frac{(p;p)^{2N}_{\infty} (q;q)^{2N}_{\infty}}{(2N+1)!}
\int_{\mathbb{T}^{2N}} \prod_{1 \leq i < j \leq 2N+1}
\frac{\Gamma(Uz_iz_j,Vz_i^{-1}z_j^{-1};p,q)}{\Gamma(z_iz_j^{-1},z_i^{-1}z_j;p,q)}
\nonumber \\  && \makebox[3em]{} \times \prod_{i=1}^{3}
\prod_{j=1}^{2N+1} \Gamma(s_iz_j,t_iz_j^{-1};p,q) \prod_{j=1}^{2N}
\frac{d z_j}{2 \pi \textup{i} z_j}, \eeqa where
$\prod_{j=1}^{2N+1}z_j=1$, and \beqa \label{s4_2}
 && I_M=
\prod_{i=1}^3 \Gamma(U^{N} s_i, V^{N} t_i;p,q)
\Gamma(U^{N-1}s_1s_2s_3,V^{N-1}t_1t_2t_3;p,q) \prod_{j=1}^{N} \Gamma((UV)^{j};p,q) \nonumber \\
&& \times \prod_{j=0}^{N-1} \Big[\prod_{i,k=1}^{3}
\Gamma((UV)^{j}s_it_k;p,q)  \prod_{1 \leq i < k
\leq 3} \Gamma(V (UV)^j s_is_k, U (UV)^j t_i t_k;p,q)\Big] ,
\end{eqnarray}
where the balancing condition reads $(UV)^{2N-1} \prod_{i=1}^3
s_it_i = pq.$

The equality $I_E=I_M$ was derived by Spiridonov in \cite{Spiridonov3}
by purely algebraic means as a consequence of other elliptic beta integrals.
In the simplest $p \rightarrow 0$ limit it reduces to
Gustafson's $q$-beta integral for the root system $A_{2N}$ \cite{gus2}.

\subsubsection{$SU(2N)$ with $T_A + \overline{T}_A + 3 \overline{f} + 3 f$}
For $N>2$ the models have the form \cite{Csaki3}: \\

\begin{center}
\begin{tabular}{|c|c|c|c|c|c|c|c|}
  \hline
   & $SU(2N)$ & $SU(3)$ & $SU(3)$ & $U(1)_1$ & $U(1)_2$ & $U(1)_B$ & $U(1)_R$ \\  \hline
  $Q$ & $f$ & $f$ & $1$ & 0 & $2N-2$ & 1 & $\frac 13$ \\
  $\widetilde{Q}$ & $\overline{f}$ & 1 & $f$ & 0 & $2N-2$ & -1 & $\frac 13$ \\
  $A$ & $T_A$ & 1 & 1 & 1 & -3 & 0 & 0 \\
  $\widetilde{A}$ & $\overline{T}_A$ & 1 & 1 & -1 & -3 & 0 & 0 \\
\hline \hline
  $Q(A\widetilde{A})^k\widetilde{Q}$ && $f$ & $f$ & 0 & $4N-4-6k$ & 0 & $\frac 23$ \\
  $\widetilde{A}(A\widetilde{A})^mQ^2$ && $T_A$ & 1 & -1 & $4N-7-6m$ & 2 & $\frac 23$ \\
  $A(A\widetilde{A})^m\widetilde{Q}^2$ && 1 & $T_A$ & 1 & $4N-7-6m$ & -2 & $\frac 23$ \\
  $A^N$ && $1$ & 1 & $N$ & $-3N$ & 0 & $0$ \\
  $\widetilde{A}^N$ && 1 & 1 & $-N$ & $-3N$ & 0 & $0$ \\
  $A^{N-1}Q^2$ && $T_A$ & 1 & $N-1$ & $N-1$ & 2 & $\frac 23$ \\
  $\widetilde{A}^{N-1}\widetilde{Q}^2$ && 1 & $T_A$ & $-N+1$ & $N-1$ & -2 & $\frac 23$ \\
  $(A\widetilde{A})^n$ && 1& 1 & 0 & $-6n$ & 0 & 0 \\ \hline
\end{tabular}
\end{center}
where $k=0,\ldots,N-1$, $m=0,\ldots,N-2$ and $n=1,\ldots,N-1$. For
$N=2$ the flavor group is enlarged to $ F= SU(3)\times SU(3) \times
SU(2) \times U(1)_2 \times U(1)_B, $
 and the fields $A$, $\tilde A$ unify to the $SU(2)$-group doublet.

The expressions for the superconformal indices are
\begin{eqnarray}\label{s5_1}
&& I_E=\frac{(p;p)^{2N-1}_{\infty} (q;q)^{2N-1}_{\infty}}{(2N)!}
\int_{\mathbb{T}^{2N-1}} \prod_{1 \leq i < j \leq 2N}
\frac{\Gamma(Uz_iz_j,
Vz_i^{-1}z_j^{-1};p,q)}{\Gamma(z_iz_j^{-1},z_i^{-1}z_j;p,q)}
\\ && \makebox[3em]{} \times
\prod_{j=1}^{2N} \prod_{i=1}^{3} \Gamma(s_iz_j,
t_iz_j^{-1};p,q)\prod_{j=1}^{2N-1} \frac{d z_j}{2 \pi \textup{i}
z_j}, \nonumber\eeqa
with $\prod_{j=1}^{2N}z_j=1$, and \beqa &&
I_M= \Gamma(U^N,V^N;p,q) \prod_{1 \leq i < k \leq 3} \Gamma(U^{N-1}
s_is_k, V^{N-1}t_it_k;p,q) \prod_{j=1}^{N} \prod_{i,k=1}^3
\Gamma((UV)^{j-1} s_i t_k;p,q)  \nonumber \\ && \makebox[3em]{}
\times \prod_{j=1}^{N-1} \Gamma((UV)^j;p,q) \prod_{j=0}^{N-2}
\prod_{1 \leq i < k \leq 3} \Gamma(V (UV)^j s_is_k,U (UV)^j t_i
t_k;p,q) ,
 \label{s5_2}\end{eqnarray}
where the balancing condition reads $(UV)^{2N-2} \prod_{i=1}^3
s_it_i = pq.$ The equality $I_E=I_M$ was also derived in
\cite{Spiridonov3} as a consequence of some other elliptic beta
integrals. In the simplest $p \rightarrow 0$ limit, it reduces to
one of Gustafson's integrals for the root system $A_{2N-1}$ \cite{gus2}.
Similar to the case of non-confining $N_f=4$ dualities
described earlier, a careful examination of the limit $V\to 1$
(or $U\to 1$) shows that the equality of superconformal indices
in this case reduces to the equality of $SP(2N)$-group confining
duality indices discussed in \cite{SV}.
This means that the elliptic Selberg integral
introduced in \cite{die-spi:elliptic}  (see integral (\ref{sp2_1})
and its evaluation (\ref{sp2_2}) below) is a limiting case of Spiridonov's
$A_n$-elliptic beta integral. This result could have been expected
since the computation of the latter integral in \cite{Spiridonov3}
used the elliptic Selberg integral.

\subsubsection{$SU(KN_f-1)$ with $N_f f + N_f \overline{f} + 1 adj$}

Taking $N \ = \ KN_f-1$ in (\ref{KSSU}) (or, $\widetilde{N}=1$), we
find the $s$-confining dual theory discussed in \cite{Csaki7}. The
field content of these theories is easily found from the tables
given in Sect. 9.1. Namely, in the electric theory one should fix
$N$ as described; on the magnetic side one should keep all the
mesons and baryons  and set $\widetilde{N}=1$ in the gauge group
part. Therefore for this case the superconformal index for the
electric theory is given by (\ref{intKSelE}), and the magnetic
superconformal index takes the form
\begin{eqnarray}
\makebox[-3em]{} && I_M=\prod_{l=1}^K \prod_{1 \leq i,j \leq N_f}
\Gamma(U^{l-1} s_i t^{-1}_j;p,q)  \prod_{i=1}^{N_f}
\Gamma(U(ST)^{\frac K2} s_i^{-1}, U(ST)^{- \frac K2} t_i;p,q),
\end{eqnarray}
 where $U=(pq)^{\frac{1}{K+1}}$, $S = \prod_{j=1}^{N_f} s_j,$
$T =  \prod_{j=1}^{N_f}t_j$, and the balancing condition reads
$ U^{2KN_f-2}ST^{-1} =(pq)^{N_f}.$

For $K=1$ one obtains the known $A_N$-root systems integral of type I from
Sect. \ref{TypeA1}. The conjecture $I_E=I_M$ for $K>1$ represents a new
elliptic beta integral requiring rigorous mathematical
justification.

\subsubsection{$SU(3KN_f-1)$ with $N_f f + N_f \overline{f} + 2adj$}
If we set $N = 3KN_f-1$ in (\ref{KS2}), then we obtain the
$s$-confinement discussed in \cite{Klein}.
The superconformal index for the electric theory is given by
(\ref{KS2adj1}), and the magnetic superconformal index takes the form
\begin{eqnarray}  \nonumber && I_M
=\prod_{L=0}^{K-1} \prod_{J=0}^2 \Gamma(U^{L+KJ/2}s_it_j^{-1};p,q)
\\   && \makebox[4em]{} \times \prod_{i=1}^{N_f} \Gamma(U^{\frac{2-K}{2}}
(ST)^{\frac{3K}{2\widetilde{N}}} s_i^{-1} , U^{\frac{2-K}{2}}
(ST)^{-\frac{3K}{2\widetilde{N}}} t_i;p,q),
\label{onemoreM}\end{eqnarray}
where $U=(pq)^{\frac{1}{K+1}}$, $ S = \prod_{j=1}^{N_f} s_j,$
$T =  \prod_{j=1}^{N_f}t_j$, and the balancing condition reads
$U^NST^{-1}=(pq)^{N_f}.$
The equality $I_E=I_M$ is a new conjectural elliptic beta integral.

\subsubsection{$SU((2K+1)N_f-4K-1)$ with $N_f f + N_f \overline{f} + 2T_A$}
If we set $N = (2K+1)N_f-4K-1$ in (\ref{KS3}), we obtain the
$s$-confinement discussed by Klein in \cite{Klein}.
The electric superconformal index is given by
(\ref{KSg1_1}), and  the magnetic superconformal index  takes the form
\begin{eqnarray}
&& I_M = \prod_{j=0}^K \prod_{k, l =1}^{N_f}
\Gamma((pq)^{\frac{j}{K+1}}s_kt_l;p,q)\prod_{k=1}^{N_f}
\Gamma((U\widetilde{U})^{\frac
12} s_k^{-1},(U\widetilde{U})^{-\frac 12} (pq)^{\frac{1}{K+1}}
t_k^{-1};p,q) \nonumber \\
 &&  \makebox[3em]{} \times \prod_{r=0}^{K-1} \prod_{1 \leq k < l \leq N_f}
\Gamma(U^{-1}(pq)^{\frac{r+1}{K+1}}s_ks_l,U(pq)^{\frac{r}{K+1}}t_kt_l;p,q),
\label{suen-1}\end{eqnarray}
where $U$ is an arbitrary parameter, $\widetilde{U}=U^{2KN_f-4K-1} ST^{-1}
(pq)^{\frac{1-KN_f+2K}{K+1}}$, $ S = \prod_{j=1}^{N_f} s_j,$
$ T =  \prod_{j=1}^{N_f}t_j$, and the balancing condition
reads $ST = (pq)^{N_f-\frac{N+2K}{K+1}}.$

For $K=0$ the parameter $U$ drops out, and one obtains the integral discussed
in Sect. \ref{TypeA1}. The general $K>0$ conjecture $I_E=I_M$ represents
another new elliptic beta integral.

\subsubsection{$SU((2K+1)N_f+4K-1)$ with $N_f f + N_f \overline{f} + 2T_S$}
If we set $N = (2K+1)N_f+4K-1$ in (\ref{KS4}), we obtain again
the $s$-confinement \cite{Klein}. The corresponding electric superconformal
index is given by (\ref{KSg2_1}), and the magnetic superconformal index
takes the form
\begin{eqnarray}
&& I_M = \Gamma(\widetilde{U} ,\widetilde{U}^{-1}
(pq)^{\frac{1}{K+1}} ;p,q)\prod_{j=0}^K \ \ \prod_{k,l =1}^{N_f} \Gamma( (pq)^{\frac{j}{K+1}}s_kt_l;p,q) \\
\nonumber && \makebox[2em]{} \times
 \prod_{r=0}^{K-1}  \ \ \prod_{1 \leq k < l \leq
N_f} \Gamma(U^{-1} (pq)^{\frac{r+1}{K+1}}s_ks_l,U
(pq)^{\frac{r}{K+1}}t_kt_l;p,q) \\
\nonumber &&  \makebox[-1em]{} \times  \prod_{k=1}^{N_f}\Big[ \prod_{r=0}^{K-1}
\Gamma(U^{-1} (pq)^{\frac{r+1}{K+1}} s_k^2,U (pq)^{\frac{r}{K+1}}
t_k^2;p,q) \Gamma((U\widetilde{U})^{\frac 12}
s_k^{-1} ,(U\widetilde{U})^{-\frac 12} (pq)^{\frac{1}{K+1}} t_k^{-1}
;p,q)\Big],
\nonumber \end{eqnarray}
where $\widetilde{U}=U^{2KN_f+4K-1} ST^{-1} (pq)^{\frac{1-KN_f-2K}{K+1}}$,
$S =\prod_{j=1}^{N_f} s_j,$ $ T = \prod_{j=1}^{N_f} t_j$,
and  the balancing condition reads $ST = (pq)^{N_f-\frac{N-2K}{K+1}}.$

Presently the conjecture $I_E=I_M$ is confirmed only for $K=0$,
which reduces again to the integral of Sect. \ref{TypeA1}.

\subsubsection{$SU((4K+3)(N_f-4)-1)$ with $N_f f + (N_f-8) \overline{f} + T_A + T_S$}
If we take $N = (4K+3)(N_f-4)-1$ in (\ref{KS5}), we obtain the
$s$-confinement \cite{Klein}. The corresponding electric superconformal index
is given by (\ref{KSg3_1}), and  the magnetic superconformal index
takes the form
\begin{eqnarray}
&& I_M = \prod_{J=0}^{2K+1} \prod_{i=1}^{N_f} \prod_{j=1}^{N_f-8}
\Gamma((pq)^{\frac{J}{2(K+1)}}s_it_j;p,q)
\\ \nonumber && \makebox[2em]{} \times \prod_{l=0}^{2K}
\prod_{1 \leq i < j \leq N_f}
\Gamma((pq)^{\frac{l+1}{2(K+1)}}U^{-1}s_is_j;p,q) \prod_{l=0}^K
\prod_{i=1}^{N_f} \Gamma((pq)^{\frac{2l+1}{2(K+1)}}U^{-1}s_i^2;p,q)
\\ \nonumber &&
\makebox[2em]{} \times \prod_{m=0}^{2K} \prod_{1 \leq i < j \leq
N_f-8} \Gamma((pq)^{\frac{m}{2(K+1)}}Ut_it_j;p,q) \prod_{m=0}^{K-1}
\prod_{i=1}^{N_f-8} \Gamma((pq)^{\frac{2m+1}{2(K+1)}}Ut_i^2;p,q) \\
\nonumber && \makebox[2em]{} \times \Gamma(\widetilde{U}^{-1}
(pq)^{\frac{1}{2(K+1)}} ;p,q) \prod_{k=1}^{N_f}
\Gamma((U\widetilde{U})^{\frac 12} s_k^{-1};p,q)
\prod_{l=1}^{N_f-8} \Gamma((U\widetilde{U})^{-\frac 12} (pq)^{\frac{1}{2(K+1)}}
t_l^{-1};p,q),
\end{eqnarray}
where $\widetilde{U}=\left( S^2 U^{N-N_f}\right)^{\frac{1}{\widetilde{N}}}$
and the balancing condition reads
$$
U^{-4}\prod_{j=1}^{N_f} s_j t_j \ = \
(pq)^{N_f-4-\frac{(4K+3)(N_f-4)+1}{2(K+1)}}.
$$
The equality $I_E=I_M$ represents another  conjectural new elliptic
beta integral.

\subsubsection{$SU(3KN_f+3)$ with $N_f f + N_f \overline{f} + adj + T_S +
\overline{T}_S$} If we take $N= 3KN_f+3$  for $K$-odd in
(\ref{B1}),  we obtain again the $s$-confinement \cite{Klein}.
The corresponding electric superconformal index is given
by (\ref{Br1_1}), and the magnetic  superconformal index is
\beqa &&
I_M = \prod_{L=0}^{K-1}
\prod_{i,j=1}^{N_f} \Gamma(U^{L+K}s_it_j^{-1},U^{L}s_it_j^{-1};p,q)
\Gamma(U^{\frac{K}{2}}
X^{N-KN_f} Y^{N},U^{\frac{K}{2}}(X^{N-KN_f} Y^{N})^{-1};p,q)
\nonumber \\ && \makebox[4em]{} \times \prod_{J=0}^{K-1} \prod_{1
\leq i < j \leq N_f} \Gamma((XY)^{-1}U^{J+K/2}s_is_j,XYU^{J+K/2}t_i^{-1}t_j^{-1};p,q) \nonumber \\
&& \makebox[4em]{} \times \prod_{J=0}^{\frac{K-1}{2}}
\prod_{i=1}^{N_f}
\Gamma((XY)^{-1}U^{2J+K/2}s_i^2,XYU^{2J+K/2}t_i^{-2};p,q) \nonumber
\\ && \makebox[4em]{} \times \prod_{i=1}^{N_f} \Gamma( U^{\frac{2-K}{2}}
X^{KN_f+2} Y^{\frac{3KN_f+4}{2}} s_i^{-1} ,
U^{\frac{2-K}{2}} X^{-(KN_f+2)} Y^{-\frac{3KN_f+4}{2}} t_i ;p,q),
\end{eqnarray}
where $U=(pq)^{\frac{1}{K+1}},$ $Y = (ST)^{1/N_f}$,
$S = \prod_{i=1}^{N_f} s_i,$   $T =\prod_{i=1}^{N_f} t_i$,
 $X$ is an arbitrary parameter, and the balancing condition reads
$U^{N-2}ST^{-1}=(pq)^{N_f}.$
Again, the proof of the general equality $I_E=I_M$ is absent.

\subsubsection{$SU(3KN_f-5)$ with $N_f f + N_f \overline{f} + adj + T_A +
\overline{T}_A$} If we take $N \ = \ 3KN_f-5$  for $K$-odd in
(\ref{B2}), we obtain  the $s$-confinement \cite{Klein}.
The corresponding electric superconformal index is given
by (\ref{Br2_1}), and the magnetic superconformal index  takes the form
\begin{eqnarray}&&I_M =
\prod_{L=0}^{K-1} \prod_{i,j=1}^{N_f}
\Gamma(U^{L+K}s_it_j^{-1},U^{L}s_it_j^{-1};p,q) \nonumber \\ &&
\makebox[4em]{} \times \prod_{J=0}^{K-1} \prod_{1
\leq i < j \leq N_f} \Gamma((XY)^{-1}U^{J+K/2}s_is_j,XYU^{J+K/2}t_i^{-1}t_j^{-1};p,q) \nonumber \\
&& \makebox[2em]{} \times \prod_{J=0}^{\frac{K-3}{2}}
\prod_{i=1}^{N_f}
\Gamma((XY)^{-1}U^{2J+1+K/2}s_i^2,XYU^{2J+1+K/2}t_i^{-2};p,q)
\\ \nonumber && \makebox[2em]{} \times \prod_{i=1}^{N_f}
\Gamma( U^{\frac{2-K}{2}} X^{KN_f-2} Y^{\frac{3KN_f-4}{2}} s_i^{-1},
U^{\frac{2-K}{2}} X^{-(KN_f-2)} Y^{-\frac{3KN_f-4}{2}} t_i;p,q),
\end{eqnarray} where
$U=(pq)^{\frac{1}{K+1}},$ $Y = (ST)^{1/N_f}$,
$ S  = \prod_{i=1}^{N_f} s_i,$  $T =\prod_{i=1}^{N_f} t_i$,
$X$ is an arbitrary parameter, and the balancing condition reads
$U^{N-2}ST^{-1}=(pq)^{N_f}.$
No proof of the equality $I_E=I_M$ is known at present.

\subsubsection{$SU(N)$ with $N_f f + (N_f-8) \overline{f} + adj + T_A +
\overline{T}_S$} If we set $N = 3K(N_f-4)-1$ in (\ref{B3}), we
obtain  the $s$-confinement \cite{Klein}. The corresponding electric
superconformal index is given by (\ref{Br3_1}), and
\begin{eqnarray}&& \makebox[-2em]{}
I_M = \prod_{L=0}^{K-1}
\prod_{i=1}^{N_f} \Big[ \prod_{j=1}^{N_f-8}
\Gamma(U^{L+K}s_it_j,U^{L}s_it_j;p,q)
\Gamma((XY)^{-1}U^{L+K/2}s_i^2;p,q)\Big]
\nonumber \\ && \makebox[0em]{} \times
\prod_{J=0}^{K-1}\Big[
\prod_{1 \leq i < j \leq N_f} \Gamma((XY)^{-1}U^{J+K/2}s_is_j;p,q)
\prod_{1 \leq i < j \leq N_f-8} \Gamma(XYU^{J+K/2}t_it_j;p,q)\Big]
\\ \nonumber && \makebox[-1em]{} \times  \prod_{i=1}^{N_f}
\Gamma(U^{\frac{2-K}{2}}Y^{\frac{3K(N_f-4)}{2 }} s_i^{-1} ;p,q)
\prod_{k=1}^{N_f-8} \Gamma(U^{\frac{2-K}{2}} Y^{-\frac{3K(N_f-4)}{2 }} t_k^{-1}
;p,q)\
\Gamma(U^{K/2}(X Y^{N})^{-1};p,q),
\nonumber\end{eqnarray}
where $U=(pq)^{\frac{1}{K+1}}$, the balancing condition reads
$U^{N}X^{-4}Y^{-4}ST=(pq)^{N_f-4}$ with $S  = \prod_{i=1}^{N_f} s_i, \
T = \prod_{i=1}^{N_f-8} t_i$, and
$$Y \ = \ X^{2} \left( ST^{-1}(pq)^{\frac{2(K-2)}{K+1}}
\right)^{\frac{1}{N_f-4}}.$$
Equality of indices defines another unproven elliptic beta integral evaluation.

\subsubsection{New confining duality}\label{NewSConf}
Let us take the electric and magnetic $\mathcal{N}=1$ superconformal
field theories described by the tables below
\begin{center}
\begin{tabular}{|c|c|c|c|c|c|}
  \hline
   & $SU(N+1)$ & $SP(2N)$ & $SU(N+3)$ & $U(1)$ & $U(1)_R$ \\  \hline
  $Q_1$ & $\overline{f}$ & $1$ & $f$ & 1 & 0 \\
  $Q_2$ & $f$ & $f$ & $1$ &$-\frac{N+3}{2}$ & 1 \\
  $X$ & $\overline{T}_A$ & 1 & 1 & $N+3$ & 0 \\
\hline \hline
  $q_1=Q_1^{N+1}$ && 1 & $\overline{T}_A$ & $N+1$ & 0 \\
  $q_2=Q_1Q_2$ && $f$ & $f$ & $-\frac{N+1}{2}$ & 1 \\ \hline
\end{tabular}
\end{center}

The dynamically generated superpotential in this case is
$W_{dyn}\propto Q_1^{N+1} (Q_1Q_2)^2.$

The indices read
\begin{eqnarray}\label{s6_1}
&& I_E=\frac{(p;p)^{N}_{\infty} (q;q)^{N}_{\infty}}{(N+1)!}
\int_{\mathbb{T}^{N}} \prod_{1 \leq i < j \leq N+1}
\frac{\Gamma(Sz_i^{-1}z_j^{-1};p,q)}{\Gamma(z_iz_j^{-1},z_i^{-1}z_j;p,q)}
\nonumber
\\&& \makebox[3em]{} \times
\prod_{j=1}^{N+1} \frac{\prod_{k=1}^{N} \Gamma(t_kz_j;p,q)
\prod_{m=1}^{N+3} \Gamma(s_mz_j^{-1};p,q)}{\prod_{k=1}^N
\Gamma(St_kz_j^{-1};p,q)}\prod_{j=1}^N \frac{d z_j}{2 \pi \textup{i} z_j},
\end{eqnarray}
where $\prod_{j=1}^{N+1}=1$, and
\begin{eqnarray}\label{s6_2} && I_M=
\prod_{k=1}^N \prod_{m=1}^{N+3}
\frac{\Gamma(t_ks_m;p,q)}{\Gamma(St_ks_m^{-1};p,q)} \prod_{1 \leq l
< m \leq N+3} \Gamma(Ss_l^{-1}s_m^{-1};p,q)
\end{eqnarray}
with the balancing condition $S= \prod_{m=1}^{N+3} s_m.$

The elliptic beta integral described by the equality $I_E=I_M$ was
discovered by the first author and Warnaar in \cite{spi-war:inversions}.
Here it defines a new pair of $\mathcal{N}=1$ supersymmetric quantum
field theories dual to each other, which was not considered earlier
in the literature. Moreover, it gives a counterexample to the
classification of $s$-confining theories in \cite{Csaki4}.
Conjecturally, there exists a symmetry
transformation for a higher order generalization of $I_E$ depending
on the bigger number of parameters. Correspondingly, there should
exist a more complicated Seiberg duality as well.

\subsection{Exceptional cases for unitary gauge groups}

\subsubsection{$SU(6)$ with $4f+4\overline{f}$}
The following pair of models was constructed in \cite{Csaki3}:
\begin{center}
\begin{tabular}{|c|c|c|c|c|c|c|}
  \hline
   & $SU(6)$ & $SU(4)$ & $SU(4)$ & $U(1)_1$ & $U(1)_2$ & $U(1)_R$ \\  \hline
  $Q$ & $f$ & $f$ & 1 & 1 & 3 & 1 \\
  $\widetilde{Q}$ & $\overline{f}$ & 1 & $f$ & -1 & 3 & 1 \\
  $A$ & $T_{3A}$ & 1 & 1 & 0 & -4 & -1 \\
\hline \hline
  $M_0=Q\widetilde{Q}$ && $f$ & $f$ & 0 & 6 & 2 \\
  $M_2=QA^2\widetilde{Q}$ && $f$ & $f$ & 0 & -2 & 0 \\
  $B_1=AQ^3$ && $\overline{f}$ & 1 & 3 & 5 & 2 \\
  $\widetilde{B}_1=A\widetilde{Q}^3$ && 1 & $\overline{f}$ & -3 & 5 & 2 \\
  $B_3=A^3Q^3$ && $\overline{f}$ & 1 & 3 & $-3$ & 0 \\
  $\widetilde{B}_3=A^3\widetilde{Q}^3$ && 1 & $\overline{f}$ & -3 & $-3$ & 0 \\
  $T=A^4$ && 1 & 1 &  0 & -16 & -4 \\ \hline
\end{tabular}
\end{center}

Their superconformal indices read
\begin{eqnarray}\label{s8_1}
&& I_E=\frac{(p;p)^{5}_{\infty} (q;q)^{5}_{\infty}}{6!}
\int_{\mathbb{T}^{5}} \frac{ \prod_{1 \leq i < j < k \leq 6}
\Gamma(Uz_iz_jz_k;p,q)}{\prod_{1 \leq i < j \leq 6}
\Gamma(z_iz_j^{-1},z_i^{-1}z_j;p,q)}
\\ \nonumber && \makebox[3em]{} \times
\prod_{j=1}^6 \prod_{k=1}^4 \Gamma(s_kz_j,t_kz_j^{-1};p,q)
\prod_{j=1}^{5} \frac{d z_j}{2 \pi \textup{i} z_j},
 \eeqa
where $\prod_{j=1}^6z_j=1$,
 and
\beqa \label{s8_2} &&\makebox[-2em]{} I_M= \Gamma(U^4;p,q) \prod_{k,
l=1}^4 \Gamma(s_kt_l,U^2s_kt_l;p,q) \nonumber \\ && \makebox[2em]{}
\times \prod_{k=1}^4 \Gamma(SUs_k^{-1},SU^3 s_k^{-1}, TU
t_k^{-1},TU^3 t_k^{-1};p,q)
\end{eqnarray}
with $S=\prod_{k=1}^4 s_k, \ T = \prod_{k=1}^4 t_k$,
and the balancing condition $ST U^6 = pq.$

There is actually a lift of this duality to interacting magnetic theories
found in \cite{Csaki2}. The theory is self-dual and is based on $SU(6)$ gauge
group and the flavor symmetry group is
$$
F=SU(6) \times SU(6) \times
U(1)_1 \times U(1)_1.$$
The matter content of the dual theories is
given in the following tables: the electric theory
\begin{center}
\begin{tabular}{|c|c|c|c|c|c|c|}
  \hline
   & $SU(6)$ & $SU(6)$ & $SU(6)$ & $U(1)_1$ & $U(1)_2$ & $U(1)_R$ \\  \hline
  $Q$ & $f$ & $f$ & 1 & $1$ & $1$ & $\frac 12$ \\
  $\overline{Q}$ & $\overline{f}$ & 1 & $f$ & $-1$ & $1$ & $\frac 12$
  \\
  $A$ & $T_{3A}$ & 1 & 1 & 0 & $-2$ & 0 \\
\hline
\end{tabular}
\end{center}
and the magnetic theory
\begin{center}
\begin{tabular}{|c|c|c|c|c|c|c|}
  \hline
   & $SU(6)$ & $SU(6)$ & $SU(6)$ & $U(1)_1$ & $U(1)_2$ & $U(1)_R$ \\  \hline
  $q$ & $f$ & $\overline{f}$ & 1 & $1$ & $1$ & $\frac 12$ \\
  $\overline{q}$ & $\overline{f}$ & 1 & $\overline{f}$ & $-1$ & $1$ & $\frac 12$
  \\
  $a$ & $T_{3A}$ & 1 & 1 & 0 & $-2$ & 0 \\
  $M_0$ & 1 & $f$ & $f$ & 0 & 2 & $1$ \\
  $M_2$ & 1 & $f$ & $f$ & 0 & $-2$ & $1$ \\
\hline
\end{tabular}
\end{center}

The electric superconformal index is
\beqa
\label{Cs6el}
&& \makebox[-3em]{}
I_E =\frac{(p;p)_{\infty}^{5}
(q;q)_{\infty}^{5} }{6!} \int_{\mathbb{T}^5} \frac{\prod_{1 \leq i <
j < k \leq 6} \Gamma(Uz_iz_jz_k;p,q)}{\prod_{1 \leq i < j \leq 6}
\Gamma(z_iz_j^{-1},z_i^{-1}z_j;p,q)}
\nonumber \\  && \makebox[2em]{}  \times
 \prod_{i=1}^6 \prod_{j=1}^6
\Gamma(s_iz_j,t_iz_j^{-1};p,q)  \prod_{j=1}^5\frac{dz_j}{2\pi \textup{i} z_j}, \eeqa
and the
magnetic index is \beqa \label{Cs6mag} && \makebox[-2em]{} I_M =
\frac{(p;p)_{\infty}^{5} (q;q)_{\infty}^{5} }{6!} \prod_{i,j=1}^6
 \Gamma(s_it_j,U^2s_it_j;p,q) \nonumber \\
&& \makebox[0em]{} \times \int_{\mathbb{T}^5} \frac{\prod_{1 \leq i
< j < k \leq 6} \Gamma(Uz_iz_jz_k;p,q)}{\prod_{1 \leq i < j \leq 6}
\Gamma(z_iz_j^{-1},z_i^{-1}z_j;p,q)} \prod_{i=1}^6 \prod_{j=1}^6
\Gamma(\frac{\sqrt[3]{S}}{s_i}z_j,\frac{\sqrt[3]{T}}{t_i}z_j^{-1};p,q)
 \prod_{j=1}^5\frac{dz_j}{2\pi \textup{i} z_j},\eeqa
where $S=\prod_{i=1}^6, \ T=\prod_{j=1}^6 t_j$,
 and the balancing condition reads $STU^6  =(pq)^3.$

\subsubsection{$SU(5)$ with $3T_A+3\overline{f}$}
Models \cite{Csaki3}:
\begin{center}
\begin{tabular}{|c|c|c|c|c|c|}
  \hline
   & $SU(5)$ & $SU(3)$ & $SU(3)$ & $U(1)$ & $U(1)_R$ \\  \hline
  $Q$ & $\overline{f}$ & 1 & $f$ & -3 & $\frac 23$ \\
  $A$ & $T_A$ & $f$ & 1 & 1 & 0 \\
\hline \hline
  $AQ^2$ && $f$ & $\overline{f}$ & -5 & $\frac 43$ \\
  $A^3Q$ && $T_{AS}$ & $f$ & 0 & $\frac 23$ \\
  $A^5$ && $T_S$ & 1 & 5 & 0 \\ \hline
\end{tabular}
\end{center}

Indices:
\begin{eqnarray}\label{s9_1}
&&  \makebox[-2em]{}
I_E=\frac{(p;p)^{4}_{\infty} (q;q)^{4}_{\infty}}{5!}
\int_{\mathbb{T}^{4}} \prod_{1 \leq i < j \leq 5}
\frac{\prod_{k=1}^3\Gamma(s_kz_iz_j;p,q)}{\Gamma(z_iz_j^{-1},z_i^{-1}z_j;p,q)}
\prod_{j=1}^5 \prod_{k=1}^3
\Gamma(t_kz_j^{-1};p,q)\prod_{j=1}^{4} \frac{d z_j}{2 \pi \textup{i} z_j},
\eeqa
with $\prod_{j=1}^5z_j=1$, and
\beqa\label{s9_2} && I_M=  \prod_{k,l=1}^3 \Gamma(T s_k
t_l^{-1};p,q) \prod_{1 \leq j < k \leq 3} \Gamma(S s_j s_k;p,q) \\
\nonumber && \makebox[3em]{} \times \prod_{j=1}^3 \Gamma(S
s_j^2;p,q) \prod_{k,j,l=1; k \neq j}^3 \Gamma(s_k^2s_jt_l;p,q)
\prod_{l=1}^3 \Gamma(St_l;p,q)^2,
\end{eqnarray}
where $S=\prod_{k=1}^3 s_k, \ T =\prod_{k=1}^3 t_k$,
and the balancing condition reads $S^3T = pq.$

\subsubsection{$SU(5)$ with $2T_A+4\overline{f}+2f$}
Models \cite{Csaki3}:
\begin{center}
\begin{tabular}{|c|c|c|c|c|c|c|c|}
  \hline
   & $SU(5)$ & $SU(2)$ & $SU(4)$ & $SU(2)$ & $U(1)_1$ & $U(1)_2$ & $U(1)_R$ \\  \hline
  $Q$ & $f$ & $1$ & 1 & $f$ & -2 & 1 & $\frac 13$ \\
  $\widetilde{Q}$ & $\overline{f}$ & 1 & $f$ & 1 & 1 & 1 & $\frac 13$ \\
  $A$ & $T_A$ & $f$ & 1 & 1 & 0 & -1 & 0 \\
\hline \hline
  $Q\widetilde{Q}$ && 1 & $f$ & $f$ & -1 & 2 & $\frac 23$ \\
  $A\widetilde{Q}^2$ && $f$ & $T_A$ & 1 & 2 & 1 & $\frac 23$ \\
  $A^2Q$ && $T_S$ & 1 & $f$ & -2 & -1 & $\frac 13$ \\
  $A^3\widetilde{Q}$ && $f$ & $f$ & 1 & 1 & -2 & $\frac 13$ \\
  $A^2Q^2\widetilde{Q}$ && 1 & $f$ & 1 &  -3 & 1 & 1 \\ \hline
\end{tabular}
\end{center}

Indices:
\begin{eqnarray}\label{s10_1}
&& I_E=\frac{(p;p)^{4}_{\infty} (q;q)^{4}_{\infty}}{5!}
\int_{\mathbb{T}^{4}} \prod_{1 \leq i < j \leq 5}
\frac{\prod_{k=1}^2
\Gamma(s_kz_iz_j;p,q) }{\Gamma(z_iz_j^{-1},z_i^{-1}z_j;p,q)} \\
\nonumber && \makebox[3em]{} \times \prod_{j=1}^5 \prod_{k=1}^4
\Gamma(t_kz_j^{-1},u_kz_j;p,q) \prod_{j=1}^{4} \frac{d z_j}{2 \pi \textup{i}
z_j},\eeqa
with $\prod_{j=1}^5z_j=1$, and \beqa \label{s10_2} && I_M= \prod_{k=1}^4 \Gamma(SU
t_k;p,q) \prod_{k=1}^4 \prod_{l=1}^2 \Gamma(t_ku_l,S t_ks_l;p,q)
\prod_{k=1}^2 \Gamma(Su_k;p,q) \\ \nonumber && \makebox[3em]{} \times
\prod_{k,l=1}^2 \Gamma(s_l^2u_k;p,q) \prod_{k=1}^2 \prod_{1 \leq l <
m \leq 4} \Gamma(s_kt_lt_m;p,q),
\end{eqnarray}
where the balancing condition reads $S^3TU =pq$ and
$S = \prod_{k=1}^2 s_k, \ T=\prod_{k=1}^4 t_k, \ U = u_1u_2.$

\subsubsection{$SU(6)$ with $2T_A+f+5\overline{f}$}
Models \cite{Csaki3}:
\begin{center}
\begin{tabular}{|c|c|c|c|c|c|c|}
  \hline
   & $SU(6)$ & $SU(2)$ & $SU(5)$ & $U(1)_1$ & $U(1)_2$ & $U(1)_R$ \\  \hline
  $Q$ & $f$ & 1 & 1 & -5 & -4 & 0 \\
  $\widetilde{Q}$ & $\overline{f}$ & 1 & $f$ & 1 & -4 & 0 \\
  $A$ & $T_A$ & $f$ & 1 & 0 & 3 & $\frac 14$ \\
\hline \hline
  $Q\widetilde{Q}$ && 1 & $f$ & -4 & -8 & 0 \\
  $A\widetilde{Q}^2$ && $f$ & $T_A$ & 2 & -5 & $\frac 14$ \\
  $A^3$ && $T_{3S}$ & 1 & 0 & 9 & $\frac 34$ \\
  $A^3Q\widetilde{Q}$ && $f$ & $f$ & -4 & 1 & $\frac 34$ \\
  $A^4\widetilde{Q}^2$ && 1  & $T_A$ & 2 & 4 & 1  \\ \hline
\end{tabular}
\end{center}

Indices:
\begin{eqnarray}\label{s11_1}
&&I_E= \frac{(p;p)^{5}_{\infty} (q;q)^{5}_{\infty}}{6!}
\int_{\mathbb{T}^{5}} \prod_{1 \leq i < j \leq 6}
\frac{\Gamma(Uz_iz_j;p,q)}{\Gamma(z_iz_j^{-1},z_i^{-1}z_j;p,q)}
\nonumber \\ && \makebox[3em]{}  \times \prod_{l=1}^2 \prod_{1 \leq
j < k \leq 6 } \Gamma(s_lz_jz_k;p,q) \prod_{j=1}^6 \prod_{k=1}^5
\Gamma(t_kz_j^{-1};p,q)\prod_{j=1}^{5} \frac{d z_j}{2 \pi \textup{i} z_j},
\eeqa
with $\prod_{i=1}^6z_i=1$, and
\beqa\label{s11_2} && I_M= \prod_{k}^5 \Gamma(Ut_k;p,q)
\prod_{k=1}^2 \prod_{j=1}^5 \Gamma(SUs_kt_j;p,q)\prod_{k=1}^2
\prod_{1 \leq j < l \leq 5} \Gamma(s_kt_jt_l;p,q)
 \\ &&  \nonumber \makebox[3em]{} \times  \prod_{1 \leq j < k \leq 5}
\Gamma(S^2t_jt_k;p,q) \prod_{j=1}^2 \Gamma(s_j^3,Ss_j;p,q),
\end{eqnarray}
where the balancing condition reads $S^4TU=pq$ and
$S=\prod_{k=1}^2 s_k, \ T =\prod_{k=1}^5 t_k.$

\subsubsection{$SU(7)$ with $2T_A+6\overline{f}$}
Models \cite{Csaki3}:
\begin{center}
\begin{tabular}{|c|c|c|c|c|c|}
  \hline
   & $SU(7)$ & $SU(2)$ & $SU(6)$ & $U(1)$ & $U(1)_R$ \\  \hline
  $Q$ & $\overline{f}$ & 1 & $f$ & -5 & $\frac 13$ \\
  $A$ & $T_A$ & $f$ & 1 & 3 & 0 \\
\hline \hline
  $AQ^2$ && $f$ & $T_A$ & -7 & $\frac 23$ \\
  $A^4Q$ && $T_S$ & $f$ & 7 & $\frac 13$ \\ \hline
\end{tabular}
\end{center}

Indices:
\begin{eqnarray}\label{s12_1}
&& I_E =\frac{(p;p)^{6}_{\infty} (q;q)^{6}_{\infty}}{7!}
\int_{\mathbb{T}^{6}}
\prod_{1 \leq i < j \leq 7} \frac{1}{\Gamma(z_iz_j^{-1},z_i^{-1}z_j;p,q)} \\
\nonumber && \makebox[3em]{} \times \prod_{k=1}^2 \prod_{1 \leq i <
j \leq 7} \Gamma(s_kz_iz_j;p,q) \prod_{k=1}^6 \prod_{j=1}^7
\Gamma(t_kz_j^{-1};p,q)\prod_{j=1}^{6} \frac{d z_j}{2 \pi \textup{i} z_j},
\eeqa
with $\prod_{i=1}^7z_i=1$, and \beqa\label{s12_2} && I_M= \prod_{k=1}^6
\Gamma(S^2t_k;p,q) \prod_{k=1}^6 \prod_{l=1}^2 \Gamma(Ss_l^2t_k;p,q)
\prod_{k=1}^2 \prod_{1 \leq l < m \leq 6} \Gamma(s_kt_lt_m;p,q),
\end{eqnarray}
where the balancing condition reads $S^5T=pq$ and
$S=\prod_{k=1}^2 s_k, \ T = \prod_{k=1}^6 t_k.$

 All the equalities of superconformal indices of dual theories, $I_E=I_M$,
described in this section represent new elliptic beta integrals
requiring  a rigorous proof (the parameter values
 are  assumed to guarantee that only sequences of poles
of the integrands converging to zero are located inside
the contour $\mathbb{T}$).

\subsection{Symplectic gauge group}

\subsubsection{$SP(2N)$ with $(2N+4)f$}\label{TypeBC1}
Models \cite{Intriligator1}:
\begin{center}
\begin{tabular}{|c|c|c|c|}
  \hline
    & $SP(2N)$ & $SU(2N+4)$ & $U(1)_R$ \\  \hline
  $Q$ & $f$ & $f$ & $2r=\frac{1}{N+2}$ \\
\hline \hline
  $Q^2$ &   & $T_A$ & $2r=\frac{2}{N+2}$ \\  \hline
\end{tabular}
\end{center}

Indices:
\begin{eqnarray}\label{sp1_1}
&& \makebox[-2em]{}
I_E=\frac{(p;p)^{N}_{\infty} (q;q)^{N}_{\infty}}{2^NN!}
\int_{\mathbb{T}^N} \prod_{1 \leq i < j \leq N}
\frac{1}{\Gamma(z_i^{\pm 1} z_j^{\pm 1};p,q)}
\prod_{j=1}^{N} \frac{\prod_{m=1}^{2N+4}
\Gamma(t_mz_j^{\pm 1};p,q)}{\Gamma(z_j^{\pm 2};p,q)}
\frac{d z_j}{2 \pi \textup{i} z_j}\eeqa
and \beqa\label{sp1_2} && I_M=
\prod_{1 \leq m < s \leq 2N+4} \Gamma(t_mt_s;p,q),
\end{eqnarray}
where the balancing condition reads $\prod_{m=1}^{2N+4} t_m = pq.$

The equality $I_E=I_M$ was introduced and partially justified by van Diejen
and the first author in \cite{die-spi:elliptic} and completely
proven in \cite{Rains} and \cite{Spiridonov3}. Its simplest
$p \rightarrow 0$ limit yields one of the Gustafson $q$-beta
integrals \cite{gus1}.

\subsubsection{$SP(2N)$ with $6f$ and $T_A$}

This duality was considered in \cite{Cho,Csaki1}.
The flavor symmetry group is $F=SU(6) \times U(1)$
and the field content is

\begin{center}
\begin{tabular}{|c|c|c|c|c|}
  \hline
    & $SP(2N)$ & $SU(6)$ & $U(1)$ & $U(1)_R$ \\  \hline
  $Q$ & $f$ & $f$ & $N-1$ & $2r=\frac{1}{3}$ \\
  $A$ & $T_A$ & 1 & -3 & 0 \\
\hline \hline
  $A^k$ &   & 1 & $-3k$ & 0 \\
  $QA^mQ$ &   & $T_A$ & $2(N-1)-3m$ & $\frac 23$ \\  \hline
\end{tabular}
\end{center}
where $k=2,\ldots,N$ and $m=0,\ldots,N-1$.

The electric superconformal index is given by the integral
\begin{eqnarray}\label{sp2_1}
&&I_E= \frac{(p;p)^{N}_{\infty} (q;q)^{N}_{\infty}}{2^NN!}
\Gamma(t;p,q)^{N-1} \int_{\mathbb{T}^N} \prod_{1 \leq i < j \leq N}
\frac{\Gamma(t z_i^{\pm 1} z_j^{\pm 1};p,q)}{\Gamma(z_i^{\pm 1}
z_j^{\pm 1};p,q)}
\\ \nonumber && \makebox[3em]{} \times \prod_{j=1}^{N}
\frac{\prod_{m=1}^{6} \Gamma(t_mz_j^{\pm 1};p,q)}{\Gamma(z_j^{\pm
2};p,q)}\prod_{j=1}^{N} \frac{d z_j}{2 \pi \textup{i} z_j} \eeqa
and the magnetic index is
\beqa\label{sp2_2} &&
I_M= \prod_{j=2}^N \Gamma(t^j;p,q)
\prod_{j=1}^N \prod_{1 \leq m < s \leq 6} \Gamma(t^{j-1}t_mt_s;p,q),
\end{eqnarray}
where the balancing condition reads $t^{2N-2}\prod_{m=1}^6 t_m = pq.$

The equality $I_E=I_M$ coincides with the elliptic Selberg integral suggested
by van Diejen and the first author in \cite{die-spi:elliptic} and proven in
\cite{die-spi:selberg} as a consequence of the $BC_n$-elliptic
beta integral of type I (its direct proof is given also in \cite{Rains}).
The Selberg integral plays a fundamental role in mathematics and
mathematical physics because of a large number of applications \cite{fw}.
Note that this exactly computable integral gives a confirmation of the KS
duality for the special values of parameters $N_f=3$, $K=N$.

\subsubsection{$SP(2M) + 4 \overline{f} + 2M f +
T_A$}\label{NewConfSP}
This new confining duality is obtained from the results of
Sect. \ref{SpSpD} by formal setting $N=0$.
The models are described in the table

\begin{center}
\begin{tabular}{|c|c|c|c|c|c|c|c|c|}
  \hline
   & $SP(2M)$ & $SU(4)$ & $SP(2l_1)$ & $SP(2l_2)$ & $\ldots$ & $SP(2l_K)$ & $U(1)$ & $U(1)_R$ \\  \hline
  $W_1$ & $f$ & $\overline{f}$ & 1 & 1 & $\ldots$ & 1 & $-\frac{M-2}{4}$ & 0 \\
  $Q_1$ & $f$ & 1 & $f$ & 1 & $\ldots$ & 1 & $-\frac{n_1}{2}$ & 1 \\
  $Q_1$ & $f$ & 1 & 1 & $f$ & $\ldots$ & 1 & $-\frac{n_2}{2}$ & 1 \\
  $\ldots$ & & & & & & & & \\
  $Q_K$ & $f$ & 1 & 1 & 1 & $\ldots$ & $f$ & $-\frac{n_K}{2}$ & 1 \\
  $X$ & $T_A$ & 1 & 1 & 1 & $\ldots$ & 1 & 1 & 0  \\
\hline \hline
  $W_1^2X^j$ & & $\overline{T}_A$ & $1$ & 1 & $\ldots$ & 1 & $j-\frac{M-2}{2}$ & 0 \\
  $W_1Q_1X^{k_1}$ & & $\overline{f}$ & $f$ & 1 & $\ldots$ & 1 & $-\frac{M-2}{4}-\frac{n_1}{2}+k_1$ & 1 \\
  $W_1Q_2X^{k_2}$ & & $\overline{f}$ & 1 & $f$ & $\ldots$ & 1 & $-\frac{M-2}{4}-\frac{n_2}{2}+k_2$ & 1 \\
  $\ldots$ & & & & & & & & \\
  $W_1Q_KX^{k_K}$ & & $\overline{f}$ & 1 & 1 & $\ldots$ & $f$ & $-\frac{M-2}{4}-\frac{n_K}{2}+k_K$ & 1
  \\ \hline
\end{tabular}
\end{center}
where $j=0,\ldots,M-1$, $k_i = 0,\ldots,n_i-1$ for any
$i=1,\ldots,K$, $n_1 \neq n_2 \neq \ldots \neq n_K$ and
$\sum_{i=1}^{K} l_i n_i \ = \ M.$

The superconformal indices have the form
\begin{eqnarray}\label{sp3_1} && I_E
=\frac{(p;p)_\infty^M (q;q)_\infty^M} {2^M M!} \Gamma(t;p,q)^{M-1}
\int_{\mathbb{T}^M} \prod_{1\le
i<j\le M} \frac{\Gamma(tz_i^{\pm1}z_j^{\pm1};p,q)}{
\Gamma(z_i^{\pm1}z_j^{\pm1};p,q)}
\\ \nonumber && \makebox[2em]{} \times
\prod_{j=1}^M\frac{\prod_{k=1}^4\Gamma(tt^{-1}_kz_j^{\pm1};p,q)
\prod_{r=1}^{K} \prod_{m=1}^{l_r} \Gamma(s_{r,m}z_j^{\pm1};p,q)}
{\Gamma(z_j^{\pm2};p,q) \prod_{r=1}^{K} \prod_{m=1}^{l_r}
\Gamma(t^{n_r}s_{r,m}z_j^{\pm1};p,q)} \frac{dz_j}{2\pi \textup{i} z_j}
\end{eqnarray}
and
\begin{eqnarray}\label{sp3_2} && I_M
= \prod_{i=0}^{M-1}\prod_{1 \le k < r \le 4}
\Gamma(t^{i+2}t_k^{-1}t_r^{-1};p,q) \prod_{k=1}^4\prod_{r=1}^{K}
\prod_{i=1}^{l_r}\prod_{k_m=0}^{n_r-1}
\frac{\Gamma(t^{k_m+1}t_k^{-1}s_{r,i};p,q)}
{\Gamma(t^{k_m}t_ks_{r,i};p,q)},
\end{eqnarray}
where the balancing condition is $\prod_{r=1}^4t_r=t^{2+M}$.
The equality $I_E=I_M$ was conjectured in \cite{rai:lit} and
proven in \cite{bult}. This duality gives another example of
$s$-confining theories missed in \cite{Csaki4}.

\subsubsection{$SP(2K(N_f-2))$ with $N_f f + T_A$}
This duality was considered in \cite{Csaki7,Klein}. From
(\ref{KSeq}) we see that the choice $N=K(N_f-2)$ yields
$\widetilde{N} = 0$, and the theory is $s$-confining. The
field content of the electric and magnetic theories is easily found
from the tables given in Sect. 10.1. For brevity we skip the electric
superconformal index given by (\ref{KSsp1_1}), and present directly
the magnetic index
\begin{eqnarray}
&&    I_M=  \Gamma(U;p,q)^{-1} \prod_{l=1}^K \prod_{1 \leq i < j
\leq 2N_f} \Gamma(U^{l-1} s_i s_j;p,q),
\end{eqnarray}
where $U=(pq)^{\frac{1}{K+1}}$ and the balancing condition reads
$U^{2KN_f-2K}\prod_{i=1}^{2N_f}s_i = (pq)^{N_f}.$
The conjecture $I_E=I_M$ represents a new elliptic beta integral. For $K=1$
it reduces to the proven relation of Sect. \ref{TypeBC1}.

\subsubsection{$SP(2(N_f-2+2KN_f))$ with $N_f f + T_S$}
Looking at (\ref{KS2eq}) and fixing $N \ = \ N_f-2+2KN_f$, we obtain
the $s$-confining theory which was considered in
\cite{Csaki7,Klein}. The corresponding electric superconformal index
is given by (\ref{KSsp2_1}), and the magnetic index takes the form
\begin{eqnarray}
&&    I_M= \prod_{l=0}^K \prod_{1 \leq i < j \leq 2N_f} \Gamma(U^l
s_i s_j;p,q)
    \\ \nonumber  &&  \makebox[3em]{}  \times \prod_{l=0}^{K-1}
\prod_{1 \leq i < j \leq 2N_f} \Gamma(U^{(2l+1)/2} s_i s_j;p,q)
    \prod_{i=1}^{2N_f} \Gamma(U^{(2l+1)/2} s_i^{\pm 2};p,q),
\end{eqnarray}
where $U=(pq)^{\frac{1}{K+1}}$  and the balancing condition reads
$U^{2N_f-2+4KN_f}\prod_{i=1}^{2N_f}s_i \ = \ (pq)^{N_f}.$
The conjecture $I_E=I_M$ represents a new elliptic beta integral.

\subsubsection{$SP(2(3KN_f-4K-2))$ with $N_f f + 2 T_A$}
Looking at (\ref{BP1}) and fixing $N \ = \ 3KN_f-4K-2$ for odd $K$,
we obtain the $s$-confining theory which was considered in
\cite{Klein}. The corresponding electric superconformal index
is given by (\ref{KSsp3_1}), and the magnetic index takes the form
\begin{eqnarray}
&& \makebox[-4em]{} I_M =      \Gamma(U,U^{\frac K2};p,q)^{-1}
    \prod_{J=0}^{K-1} \prod_{L=0}^2 \prod_{1 \leq i < j \leq 2N_f} \Gamma(U^{J+\frac{KL}{2}} s_i
    s_j;p,q)\prod_{J=0}^{\frac{K-1}{2}} \prod_{j=1}^{2N_f} \Gamma(U^{2J+1+\frac K2} s_j^2;p,q),
\end{eqnarray}
where $U=(pq)^{\frac{1}{K+1}}$
and the balancing condition reads $U^{3KN_f-2K-1}\prod_{i=1}^{2N_f}s_i=(pq)^{N_f}.$
Rigorous justification of the expected equality $I_E=I_M$ is  absent  at the moment.

\subsubsection{$SP(2(3KN_f-4K+2))$ with $N_f f + T_S + T_A$}
Looking at (\ref{BP2}) and fixing $N = 3KN_f-4K+2$ for $K$ odd,
we obtain the $s$-confining theory which was considered in
\cite{Klein}. The corresponding electric superconformal index
is given by (\ref{KSsp4_1}), and the magnetic index has the form
\begin{eqnarray}
  && \makebox[-2em]{}  I_M =     \Gamma(U;p,q)^{-1}
     \prod_{J=0}^{K-1} \prod_{L=0}^2 \prod_{1 \leq i < j \leq 2N_f} \Gamma(U^{J+\frac{KL}{2}} s_i
    s_j;p,q) \prod_{J=0}^{\frac{K-1}{2}} \prod_{j=1}^{2N_f} \Gamma(U^{2J+\frac K2} s_j^2;p,q),
\end{eqnarray}
where $U=(pq)^{\frac{1}{K+1}}$
and the balancing condition reads
$U^{3KN_f-2K+1}\prod_{i=1}^{2N_f}s_i= (pq)^{N_f}.$
The conjectural equality $I_E=I_M$ is our last new elliptic beta integral
for classical root systems.

\section{Exceptional $G_2$ group}

{\bf $G_2$ with 5 flavors.}  This $s$-confining duality was discussed
in \cite{Exceptional2,Exceptional1}. The electric theory with
the gauge group $G_2$ and its magnetic dual are described in the table below

\begin{center}
\begin{tabular}{|c|c|c|c|}
  \hline
    & $G_2$ & $SU(5)$ & $U(1)_R$ \\  \hline
  $Q$ & $7$ & $f$ & $2r=\frac{1}{5}$ \\
\hline \hline
  $Q^2$ &   & $T_S$ & $\frac 25$ \\
  $Q^3$ &   & $\overline{T}_A$ & $\frac 35$ \\
  $Q^4$ &   & $\overline{f}$ & $\frac 45$ \\  \hline
\end{tabular}
\end{center}

The superconformal indices are
\begin{eqnarray}\label{G2E}
 I_E &=& \frac{(p;p)_\infty^2 (q;q)_\infty^2}{2^23}
\prod_{m=1}^5 \Gamma(t_m;p,q) \int_{{\mathbb
T}^2} \frac{\prod_{k=1}^3\prod_{m=1}^5\Gamma(t_mz_k^{\pm1};p,q)}
{\prod_{1\leq j<k\leq3}
\Gamma(z_j^{\pm1}z_k^{\pm1};p,q)}\prod_{j=1}^2\frac{dz_j}{2\pi \textup{i}z_j},
\end{eqnarray}
where $z_1z_2z_3=1, \ |t_m|<1,$ and
\begin{eqnarray}\label{G2M}
I_M&=&
\prod_{m=1}^5\frac{\Gamma(t_m^2;p,q)} {\Gamma((pq)^{1/2}t_m;p,q)}
\prod_{1\leq l<m\leq 5}
\frac{\Gamma(t_lt_m;p,q)}{\Gamma((pq)^{1/2}t_lt_m;p,q)}
\label{G2}\end{eqnarray}
with the balancing condition $\prod_{m=1}^5 t_m=(pq)^{1/2}.$

The conjecture $I_E=I_M$ describes the first elliptic beta
integral for exceptional root systems (it was mentioned in \cite{SV}
and proposed also earlier by M. Ito). Substituting $t_5=(pq)^{1/2}/(t_1t_2t_3t_4)$
in (\ref{G2E})  and (\ref{G2}),  and taking the limit $p\to 0$,
one obtains the four parameter $q$-beta integral on the $G_2$
root system derived in \cite{gus-trans}.

{\bf $G_2$ with $5 < N_f < 12$ flavors.} This duality was discovered
by Pouliot in \cite{P2}. The electric theory has gauge group $G_2$, but its
magnetic dual has $SU(N_f-3)$ gauge group. Their field content is
presented in the tables below.

The electric theory (the vector superfield $V$ is omitted):
\begin{center}
\begin{tabular}{|c|c|c|c|}
  \hline
    & $G_2$ & $SU(N_f)$ & $U(1)_R$ \\  \hline
  $Q$ & $7$ & $f$ & $2r=1-\frac{4}{N_f}$ \\
\hline
\end{tabular}
\end{center}
and the magnetic theory (the vector superfield $\widetilde V$ is omitted):
\begin{center}
\begin{tabular}{|c|c|c|c|} \hline
  & $SU(N_f-3)$ & $SU(N_f)$ & $U(1)_R$ \\ \hline
  $q$ & $\overline{f}$  & $\overline{f}$ & $2r_q=\frac{3}{N_f}(1-\frac{1}{N_f-3})$ \\
  $q_0$ &  $\overline{f}$  & 1 & $2r_{q_0}=1-\frac{1}{N_f-3}$ \\
  $s$ & $T_S$  & 1 & $2r_s=\frac{2}{N_f-3}$ \\
  $M$ &  1  & $T_S$ & $2r_M=2-\frac{8}{N_f}$ \\
\hline
\end{tabular}
\end{center}

Corresponding superconformal indices are described by the integrals
\begin{equation}\label{G2S}
I_E=\frac{(p;p)_\infty^2 (q;q)_\infty^2}{2^23} \prod_{m=1}^{N_f}
\Gamma(t_m;p,q) \int_{{\mathbb T}^2} \frac{\prod_{k=1}^3
\prod_{m=1}^{N_f} \Gamma(t_mz_k^{\pm1};p,q)} {\prod_{1\leq j<k\leq3}
\Gamma(z_j^{\pm1}z_k^{\pm1};p,q)} \prod_{k=1}^2\frac{dz_k}{2\pi \textup{i} z_k},
\end{equation}
where $z_1z_2z_3=1,$ and
 \begin{eqnarray}\label{G2SM} &&
I_M =\frac{(p;p)_{\infty}^{N_f-4}
(q;q)_{\infty}^{N_f-4}}{(N_f-3)!} \prod_{1 \leq j < k
\leq N_f} \Gamma(t_jt_k;p,q) \prod_{j=1}^{N_f} \Gamma(t_j^2;p,q) \\
\nonumber && \makebox[4em]{} \times \int_{{\mathbb T}^{N_f-4}}
 \prod_{1 \leq j < k \leq N_f-3}
\frac{\Gamma((pq)^{r_s}z_jz_k;p,q)}{\Gamma(z_j^{-1}z_k,z_jz_k^{-1};p,q)}
\prod_{j=1}^{N_f-3} \Gamma((pq)^{r_s}z_j^2;p,q)
\\ \nonumber && \makebox[6em]{} \times
\prod_{j=1}^{N_f-3} \Gamma((pq)^{(1-r_s)/2}z_j^{-1};p,q)
\prod_{k=1}^{N_f}
\Gamma((pq)^{(1-r_s)/2}t_k^{-1}z_j^{-1};p,q)\prod_{j=1}^{N_f-4}
\frac{dz_j}{2 \pi \textup{i} z_j},
\end{eqnarray}
where $\prod_{j=1}^{N_f-3}z_j=1,$
and the balancing condition reads
$\prod_{m=1}^{N_f} t_m=(pq)^{(N_f-4)/2}.$
The equality $I_E=I_M$ represents a new symmetry transformation
formula for general elliptic hypergeometric integrals on the $G_2$
root system. Independently, it was also considered earlier by F. A. Dolan.

For $N_f=5$ the integral $I_M$ takes the form
\begin{eqnarray}
&& I_M =\frac{(p;p)_{\infty} (q;q)_{\infty}}{2} \prod_{1 \leq j < k
\leq 5} \Gamma(t_jt_k;p,q) \prod_{j=1}^{5} \Gamma(t_j^2;p,q) \\
\nonumber && \makebox[4em]{} \times \int_{{\mathbb T}} \frac{
\Gamma((pq)^{1/4}z_j^{\pm 1};p,q) \prod_{k=1}^{5}
\Gamma((pq)^{1/4}t_k^{-1}z_j^{\pm1};p,q)}{\Gamma(z^{\pm2};p,q)}
\frac{dz}{2 \pi \textup{i} z}.
\end{eqnarray}

Using the univariate elliptic beta integral, one can compute this
$I_M$ and find the index coinciding with (\ref{G2}).
As to the general $G_2$-transformation $I_E=I_M$,
it should be a consequence of the original $SU(3)$-gauge group
Seiberg duality. Indeed, let us take $N=3$ and set
$t_i^{-1}=s_i$ in the electric index (\ref{IE-seiberg}). Then, if we impose
the constraint $s_{N_f}=pq$, we obtain the $G_2$-group electric index (\ref{G2S})
with $N_f$ and $t_i$ replaced by $N_f-1$ and $s_i$, respectively. Therefore
it is expected that the $G_2$-magnetic index  can be obtained after
appropriate restrictions on $I_M$ in (\ref{IM-seiberg}). A
difficulty lies in computing the limit $s_{N_f}\to pq$,
since it leads to a diverging integral multiplied by a vanishing coefficient.
This limit is currently under investigation.

\section{'t Hooft anomaly matching conditions}

In this section we describe the standard 't Hooft anomaly matching conditions
\cite{Hooft,Terning} for some of the new dualities. Needed Casimir operators
for unitary and symplectic groups can be found in Appendix C. There
exist also the discrete anomalies matching conditions \cite{Csaki7},
but we skipped their consideration in the present work.

{\bf Multiple $SP(2N)$ duality.}
Let us begin with the multiple duality for $SP(2N)$ gauge group
found in \cite{SV} and discussed in Sect. \ref{SPDual}. Coincidence of the anomalies
is checked for the smaller flavor groups of dual theories. The subgroup
$SU(4) \times SU(4) \times U(1)_B \times U(1) \times U(1)_R$ of the
electric theory has the following triangle anomaly coefficients:

\beqa SU(4)^3_L &\ & 2N, \ \ \ \ \ \ \ \ \ \ \ \ \ \ \ \ \ \ \ \ \ \
\ \ \ \ \ \ \ \
SU(4)^2_L \times U(1)_R \ \ \ \ \ -2N^2+1 \nonumber \\
SU(4)^2_L \times U(1) &\ & \frac{3N^2-2N-1}{2}, \ \ \ \ \ \ \ \ \ \
\ \ \ \ \
SU(4)^2_L \times U(1)_B \ \ \ \ \ \ 2N \nonumber \\
U(1)_B^2 \times U(1) &\ & -4N(N-1), \ \ \ \ \ \ \ \ \ \ \ \ \ \ \ \
\ U(1)_B^2 \times U(1)_R \ \ \ \ \ \ \ 0 \\   U(1)^2 \times U(1)_B
&\ & 0, \ \ \ \ \ \ \ \ \ \ \ \ \ \ \ \ \ \ \ \ \ \ \ \ \ \ \ \ \ \
\ \ U(1)^2 \times U(1)_R \ \ \ \ \ \ \ -\frac{N^3-1}{2}  \nonumber
\\ \nonumber U(1)_R  &\ & -(2N^2+7N+1), \ \ \ \ \ \ \ \ \ \ U(1)_R^3
\ \ \ \ \ \ \ \ \ \ \ \ \ \ \ \ \ \ -(2N^2+N+1). \eeqa

We have verified that all three dual magnetic theories have the
same anomaly coefficients. Also it is easy to check that the real anomaly is
equal to zero in the electric and magnetic theories. Explicitly,
for the electric theory one has: $2N+2-\frac 12 8 - (2N-2) = 0.$

{\bf $SP \leftrightarrow SP$ groups duality.} Here we discuss the
duality of Sect. \ref{SpSpD}. In the electric theory, anomaly
coefficients for $SU(4) \times SP(2l_1) \times SP(2l_2) \times
\ldots \times SP(2l_K) \times U(1) \times U(1)_R$ global symmetry
group have the values \beqa SU(4)^3 &\ & -2M , \ \ \ \ \ \ \ \ \ \ \
\ \ \ \ \ \ \ \ \ \ \ \ \ SU(4)^2 \times U(1) \ \ \ \ -\frac{1}{2} M
(M-N-2) \nonumber \\ SP(2l_i)^2 \times U(1) &\ & -Mn_i, \ \ \ \ \ \
\ \ \ \ \ \ \ \ \ \ \ \ \ \ \ \
SU(4)^2 \times U(1)_R \ \ \ -2M \nonumber \\
SP(2l_i)^2 \times U(1)_R &\ & 0, \ \ \ \ \ \ \ \ \ \ \ \ \ \ \ \ \ \
\ \ \ \ \ \ \ \ \ \ \ U(1)_R \ \ \ \ \ \ \ \ \ \ \ \ \ \ \ \ \ 1-6M
\nonumber \\ U(1)_R^3  &\ & 1-6M \nonumber \nonumber \\
U(1)^2 \times U(1)_R &\ & \frac 12 \left(
-M^3+2NM^2-MN^2-4MN-2M+2\right) \eeqa coinciding with the
coefficients in the magnetic theory. Computation of the real gauge
anomaly coefficient yields: $-4-(2M-2)+2M+2 = 0.$

{\bf Multiple $SU(2N)$ duality.} The electric theory of Sect.
\ref{Mult2N} has the following anomaly coefficients for $SU(4)
\times SU(4) \times U(1)_1 \times U(1)_2 \times U(1)_B \times
U(1)_R$ global symmetry group:

\beqa SU(4)^3 \ && \  2N,\ \ \ \  \ \ \ \ \ \ \ \ \ \ \ \ \ \ \ \ \ \ \ SU(4)^2 \times U(1)_1 \ \ \ \ \ \ \ 0 \nonumber \\
SU(4)^2 \times U(1)_B \ && \  2N,\ \ \ \ \ \ \ \ \ \ \ \ \ \ \ \ \ \
\ \ \ \  \  SU(4)^2 \times U(1)_2 \ \ \ \ \ \ \ 4N(N-1) \nonumber \\
SU(4)^2 \times U(1)_R \ && \ -N,\
\ \ \ \ \ \ \ \ \ \ \ \ \ \ \ \ \ \ \ \ \ U(1)_1^2 \times U(1)_B \ \ \ \ \ \ \ \ \ 0 \nonumber \\
U(1)_1^2 \times U(1)_2 \ && \
-8N(2N-1),\ \ \ \ \ \ \ \ \ U(1)_1^2 \times U(1)_R \ \ \ \ \ \ \ -2N(2N-1) \nonumber \\
U(1)_B^2 \times
U(1)_1 \ && \  0,\ \ \ \ \ \ \ \ \ \ \ \ \ \ \ \ \ \ \ \ \ \ \ \ \ \ U(1)_B^2 \times U(1)_2 \ \ \ \ \ \ \ \ 32N(2N-1) \nonumber \\
U(1)_B^2 \times U(1)_R \ && \ -8N,\ \ \ \ \ \ \ \ \ \ \ \ \ \ \ \ \
\ \ \ \ U(1)_2^2 \times U(1)_1 \ \ \ \ \ \ \ \ 0 \nonumber
\\ U(1)_2^2 \times U(1)_B \ && \ 0,\ \ \ \ \ \ \ \ \ \ \ \ \ \ \ \ \ \ \ \ \ \ \ \ \ \
U(1)_2^2 \times U(1)_R \ \ \ \ \ \ \ -32N(N-1)^2 \nonumber \\
U(1)_R \ && \ -8N+4N^2-1, \ \ \ \ \ \ \ U(1)_R^3  \ \ \ \ \ \ \ \ \
\ \ \ \ \ \ \ \ -2N+4N^2-1\eeqa
which coincide with the anomaly coefficients
in all three dual magnetic theories.
Calculation of the real gauge anomaly yields:
$-\frac 12 8 -2 (2N-2) + 4N = 0. $

{\bf New confining duality.} The electric theory of Sect.
\ref{NewSConf} has the following anomaly coefficients for $SP(2N)
\times SU(N+3) \times U(1)_1 \times U(1)_2 \times U(1)_R$ global
symmetry group \beqa SU(N+3)^3 &\ & N+1,\ \ \ \ \ \ \ \ \ \ \ \ \ \
\ \ \ \ \ \ \ \ \ \ \  U(1)_R^3 \ \ \ \
\ \ \ \ \ \ \ \ \ \ \ \ \ \ \ \ -\frac 12 (N+2)(N+3)\nonumber \\
SU(N+3)^2 \times U(1)_R &\ & -(N+1),\ \ \ \ \ \ \ \ \ \ \ \ \ \ \ \
\ \ \ \
SP(2N)^2 \times U(1) \ \ \ \ \ \  -\frac{(N+1)(N+3)}{2} \nonumber \\
SU(N+3)^2 \times U(1) &\ & N+1,\ \ \ \ \ \ \ \ \ \ \ \ \ \ \ \ \ \ \
\ \ \ \ \ \       \nonumber \\
U(1)^2 \times U(1)_R &\ &  -\frac 12 (N+1)^2(N+2)(N+3) \nonumber \\
U(1)_R &\ & -\frac 12 (N+2)(N+3),\ \ \ \ \ \ \ \  SP(2N)^2 \times
U(1)_R \ \ \ \ \ \ 0\eeqa and the same picture holds for the
magnetic partner. Calculation of the real gauge anomaly yields:
$-(N+3)-(N-1)+2(N+1) \ = \ 0.$

{\bf $SU \leftrightarrow SP$ groups duality.} The anomaly matching
for the common global group $SU(N+3) \times SU(N+3) \times U(1)_B
\times U(1)_R$ of the duality described in Sect. \ref{Rains1D} is
checked and yields: \beqa SU(N+3)^3_L &\ & N+1,\ \ \ \ \ \ \ \ \ \ \
\ \ \ \ \ \ \ \ \ \ \ \ \ SU(N+3)^2_L \times U(1)_R \ \ \ \
-\frac{(N+1)^2}{N+3} \nonumber \\ SU(N+3)^2_L \times U(1)_B &\ &
2(N+1),\ \ \ \ \ \ \ \ \ \ \ \ \ \ \ \ \ \ \ \
U(1)_B^2 \times U(1)_R \ \ \ \ \ \ \ \ \ \ \ \ -8(N+1)^2 \nonumber \\
U(1)_R &\ & -(N^2+2N+2),\ \ \ \ \ \ \ \ \ \ U(1)_R^3 \ \ \
-\frac{N^4-9N^2-10N+2}{(N+3)^2}.\eeqa

{\bf $SU \leftrightarrow SU$ groups duality.}
Here we consider the dualities of Sect. \ref{Rains2D}.
The anomaly matching is checked for the  global group $SU(K)_L
\times SU(N+3-K)_L \times U(1)_1 \times SU(K)_R \times SU(N+3-K)_R
\times U(1)_2 \times U(1)_B \times U(1)_R$ yielding

\beqa SU(K)^3_L &\ & N+1,\ \ \ \ \ \ \ \ \ \ \ \ \ \ \ \ \ \ \
SU(K)^2_L \times U(1)_R \ \ \ \ \ \ -\frac{(N+1)^2}{N+3} \nonumber \\
SU(K)^2_L \times U(1)_B &\ & (N+1),\ \ \ \ \ \ \ \ \ \ \ \ \ \ \ \ \
SU(K)^2_L \times U(1)_1 \ \ \ \ \ \ (N+1)(N+3-K)  \nonumber \\
U(1)_B^2 \times U(1)_1 &\ & 0,\ \ \ \ \ \ \ \ \ \ \ \ \  \ \ \ \ \ \
\ \ \ \ \ \ \
U(1)_B^2 \times U(1)_R \ \ \ \ \ \ \ \  -2(N+1)^2 \nonumber \\
U(1)_1^2 \times U(1)_B &\ & (N+1)(N+3)K(N+3-K) \nonumber \\
U(1)_1^2 \times U(1)_R &\ & -(N+1)^2K(N+3-K) \\
U(1)_R &\ & -(N^2+2N+2) \ \ \ \ \ \ \ \ \
U(1)_R^3 \ \ \ \ \ \ \ \ \ \ -\frac{N^4-9N^2-10N+2}{(N+3)^2}.
 \nonumber \eeqa
Computation of the real gauge anomaly yields: $-2(N+1)+2(N+1) = 0.$

Comparing the 't Hooft anomaly matching conditions for all dualities
described in this paper and the analysis
of total ellipticity of the elliptic hypergeometric terms lying
behind the equalities of superconformal indices, we come to
the following conjecture.

\begin{conjecture}\label{conjecture2}
The condition of total ellipticity
for an elliptic hypergeometric term is necessary and sufficient
for validity of the 't Hooft anomaly matching conditions for dual superconformal
field theories whose superconformal indices are determined
by this term.
\end{conjecture}

For proving this hypothesis it is necessary to use the formal mathematical
definition of anomalies as cocycles of the gauge groups (see, e.g.,
\cite{RSF}). For dual
theories we have two, in general different, gauge groups. Therefore the
anomaly matching condition looks like an equality of Chern classes of
dual theories, and the conditions of total ellipticity---as a
condition of vanishing of the combined Chern classes. This
problem deserves a separate detailed discussion.

\section{Conclusion}

To summarize our results, on the mathematical side  we have conjectured
many new symmetry transformations for elliptic hypergeometric integrals
or exact evaluation formulas. On the physical side, we have found many
new Seiberg dualities. Sections 6, \ref{SpSpD}, 8,  and \ref{NewDual}
contain new electric-magnetic dualities for $\mathcal{N}=1$
SYM theories based on unitary and symplectic gauge groups with specific
matter content. Sections \ref{NewSConf} and \ref{NewConfSP} contain
new examples of $S$-confining theories derived from known identities
for elliptic hypergeometric integrals.

It should be clear that this paper does not contain a description of all known dual
superconformal field theories. We have limited ourselves only to {\em
simple} gauge groups $G=SU(N), SP(2N),$  and $G_2$. First, there are other
simple groups $G=SO(N), F_4, E_6, E_7,E_8$ consideration of which we
have skipped. The situation with the dualities for the
exceptional groups \cite{Distler:1996ub,Karch,Exc1} is not clear in general
(except of the $G_2$-cases described above) due to the complexity of the
invariants of these groups
\cite{Cho1,P1}. There are very many dualities involving orthogonal
groups $SO(N)$. Originally we hoped to tackle them as well, but
their amount is very big, and it was decided to consider them
separately. It is known that many group-theoretical objects for
the $SO(N)$ groups can be obtained as reductions of the $SP(2N)$-group
constructions. Some of such reductions were considered by Dolan and
Osborn at the level of superconformal indices \cite{Dolan}. However,
there are many dualities that they did not analyze. Many elliptic
hypergeometric integrals for the $B_N$ (i.e., $SO(2N+1)$ groups) and
$D_N$ (i.e., $SO(2N)$ groups) root systems can be obtained by
special restriction of the $BC_N$-integrals (cf. the forms of the
corresponding invariant measures given in Appendix B). However,
it is not clear at the moment whether superconformal indices of {\em
all} known $SO(N)$-group theories and their duals can be obtained in this way.
There are also other types of reduction of the indices and dualities,
e.g., those leading to dualities outside the
conformal windows \cite{SV_CW}.

Second, we have deliberately skipped consideration of the superconformal
indices for extended $\mathcal{N}>1$ supersymmetric field theories
\cite{BDHO,Kinney}. The best known examples correspond to the
Seiberg-Witten $\mathcal{N}=2$ theories \cite{SW,SW1}.
Consider the following electric and magnetic theories
\begin{center}
\begin{tabular}{|c|c|c|c|}
  \hline
      & $SO(3)$ & $SU(3)$ & $U(1)_R$ \\  \hline
  $Q$ & $f$ & $f$ & $\frac 23$ \\
\hline
\end{tabular}
\ \ \ \
\begin{tabular}{|c|c|c|c|}
  \hline
      & $SO(4)$ & $SU(3)$ & $U(1)_R$ \\  \hline
  $q$ & $f$ & $\overline{f}$ & $\frac 13$ \\
  $M$ & $1$ & $T_S$ & $\frac 43$ \\
\hline
\end{tabular}
\end{center}
As discussed by Intriligator and Seiberg \cite{IN4,IN3,IN2} (see also
\cite{gprr}), the $SO(3)$ Seiberg duality electric model becomes the
$SU(2)$ group $\mathcal{N}=4$ super-Yang-Mills theory in the infrared
region after introducing the tree level superpotential $W_{tree}\propto
\det Q$.  Superconformal indices have the form
\begin{equation}\label{E}
I_E \ = \ \frac{(p;p)_{\infty}(q;q)_{\infty}}{2} \prod_{j=1}^3
\Gamma((pq)^{1/3}s_j;p,q) \int_{\mathbb{T}} \frac{\prod_{j=1}^3
\Gamma((pq)^{1/3}s_jz^{\pm 1};p,q)}{\Gamma(z^{\pm1};p,q)}\frac{dz}{2
\pi \textup{i} z},
\end{equation}
where $\prod_{j=1}^3 s_j=1$, and
\begin{eqnarray}\label{M}
&& I_M \ = \ \frac{(p;p)^2_{\infty}(q;q)^2_{\infty}}{4} \prod_{1
\leq i < j \leq 3} \Gamma((pq)^{\frac 23} s_is_j;p,q) \prod_{i=1}^3
\Gamma((pq)^{\frac 23} s_i^2;p,q) \nonumber \\
&& \makebox[3em]{} \times \int_{\mathbb{T}^2}
\frac{\prod_{j=1}^2\prod_{i=1}^3 \Gamma((pq)^{1/6}s^{-1}_iz_j^{\pm
1};p,q)}{\Gamma(z_1^{\pm1}z_2^{\pm1};p,q)}\prod_{j=1}^2\frac{dz_j}{2 \pi \textup{i} z_j}.
\eeqa
By a change of integration variables $y_1 = \sqrt{z_1z_2}, \ y_2 = \sqrt{z_1/z_2}$
in $I_M$, one of the
integrations can be taken explicitly with the help of univariate elliptic beta
integral, which shows that (\ref{M}) is equal to (\ref{E}). This
equality can be obtained as a reduction of the $BC_N$-relations as
well \cite{Dolan}. We suppose therefore that it is necessary to
consider first all possible $SP(2N)$-group identities for integrals
and then try to reduce them to the relations for superconformal
indices of extended supersymmetric dual theories with $SO(N)$ gauge groups.

Third, we skipped the quiver gauge group cases, when there is more than one
simple gauge group (which corresponds  also to the deconfinement phenomenon
\cite{deconf}). Is is expected that equalities
of the superconformal indices for them are mere consequences of the
so-called Bailey-type chains (forming a tree) of symmetry transformations
for integrals discovered by the first author in \cite{spi:bailey2} and extended
in \cite{spi-war:inversions} to root systems. Within this context, the
duality transformation acquires a simple meaning of the integral
transform whose properties resemble the
classical Fourier transformation, see \cite{spi-war:inversions}.

Let us list some other possible applications of our results.
Counting of the gauge invariant operators for a number
of supersymmetric gauge theories was considered in detail in
\cite{Hanany1,Hanany2}. It is not difficult to see that the
corresponding generating functions are obtained from our
superconformal indices by taking the limits $p,q\to 0$. To take the
simplest possible limit $p\to 0$ one needs first to get rid of the balancing
conditions by multiplying a number of parameters by integer powers
of $p$ and applying the reflection formula for the elliptic gamma
function, see \cite{Spiridonov1}. However, in the present work we have a much
larger list of theories where this gauge invariant operators counting
technique is applicable (in particular, this concerns the theories
described in Sects. 7, 8, 9.2-9.6, 10.2-10.4, 11-13).
The limit $p\to 0$ in all these theories leads to
$q$-hypergeometric functions, the meaning of which is not clarified yet
from the superconformal index point of view.
The subsequent limit $q\to 0$ can be replaced by $q\to 1$
yielding the plain hypergeometric functions, which also should have thus some
topological meaning in gauge field theories. Similar clarification is needed
for the situations when the elliptic hypergeometric integrals are reduced
to terminating elliptic hypergeometric series by some special choices
of the parameters, or for the relations between integrals with different
powers of $p$  and $q$.

In \cite{Spiridonov3,spi:cont}, the first author has constructed univariate
biorthognal functions associated with the elliptic beta integral.
Naturally, it was conjectured there that some multivariable
biorthogonal functions exist for all known elliptic beta integrals
(which serve as the orthogonality measures). The first family of such functions was
constructed by Rains in \cite{Rains,rai:abelian}. As a consequence of our work,
the expected number of similar families of multivariable biorthogonal
functions has now increased essentially.

In  \cite{spi:thesis,spi:cs}, it was
shown that some of the $BC_N$ elliptic hypergeometric integrals can
be associated with the relativistic Calogero-Sutherland type models.
It was conjectured there that other models of such type can be built out of all
other existing elliptic beta integrals and their appropriate generalizations.
Because we have now the interpretation of the elliptic hypergeometric integrals
as superconformal indices of supersymmetric field theories, we
come to the natural conjecture that behind each $\mathcal{N}=1$ superconformal
field theory there is a Calogero-Sutherland type model for which
these integrals serve either as the topological indices or
the wave functions normalizations, respectively.
We would like to mention in this context the known
appearance of the usual elliptic Calogero-Sutherland models
within the $\mathcal{N}=2$ Seiberg-Witten theories \cite{NO}.

The group-theoretical interpretation of the elliptic hypergeometric
integrals discussed in \cite{Romelsberger2,Dolan,SV} and the present paper
opens possibilities for general structural theorems on the integrals
themselves. It may play a key role in the classification of
such integrals on root systems.
In particular, it naturally leads to the conjecture that there exist
infinitely (countably) many dualities and related elliptic
hypergeometric integral identities.
All the problems mentioned above deserve detailed
investigation either in relation to supersymmetric dualities or
on plain mathematical grounds. As to the proofs of many new hypergeometric
identities conjectured in this paper, we refer to known methods described in
\cite{die-spi:elliptic,die-spi:selberg,Rains,RS,Spiridonov2,Spiridonov3,
spi:bailey2,spi-war:inversions} (or indicated above in some cases)
which are available for their treatment. We plan to consider them case
by case depending on their tractability.

{\bf Acknowledgments.} We would like to thank A.~A.~Belavin, F.~A.~Dolan,
D.~I.~Kazakov, A.~Khmelnitsky, H.~Osborn, A.~F.~Oskin, V.~A.~Rubakov,
A.~Schwimmer, M.~A.~Shifman and S.~Theisen for valuable discussions,
comments and remarks.
The first author is indebted to L.~D.~Faddeev for discussions on
elliptic hypergeometric functions and their applications,
which inspired the choice of present paper's title
(it matches in spirit with \cite{vol:2005},
and it is expected that there exists a noncommutative extension of the
elliptic hypergeometry, formulation of which is a rather difficult task).
He is also grateful to A.~M.~Vershik for persistent support and
encouragement, which was the main driving force for writing the survey
\cite{Spiridonov1}. The second author would like to thank BLTP JINR for its
creative atmosphere due to which he was able to join this project.

The first author was partially supported by RFBR grants
no. 09-01-00271, 09-01-93107-NCNIL-a and, at the final stages of work,
by CERN TH (Geneva) and MPIM (Bonn). The second author was partially supported
by the Dynasty foundation, RFBR grant no. 08-02-00856-a and grant of
the Ministry of Education and Science of the Russian Federation no.
1027.2008.2.

\appendix
\section{Characters of representations of classical
groups}

Here we present general results for characters of the Lie group representations
used in the paper.
For the $SU(N)$ group  representations, the characters, depending on
$x=(x_1,\dots , x_N)$ subject to the constraint $\prod_{i=1}^N x_i=1$,
are the well known Schur polynomials
\begin{equation}
s_{\underline{\lambda}}(x) \ = \
s_{(\lambda_1,\ldots,\lambda_N)}(x) \ = \ \frac{det \left[
x_i^{\lambda_j+N-j} \right]}{det \left[ x_i^{N-j} \right]},
\end{equation}
where $\underline{\lambda}$ is the partition ordered so that
$\lambda_1 \ge \lambda_2 \ge \ldots \ge \lambda_N$.
They obey the property
${s}_{(\lambda_1,\ldots, \lambda_N)}(x) = {s}_{(\lambda_1+c,\dots,
\lambda_N+c)}(x)$, where $c \in \mathbb{Z}$. Therefore one can assume
that $\lambda_N=0$ without loss of generality.

Let us list explicitly the simplest characters. The fundamental and
antifundamental representations:
$$\chi_{SU(N),f}(x) \ = \ s_{(1,0,\ldots,0)}(x) \ = \ \sum_{i=1}^N x_i, \ \ \ \ \ \chi_{SU(N),\overline{f}} \ = \ s_{(1,\ldots,1,0)}(x) \ = \ \chi_{SU(N),f}(x^{-1}).$$
The adjoint representation:
$$\chi_{SU(N),adj}(x) \ = \ s_{(2,1,\ldots,1,0)}(x) \ = \ \sum_{1 \leq i,j \leq N} x_ix_j^{-1}-1.$$
The absolutely anti-symmetric tensor representation of rank two:
$$\chi_{SU(N),T_A}(x) \ = \ s_{(1,1,0,\ldots,0)}(x) \ = \ \sum_{1 \leq i < j \leq N} x_i x_j,\ \ \ \ \   \chi_{SU(N),\overline{T}_A} \ = \ \chi_{SU(N),T_A}(x^{-1}).$$
The symmetric representation:
$$\chi_{SU(N),T_S}(x) \ = \ s_{(2,0,\ldots,0)}(x) \ = \ \sum_{1 \leq i < j \leq N} x_i x_j + \sum_{i=1}^N x_i^2, \ \ \ \ \ \chi_{SU(N),\overline{T}_S}(x) \ = \  \chi_{SU(N),T_S}(x^{-1}).$$
The absolutely anti-symmetric tensor representation of rank three:
$$\chi_{SU(N),T_{3A}}(x) \ = \ s_{(1,1,1,0,\ldots,0)}(x) \ = \ \sum_{1 \leq i < j < k \leq N} x_i x_jx_k.$$
The absolutely symmetric tensor representation of rank three:
$$\chi_{SU(N),T_{3S}}(x) \ = \ s_{(3,0,\ldots,0)}(x) \ = \
\sum_{1 \leq i < j < k \leq N} x_i x_j x_k +
\sum_{i, j=1, i \neq j}^N x_i^2 x_j + \sum_{i=1}^N x_i^3.$$
In the mixed case, we have
$$
\chi_{SU(N),T_{AS}}(x) \ = \ s_{(2,1,0,\ldots,0)}(x) \ = \
2 \sum_{1 \leq i < j < k \leq N} x_i x_j x_k +
\sum_{i, j=1; i \neq j}^N x_i^2 x_j.$$

The Weyl characters for $SP(2N)$ group are given by the
determinant
\begin{equation}
\widetilde{s}_{(\lambda_1,\ldots,\lambda_N)}(x) \ = \ \frac{det
\left[ x_i^{\lambda_j+N-j+1} - x_i^{-\lambda_j-N+j-1} \right]}{det
\left[ x_i^{N-j+1} - x_i^{-N+j-1} \right]},
\end{equation}
with $\lambda_1 \ge \lambda_2 \ge \ldots \ge \lambda_N\ge0$.
For the fundamental and antifundamental representations
$$
\chi_{SP(2N),f}(x) \ = \chi_{SP(2N),\overline{f}}(x) \ =
\ \widetilde{s}_{(1,0,\ldots,0)}(x) \
= \ \sum_{i=1}^N (x_i + x_i^{-1}),$$
For the adjoint representation
\begin{eqnarray*}
&& \chi_{SP(2N),adj}(x) = \widetilde{s}_{(2,0,\ldots,0)}(x)
\\ && \makebox[2em]{}
= \sum_{1 \leq i < j \leq N} (x_ix_j + x_ix_j^{-1} + x_i^{-1}x_j + x_i^{-1}x_j^{-1})
+ \sum_{i=1}^N(x_i^2+x_i^{-2}) + N.
\end{eqnarray*}
For the absolutely anti-symmetric representation
$$
\chi_{SP(2N),T_A}(x) = \widetilde{s}_{(1,1,0,\ldots,0)}(x)
= \sum_{1 \leq i < j \leq N}(x_ix_j + x_ix_j^{-1}
+ x_i^{-1}x_j + x_i^{-1}x_j^{-1}) + N-1.
$$

As to the exceptional group $G_2$, its fundamental representation has the character
$$\chi(z_1,z_2,z_3) = 1 + \sum_{i=1}^3 \left( z_i + z_i^{-1} \right),$$
where $z_1z_2z_3 = 1.$
The character for the adjoint representation of $G_2$ group is
$$\chi(z_1,z_2,z_3) = 2 + \sum_{1 \leq i < j \leq 3}
\left( z_iz_j + z_i^{-1}z_j + z_iz_j^{-1} + z_i^{-1}z_j^{-1} \right).$$

\section{Invariant matrix group measures}

Here we describe the invariant measures for
integrals over classical Lie groups and the exceptional
group $G_2$. Such a measure for the unitary group $SU(N)$
with any symmetric function $f(z)$, where
$z=(z_1,\ldots,z_N), \ \prod_{j=1}^Nz_j=1$, has the form
\begin{equation}\label{invSU}
\int_{SU(N)} f(z) d\mu(z) \ = \   \frac{1}{N!} \int_{\mathbb{T}^{N-1}}
\Delta(z) \Delta(z^{-1}) f(z) \prod_{j=1}^{N-1} \frac{dz_j}{2 \pi \textup{i}
z_j},
\end{equation}
where $\Delta(z)$ is the Vandermonde determinant
$$\Delta(z) \ = \ \prod_{1 \leq i < j \leq N} (z_i-z_j).$$

The invariant measure for the symplectic group $SP(2N)$ with
any symmetric function $f(z),\ z=(z_1,\ldots,z_N)$, has the form
\begin{equation}\label{invSP}
\int_{SP(2N)} f(z) d\mu(z) \ = \ \frac{(-1)^N}{2^NN!}
\int_{\mathbb{T}^{N}} \prod_{j=1}^N (z_j-z_j^{-1})^2
\Delta(z+z^{-1})^2 f(z) \prod_{j=1}^{N} \frac{dz_j}{2 \pi \textup{i} z_j},
\end{equation}

For the invariant measures over the orthogonal group $SO(N)$ and
any symmetric function $f(z),\ z=(z_1,\ldots,z_N)$, one has to
distinguish the cases of odd and even $N$:
\begin{equation}\label{invSOeven}
\int_{SO(2N)} f(z) d\mu(z) \ = \ \frac{1}{2^{N-1}N!}
\int_{\mathbb{T}^{N}} \Delta(z+z^{-1})^2 f(z)
\prod_{j=1}^{N} \frac{dz_j}{2 \pi \textup{i} z_j},
\end{equation}
and
\begin{equation}\label{invSOodd}
\int_{SO(2N+1)} f(z) d\mu(z) \ = \ \frac{(-1)^N}{2^{N}N!}
\int_{\mathbb{T}^{N}} \prod_{j=1}^N \left( z_j^{\frac
12} - z_j^{-\frac 12} \right)^2 \Delta(z+z^{-1})^2 f(z)
\prod_{j=1}^{N} \frac{dz_j}{2 \pi \textup{i} z_j},
\end{equation}

The invariant measure for the exceptional group $G_2$ and any
symmetric function $f(z),\ z=(z_1,z_2,z_3)$, where $z_1z_2z_3=1$, has the form
\begin{equation}\label{invG2}
\int_{G_2}  f(z)d\mu(z) \ = \ \frac{1}{2^23} \int_{\mathbb{T}^2}
\Delta(z+z^{-1})^2 f(z) \prod_{j=1}^{2} \frac{dz_j}{2 \pi \textup{i} z_j}.
\end{equation}

\section{Relevant Casimir operators}

Commutators of the generators $T^a$, $a=1,\dots, \dim\, G,$
of some classical Lie group $G$
are defined with the help of structure constants $f^{abc}$
\begin{equation}
[T^a,T^b] \ = \ i f^{abc}T^c.
\end{equation}

It is straightforward to obtain the Casimir operators \cite{Terning}
\begin{eqnarray}
\sum_{a,l}(T_{\textbf{r}}^a)^m_l (T_{\textbf{r}}^a)^l_n
= C_2(\textbf{r}) \delta^m_n, \qquad
\sum_{n,m}(T_{\textbf{r}}^a)^m_n (T_{\textbf{r}}^b)^n_m = T(\textbf{r})
\delta^{ab},
\end{eqnarray}
where $\textbf{r}$ is some irreducible representation. These Casimir
operators and the dimension of the representation $d(\textbf{r})$
are connected through the adjoint representation $\textbf{adj}$,
\begin{equation}
d(\textbf{r})C_2(\textbf{r}) \ = \ d(\textbf{adj}) T(\textbf{r}).
\end{equation}

For checking the 't Hooft anomaly matching conditions we need the
triple Casimir operator which comes from the trace
\begin{equation}
A^{abc} \ = \ Tr[T^a\{T^b,T^c\}].
\end{equation}
Then it is convenient to define the operator $A(\textbf{r})$
relative to the fundamental representation
\begin{equation}
A^{abc}(\textbf{r}) \ = \ A(\textbf{r}) \textbf{A}^{abc},
\end{equation}
where $\textbf{A}^{abc}=Tr[T_F^a\{T_F^b,T_F^c\}]$ and $T_F^a$ are
the generators in the fundamental representation.

In the tables below we give the dimensions $d(\textbf{r})$, the
Casimir operators $2T(\textbf{r})$, and $A(\textbf{r})$ for the unitary
group and the dimensions $d(\textbf{r})$,
the Casimir operators $T(\textbf{r})$ for symplectic and $G_2$ groups.
Note that in the verification of the anomaly matchings for unitary
groups we use $2T(\textbf{r})$.
\begin{center}
$SU(N)$ group:
\begin{tabular}{|c|c|c|c|} \hline
  Irrep  $\textbf{r}$ & $d(\textbf{r})$ & $2T(\textbf{r})$ & $A(\textbf{r})$ \\ \hline
  $f$ & $N$ & 1 & 1 \\
  $\textbf{adj}$ & $N^2-1$ & $2N$ & 0 \\
  $T_A$ & $\frac 12 N(N-1)$ & $N-2$ & $N-4$ \\
  $T_S$ & $\frac 12 N(N+1)$ & $N+2$ & $N+4$ \\
  $T_{3A}$ & $\frac 16 N(N-1)(N-2)$ & $\frac 12 (N-3)(N-2)$ & $\frac 12 (N-6)(N-3)$ \\
  $T_{3S}$ & $\frac 16 N(N+1)(N+2)$ & $\frac 12 (N+3)(N+2)$ & $\frac 12 (N+6)(N+3)$ \\
  $T_{AS}$ & $\frac 13 N(N-1)(N+1)$ & $N^2-3$ & $N^2-9$ \\
\hline
\end{tabular}
\end{center}

\begin{center}
$SP(2N)$ group:
\begin{tabular}{|c|c|c|} \hline
  Irrep $\textbf{r}$ & d(\textbf{r}) &  T(\textbf{r}) \\ \hline
  $f$ & $2N$ & 1 \\
  $\textbf{adj}=T_S$ & $N(2N+1)$ & $2N+2$ \\
  $T_A$ & $N(2N-1)-1$ & $2N-2$ \\
\hline
\end{tabular}
\end{center}

\begin{center}
$G_2$ group:
\begin{tabular}{|c|c|c|} \hline
  Irrep $\textbf{r}$ & d(\textbf{r}) &  T(\textbf{r}) \\ \hline
  $f$ & $7$ & 2 \\
  $\textbf{adj}$ & $14$ & 8 \\
\hline
\end{tabular}
\end{center}

\section{Total ellipticity for the KS duality indices}

In order to illustrate the work hidden behind our conjectures,
we briefly describe in this Appendix verification of the total ellipticity for the
transformation identity for elliptic hypergeometric integrals
associated with the Kutasov-Schwimmer duality from Sect. \ref{KSel}.

First, we change the integration variables $\z$ in
(\ref{intKSelM}) to $\z = U^{-1} (ST)^{-\frac{K}{2\widetilde{N}}} \y$
 and  assume that the contours of integration in $y$-variables can be
deformed back to $\mathbb{T}$ without crossing the poles.
Then the equality of integrals (\ref{intKSelE}) and (\ref{intKSelM})
is rewritten in the following form
\begin{eqnarray}\label{SUKSIntn}
&& \kappa_N \int_{\mathbb{T}^{N-1}} \Delta_E(\z,\t,\s)
\prod_{j=1}^{N-1} \frac{d z_j}{2 \pi \textup{i} z_j} = \kappa_{\widetilde{N}}
\int_{\mathbb{T}^{\widetilde{N}-1}} \Delta_M(\y,\t,\s)
\prod_{j=1}^{\widetilde{N}-1} \frac{d y_j}{2 \pi \textup{i} y_j},
\end{eqnarray}
where $\widetilde{N} \ = \ K N_f - N$ and
$$
\kappa_N \ = \ \frac{(p;p)_{\infty}^{N-1}
(q;q)_{\infty}^{N-1}}{N!} \Gamma(U;p,q)^{N-1}
$$
with $U=(pq)^{\frac{1}{K+1}}$.
The kernels of the elliptic hypergeometric integrals are
\begin{eqnarray} &&
\Delta_E(\z,\t,\s) =
\prod_{1 \leq i < j \leq N} \frac{
\Gamma(Uz_iz_j^{-1},Uz_i^{-1}z_j;p,q)}{
\Gamma(z_iz_j^{-1},z_i^{-1}z_j;p,q)} \prod_{i=1}^{N_f} \prod_{j=1}^N \Gamma(s_iz_j,t_i^{-1}z_j^{-1};p,q), \\
&& \Delta_M(\y,\t,\s)= \prod_{l=1}^K \prod_{i,j=1}^{N_f}
\Gamma(U^{l-1}s_it_j^{-1};p,q) \prod_{1 \leq i < j \leq
\widetilde{N}} \frac{ \Gamma(Uy_iy_j^{-1},Uy_i^{-1}y_j;p,q)}{
\Gamma(y_iy_j^{-1},y_i^{-1}y_j;p,q)} \nonumber \\
&&  \makebox[6em]{} \times \prod_{i=1}^{N_f}
\prod_{j=1}^{\widetilde{N}} \Gamma(s_i^{-1}y_j,U^{2}
t_iy_j^{-1};p,q), \nonumber
\end{eqnarray}
where $\prod_{i=1}^N z_i=1$, $\prod_{i=1}^{\widetilde{N}}y_i=U^{\widetilde{N}+NK}
(pq)^{-\frac{1}{2}N_fK} S^K$, and $  U^{2N}ST^{-1} =(pq)^{N_f}.$

\begin{theorem} The function
$$
\rho(\z,\y,\t,\s) \ = \
\frac{\Delta_E(\z,\t,\s)}{\Delta_M(\y,\t,\s)}
$$
is a totally elliptic hypergeometric term.
\end{theorem}

{\bf Ellipticity of the $z$-variables $q$-certificates.}
As described in Sect. 2, we should consider the ratios
\begin{eqnarray}\label{KSz0}
&& h_{\I}^z(\z,\y,\t,\s,q) = \frac{\rho(\z,\y,\t,\s)|_{ z_{\I}
\rightarrow qz_{\I}, z_N \rightarrow q^{-1} z_N}}{\rho(\z,\y,\t,\s)} \\
\nonumber && \makebox[2em]{} = \prod_{j=1, j \neq {\I}}^{N-1}
\frac{\T(Uz_{\I}z_j^{-1},Uz_jz_N^{-1},q^{-1}z_{\I}^{-1}z_j,q^{-1}z_j^{-1}z_N;p)}{\T(Uq^{-1}z_{\I}^{-1}z_j,Uq^{-1}z_j^{-1}z_N,z_{\I}z_j^{-1},z_jz_N^{-1};p)}
\\ \nonumber && \makebox[4em]{} \times
\frac{\T(Uqz_{\I}z_N^{-1},Uz_{\I}z_N^{-1},q^{-2}z_{\I}^{-1}z_N,q^{-1}z_{\I}^{-1}z_N;p)}{\T(Uq^{-1}z_{\I}^{-1}z_N,Uq^{-2}z_{\I}^{-1}z_N,qz_{\I}z_N^{-1},z_{\I}z_N^{-1};p)}
\prod_{i=1}^{N_f} \frac{\T(s_iz_{\I};p)}{\T(q^{-1}s_iz_N;p)}
\frac{\T( t_i^{-1}z_N^{-1};p)}{\T( q^{-1}t_i^{-1}z_{\I}^{-1};p)},
\end{eqnarray}
where $ a=1,\dots, N-1$,
and check that these are totally elliptic functions.
Indeed, $h_{\I}^z(\z,\y,\t,\s,q)$ functions are automatically invariant
under the transformations
1)  $s_{\J} \rightarrow ps_{\J}, s_{N_f} \rightarrow p^{-1}
s_{N_f}$, 2) $t_{\J} \rightarrow pt_{\J}, t_{N_f}
\rightarrow p^{-1} t_{N_f},$ 3) $y_{\J} \rightarrow
py_{\J}, y_{\widetilde{N}}\rightarrow p^{-1} y_{\widetilde{N}}$.
Whereas the invariance with
respect to the substitutions 4) $z_{\K} \rightarrow pz_{\K}, z_N
\rightarrow p^{-1} z_N$ for $\K \neq\I$ or 5) $z_{\I} \rightarrow
pz_{\I}, z_N \rightarrow p^{-1} z_N$ uses the balancing condition.
Similarly, one checks the invariance with respect to the mixed
transformations $s_{\J}
\rightarrow p s_{\J}, t_{\K} \rightarrow p t_{\K}$
and $y_{\L} \rightarrow p^K y_{\L}$.

The most complicated part of the work consists in establishing
ellipticity in the variable $q$. The difficulty comes from
the presence of fractional powers of $q$ entering (\ref{KSz0})
through the variable $U$. Because of that
one should scale $q$ by such a power of $p$ that there will be
a match with the period of the elliptic functions $h_{\I}^z(\z,\y,\t,\s,q)$.
Simultaneously, we should preserve the balancing condition and all
other constraints on the parameters we have.
This is reached by the following transformation of the parameters
\begin{eqnarray}
6) \ \  q \rightarrow p^{K+1}q, \ \ \
t_{N_f}^{-1} \rightarrow p^{(K+1)N_f-2N} t_{N_f}^{-1}, \ \ \
  y_{\widetilde{N}} \rightarrow
p^{\widetilde{N}+NK-N_fK(K+1)/2}y_{\widetilde{N}},
\eeqa
which leads to $U \rightarrow pU$, as required.
It is a matter of a neat computation (at the intermediate steps there
appears a very cumbersome expression) to show that
$h_{\I}^z(\z,\y,\t,\s,q)$ do not change under these substitutions.

{\bf Ellipticity of the $y$-variables $q$-certificates.} Now we
consider the ratios
\begin{eqnarray}
&& h_{\I}^y(\z,\y,\t,\s,q) = \frac{\rho(\z,\y,\t,\s)|_{ y_{\I}
\rightarrow qy_{\I}, y_{\widetilde{N}} \rightarrow q^{-1} y_{\widetilde{N}}}}{\rho(\z,\y,\t,\s)} \\
\nonumber && \makebox[2em]{} = \prod_{j=1, j \neq
{\I}}^{\widetilde{N}-1} \frac{
\T(Uq^{-1}y_{\I}^{-1}y_j,Uq^{-1}y_j^{-1}y_{\widetilde{N}},y_{\I}y_j^{-1},y_jy_{\widetilde{N}}^{-1};p)}{
\T(Uy_{\I}y_j^{-1},Uy_jy_{\widetilde{N}}^{-1},q^{-1}y_{\I}^{-1}y_j,q^{-1}y_j^{-1}y_{\widetilde{N}};p)}
\\ \nonumber && \makebox[4em]{} \times
\frac{
\T(Uq^{-1}y_{\I}^{-1}y_{\widetilde{N}},Uq^{-2}y_{\I}^{-1}y_{\widetilde{N}},qy_{\I}y_{\widetilde{N}}^{-1},y_{\I}y_{\widetilde{N}}^{-1};p)}{
\T(Uqy_{\I}y_{\widetilde{N}}^{-1},Uy_{\I}y_{\widetilde{N}}^{-1},q^{-2}y_{\I}^{-1}y_{\widetilde{N}},q^{-1}y_{\I}^{-1}y_{\widetilde{N}};p)}
\prod_{i=1}^{N_f}
\frac{\T(q^{-1}s_i^{-1}y_{\widetilde{N}};p)}{\T(s_i^{-1}y_{\I};p)}
\frac{\T(U^2q^{-1}t_iy_{\I}^{-1};p)}{\T(U^2t_iy_{\widetilde{N}}^{-1};p)},
\end{eqnarray}
where  $ a=1,\dots, \widetilde{N}-1$.
Again, these are the totally elliptic functions.
They are automatically invariant under the transformations
1) $s_{\J} \rightarrow ps_{\J}, s_{N_f} \rightarrow p^{-1} s_{N_f}$,
2) $t_{\J} \rightarrow pt_{\J}, t_{N_f} \rightarrow p^{-1}t_{N_f}$,
3) $z_{\J} \rightarrow pz_{\J}, z_{N} \rightarrow p^{-1} z_{N}$.
The invariance with respect to the substitutions 4) $y_{\J} \rightarrow py_{\J},
y_{\widetilde{N}} \rightarrow p^{-1} y_{\widetilde{N}}$, $\J \neq\I$,
or 5) $y_{\I} \rightarrow py_{\I},
y_{\widetilde{N}} \rightarrow p^{-1}
y_{\widetilde{N}}$ uses the balancing condition.
The most difficult part is the verification of the invariance
with respect to the transformations
$$
6) \ \  q \rightarrow p^{K+1}q, \ \ \  U \rightarrow pU, \ \ \
t_{N_f}^{-1} \rightarrow p^{(K+1)N_f-2N} t_{N_f}^{-1}, \ \ \
 y_{\widetilde{N}} \rightarrow
p^{\widetilde{N}+NK-N_fK(K+1)/2}y_{\widetilde{N}}.
$$

{\bf Ellipticity of the $t$-parameters $q$-certificates.}
Now we need to investigate the functions
\begin{eqnarray}
&& h_{\I}^t(\z,\y,\t,\s,q) = \frac{\rho(\z,\y,\t,\s)|_{ t_{\I}
\rightarrow qt_{\I}, t_{N_f} \rightarrow q^{-1} t_{N_f}}}{\rho(\z,\y,\t,\s)} \\
\nonumber && \makebox[2em]{} = \prod_{l=1}^K \prod_{i=1}^{N_f}
\frac{\T(U^{l-1}q^{-1}s_it_{\I}^{-1};p)}{\T(U^{l-1}s_it_{N_f}^{-1};p)}
\prod_{j=1}^N
\frac{\T(t_{N_f}^{-1}z_j^{-1};p)}{\T(q^{-1}t_{\I}^{-1}z_j^{-1};p)}
\prod_{i=1}^{\widetilde{N}}
\frac{\T(U^{2}q^{-1}t_{N_f}y_j^{-1};p)}{\T(U^{2}t_{\I}y_j^{-1};p)},
\end{eqnarray}
where  $ a=1,\dots, N_f-1$,
and show that they are totally elliptic. Again, invariance under
1) $y_{\J} \rightarrow py_{\J},
y_{\widetilde{N}} \rightarrow p^{-1}
y_{\widetilde{N}}$,
2) $s_{\J} \rightarrow ps_{\J},
s_{N_f} \rightarrow p^{-1} s_{N_f}$, and
3) $z_{\J} \rightarrow
pz_{\J}, z_{N} \rightarrow p^{-1} z_{N}$ transformations is automatic.
The balancing condition is needed for symmetries 4) $t_{\K}
\rightarrow pt_{\K}, t_{N_f} \rightarrow p^{-1} t_{N_f}$, $\K\neq\I$,
and 5) $t_{\I} \rightarrow pt_{\I}, t_{N_f} \rightarrow p^{-1} t_{N_f}$
Computations during the verification of invariance under the transformations
$$
6) \ \  q \rightarrow p^{K+1}q, \ \ \  U \rightarrow pU, \ \ \
t_{N_f-1}^{-1} \rightarrow p^{(K+1)N_f-2N}
t_{N_f-1}^{-1},\ \ \  y_{\widetilde{N}} \rightarrow
p^{\widetilde{N}+NK-N_fK(K+1)/2}y_{\widetilde{N}}
$$
are very lengthy and require a lot of attention to reach the
needed statement. Consideration of the  $s$-parameters certificates
is equivalent to the $t$-variables case because of the symmetries of
the initial integral kernels.

\end{document}